\documentclass[12pt,a4paper, captions=figureheading]{scrartcl}
\usepackage[usenames,dvipsnames,table,xcdraw]{xcolor}
\usepackage[utf8]{inputenc}
\usepackage[T1]{fontenc}
\usepackage{lmodern}
\usepackage{adjustbox}
\usepackage{tabularx}
\usepackage{booktabs}
\usepackage[normalem]{ulem}
\useunder{\uline}{\ul}{}
\usepackage[affil-it]{authblk}
\usepackage{pdfpages}
\setkomafont{disposition}{\normalcolor\bfseries\rmfamily}

\usepackage[style=authoryear, citetracker=true, maxcitenames=2, hyperref=true, url=true, isbn=false, doi=true, natbib=true, 
bibstyle=authoryear, maxbibnames=30, sorting=nyt, uniquename=full, uniquelist=false]{biblatex}

\usepackage{csquotes, color, amsmath}
\usepackage[titletoc,toc,title]{appendix}
\usepackage[colorlinks=true, allcolors=blue]{hyperref}
\usepackage{setspace, float, placeins, verbatim, subcaption, ragged2e}
\usepackage{gensymb}
\usepackage{dsfont}
\usepackage{subcaption}
\usepackage{pdflscape}
\usepackage{tocloft}
\makeatletter
\renewcommand{\numberline}[1]{{\@cftbsnum #1\@cftasnum~}\@cftasnumb}
\makeatother
\setkomafont{caption}{\small}
\setkomafont{captionlabel}{\usekomafont{caption}}

\newcommand{\KH}[1]{\textcolor{black}{#1}}

\definecolor{bronze}{rgb}{0.8,0.5,0.2}

\makeatletter
\renewenvironment{quotation}
               {\list{}{\listparindent 1.5em%
                        \rightmargin   \leftmargin
                        \parsep        \z@ \@plus\p@}%
                \item\relax}
               {\endlist}
\makeatother

\usepackage{threeparttable}
\makeatletter
\protected\def\ccell#1#{%
  \kern-\fboxsep
  \@ccell{#1}%
}
\def\@ccell#1#2#3{%
  \colorbox#1{#2}{#3}%
  \kern-\fboxsep
}
\makeatother

\title{Measuring artificial intelligence: a systematic assessment and implications for governance}

\author{Kerstin H\"otte,$^{1}$\footnote{Corresponding author: kerstin.hotte@kedgebs.com\\ORCiD: KH: 0000-0002-8633-4225; VV: 0000-0001-7979-3450} 
Taheya Tarannum,$^{2}$ Vilhelm Verendel,$^{3,4}$ Lauren Bennett$^{5}$
\\
\scriptsize{$^{1}$ Kedge Business School, Paris, FR}\\

\scriptsize{$^{2}$ Oxford Martin Programme on Technological and Economic Change, 
University of Oxford, UK}\\
\scriptsize{$^{3}$ Institute for New Economic Thinking at the Oxford Martin School, University of Oxford, UK}\\
\scriptsize{$^4$} Department of Physics, Chalmers University of Technology, SE\\
\scriptsize{$^5$} The Leverhulme Doctoral Training Programme for the Ecological Study of the Brain, University College London, UK
}

\date{\normalsize\today}
\begin{document}
\maketitle
\begin{abstract}
Governing artificial intelligence (AI) inventions is a major policy concern, yet definitions and measurement remain contested. We compare four patent-based approaches reflecting distinct understandings of AI. Using US patents (1990-2019), we assess the degree to which each approach classifies AI as a general-purpose technology (GPT) and examine patent concentration--two central policy--relevant dimensions. The approaches overlap in just 1.37\% of patents, defining between 3–17\% of all US patents in 2019 as AI. All approaches confirm AI’s GPT characteristics, with the smallest keyword-based set exhibiting the highest growth and generality. High GPTness indicates public good characteristics, justifying public support. Across methods, AI patents concentrate among a few firms, highlighting market power and regulatory challenges. Policy implementation, thus, requires careful consideration of multiple classification methods to ensure robust, inclusive, and effective AI governance.
 
\bigskip
\footnotesize
 \noindent \textbf{JEL Classifications:} O31, O33, O34\\  
 \noindent \textbf{Keywords:} Artificial Intelligence, Governance, General-Purpose Technology, Concentration, Patent, Classification

\end{abstract}

\newpage
\section{Introduction}
Artificial intelligence (AI) governance is a significant policy concern, reflecting the aim to mitigate risks and maximize societal benefits \citep{jelinek2021policy, mazzucato2022governing, schmitt2022mapping}.
Governance requires a clear identification of AI, but so far a consensual definition is lacking \citep{krafft2020defining}. A wide audience is interested in understanding recent trends in AI development but the lack of clear identifying methods can hamper assessments of AI impacts and the effectiveness of regulatory policy \citep{dafoe2018ai}. Ambiguities in defining and measuring AI contribute to empirical controversies concerning its impacts on labour markets, technological leadership, and productivity \citep{cazzaniga2024gen, brynjolfsson2021productivity, bresnahan2023innovation, alderucci2020quantifying, babina2022artificial}.
In this paper, we use patents as a quantitative and qualitative record of AI inventions. Comparing differences across four different AI classification methods, we investigate how empirical conclusions about the characteristics and concentration of AI can be sensitive to the method. 

AI is widely seen as a general-purpose technology (GPT), expected to profoundly transform production processes, labour markets, and economic leadership at national and global levels \citep{agrawal2019economic,cockburn20194, brynjolfsson2021productivity, cockburn2018impact, valdes2017patents, webb2019impact, alderucci2020quantifying}. 
Testing these predictions requires robust empirical measurement; yet ex-ante measurement of radical novelty remains challenging \citep{schumpeter2005joseph}. AI is still widely regarded as being in the early stages of development \citep{brynjolfsson2021productivity}, and its long-run effects have not yet manifested and may be shaped by decisions undertaken today \citep{jacobides2021evolutionary, petit2017antitrust, fanti2022heron}.

This paper does not aim to identify the best AI definition, but rather examines agreements and differences across measurement methods and their implications for policy and research. 
We do so by comparing four samples of AI patents, each reflecting a different way to understand AI. The samples are identified by: 
\begin{enumerate}
    \item keywords focusing on trends in neural networks, robotics, and natural language processing (NLP); 
    \item scientific citations reflecting the academic origins of AI; 
    \item the World Intellectual Property Organization (WIPO) classification method, accounting for both the hardware and software dimensions of AI; and 
    \item the United States Patent and Trademark Office (USPTO) approach capturing the widespread use of AI in other inventions. 
\end{enumerate}
In the union of all samples, we identify 732k AI patents from 1990 to 2019. The individual samples vary greatly by scale: 54k patents are captured by the Keyword, 178k by the Science, 159k by the WIPO, and 595k by the USPTO approach. Strikingly, all four methods agree on only 1.37\% of AI patents, with pairwise overlaps of 10-20\% or less throughout the entire period. Further, the four approaches reflect disparate time trends, with the Science and USPTO sample showing an AI slowdown in recent years, while Keyword and WIPO patents tell a story of accelerating growth since the 2000s.

We evaluate whether the four samples reflect the beliefs of AI being a GPT, by assessing three GPT characteristics established in the literature \citep{bresnahan1995general, hall2006uncovering, petralia2020mapping}: 
\begin{enumerate}
    \item Growth: GPTs are engines of growth with continued technological improvements \citep{petralia2020mapping}. We measure this feature by the growth rates of each AI sample, and patents that rely on them, as indicated by citations.
    \item Generality: GPTs can be used across a wide range of products and processes. We examine the technological diversity of patent citations. As GPTs often experience long delays before being widely taken up \citep{comin2018if}, we measure citation lags between AI and subsequent inventions \citep{hall2006uncovering}. 
    \item Complementarity: GPTs complement technologies in many fields \citep{petralia2020mapping}, being reflected in a high technological diversity. We quantify the diversity of co-classifications of AI patents across technology groups. 
\end{enumerate}
Evaluating the ``GPTness'' of AI is relevant because it indicates the potential for continued macroeconomic growth with large social benefits in the long run \citep{lipsey2005economic}, and the public good characteristics of GPTs can justify public support \citep{bresnahan1995general}. 

Whether AI is a GPT or not can not be taken as empirically given: If AI inventions have little or no GPT characteristics, arguments claiming that AI needs public support would be undermined. However, an observed lack of GPTness could also be a matter of classification or barriers to the realisation of AI benefits, which do not necessarily arise by themselves: commercial actors may maximise the private rather than the social value of an invention, leading to a pre-mature lock-in to an inferior pathway of AI development \citep{klinger2022narrowing, bresnahan2023innovation}.
This consideration motivates an additional analysis of the key actors in AI development and the concentration of inventive activities. 

Our results indicate that all methods classify AI as a GPT, yet the AI inventions identified by Keywords and the WIPO method show the highest levels of GPTness. 
Inventive activities by firms appear least concentrated in technology areas identified by Keywords. 
Relying on GPTness and diversity in the market of AI development, one may conclude that AI as captured by Keywords bear the greatest potential for public benefits, compared to the other methods. This could guide policy aimed at supporting AI inventions for the public good.  

Our patent-based approach captures only a segment of the AI ecosystem and evaluates a narrow set of policy-relevant criteria \citep{jacobides2021evolutionary, roche2023ethics}. 
\KH{Still, within this scope, we find concentration levels among the Big Tech companies to be highest in electronics and computing, while invention in AI edge applications appears to be more widely spread.}
Attempts to empirically assess the impact of AI and policy should therefore rely on a variety of methods to identify AI before drawing conclusions. 
Our systematic comparison can guide methodological choices in patent-based research by highlighting the implications and conceptual considerations when measuring AI. 
Further, our results inform discussions on AI definitions by showing consensual features of AI and quantifying nuanced differences in the types of industries, technology fields, and key players involved in AI development.  

The relevance of this research ranges from researchers (who reflect on AI definitions or rely on classifications in applied research) to policymakers (who rely on AI definitions and empirical insights on AI impacts in everyday policy). A deeper reflection on AI definitions and their implementation may help clarify some of the empirical controversies about AI impacts on society.

The remainder of our paper is structured as follows: Section \ref{sec:lit} introduces GPTs and AI in patent data;  Section \ref{sec:approach} describes the methods, followed by the results (Section \ref{sec:results}). In Section \ref{sec:discussion}, we discuss the findings, and Section \ref{sec:conclusions} concludes. 
\section{Background}
\label{sec:lit}
 Here, we review (1) the arguments for AI being a GPT (Section \ref{subsec:Literature_GPT}), (2) the history of AI and its definitions (Section \ref{subsec:Literature_AI_history}) and (3) patents as a data source (Section \ref{subsec:Literature_patent_data}).

\subsection{What is a GPT, and does it matter for AI?}
\label{subsec:Literature_GPT}
\citet{bresnahan1995general} define GPTs as technologies that pervade the economy and spur inventions via complementarities. Well-established GPTs, such as electricity and Information and Communication Technologies (ICTs), are widely considered as long-term drivers of societal value, growth and technological progress. Quantitative research identifies GPTs by three characteristics: rapid intrinsic improvement (growth), economic pervasiveness (generality), and productivity spillovers across sectors (complementarity) \citep{bresnahan1995general, lipsey2005economic}.

The first criterion refers to GPTs' inherent capacity to rapidly improve. If the technology is sufficiently mature to be valuable for many uses, this should be reflected in high growth rates. 
The second criterion refers to GPTs' ability to engender new methods of production or innovation. Due to their pervasiveness, GPTs inspire a wave of technological inventions as they embody new universal tools for production and research. Other impactful technologies, such as nuclear power and fMRI, lack the generality required to pervade a significant number of sectors \citep{agrawal2018finding, brynjolfsson2019artificial}. 
The third criterion highlights how GPTs generate productivity spillovers by complementing existing products and processes. This often comes with creative destruction and technological discontinuities rendering established technologies and jobs obsolete \citep[][]{lipsey2005economic}. 

Together, these criteria require longer time for GPTs to evolve, diffuse and realise their full economic impact \citep{lipsey2005economic}. Once the inventions have evolved and spread sufficiently, GPTs depend on complementary infrastructure and secondary innovations to transform organizations and reshape production processes across the economy \citep{bresnahan2023innovation}. 

While much of the literature frames AI as a GPT \citep[e.g.][]{agrawal2018finding, cockburn2018impact, brynjolfsson2021productivity, cockburn20194, valdes2017patents, webb2019impact, alderucci2020quantifying}, some scholars argue that the GPT concept may be too simplistic for understanding AI. Recent work has raised concerns about conceptualising AI as a GPT, particularly from a policy perspective \citep{jacobides2021evolutionary}. Others caution that the realisation of GPT-related benefits cannot be taken for granted \citep{bresnahan2023innovation}. 
GPTs may generate market failures due to their public good characteristics, which can lead to suboptimal levels of private investment \citep{bresnahan1995general}. However, it remains unclear whether AI constitutes a single GPT or an amalgamation of multiple technologies that perform diverse functions, including data provision, prediction, classification, soft- and hardware integration, and edge applications \citep{jacobides2021evolutionary}. 

Certain AI ecosystem components are more essential than others, and countries and companies show different specialisations in sub-areas of AI, but those that lack access to the critical components run the risk of being cut off from realising the benefits of AI \citep{franco2023producing, klinger2022narrowing, cazzaniga2024gen, jacobides2021evolutionary, bresnahan2023innovation, babina2022artificial}. 

Therefore, if AI is understood as a system of interdependent technologies and resources--potentially organised in a hierarchical structure--the study of AI should be broadened to include the key actors shaping its technological evolution. The growing concentration of AI development among a few dominant firms \citep{klinger2022narrowing, babina2022artificial} has raised concerns about the influence of Big Tech \citep{jacobides2021evolutionary} and geographic or institutional hubs \citep{klinger2022narrowing}. This concentration also casts doubt on the case for public funding of AI, as advocated from a GPT perspective \citep{jacobides2021evolutionary, bresnahan1995general}. On the contrary, additional public support--if not accompanied by competition policy and regulatory oversight--may reinforce existing asymmetries in the AI ecosystem \citep{petit2017antitrust, hennemann2020artificial}.

In this paper, we analyse whether the GPT framing is consistent with the four different AI classification approaches. Our analysis is informative along three main dimensions: (1) timescale and history: different definitions of AI affect the timescale on which it is viewed and the perceived history of its development; (2) concentration and key actors: theoretical approaches differ in which key actors that are identified to shape AI innovation and the implications for market power; (3) GPT characteristics: the choice of definition may suggest different guidelines for the direction supporting AI technologies, depending on the degree to which AI satisfies the public goods of a GPT. Our analysis investigates the methodological basis of empirical research on AI and its impact, showing how empirical controversies can arise from different approaches to defining and measuring AI. 

\subsection{AI history and different definitions}
\label{subsec:Literature_AI_history}
AI broadly refers to technologies that perform tasks requiring intelligence. Its conceptual roots trace back to \citet{machinery1950computing}. Subsequent decades saw sustained R\&D investment that led to advances in using computers as problem-solving machines. However, unmet high expectations, combined with limited computing power and reduced funding, resulted in stagnation periods (``AI winters'') in the 1970s and 1990s \citep{stuart2003artificial}.

Modern AI--since at least the mid-2000s--has been largely driven by methods from machine learning (ML), which typically involve computational techniques for detecting and modelling patterns in diverse data sources \citep{mitchell2007machine}. ML draws on foundations from computer science, statistics, logic, probability, and optimisation. In many cases, ML software is integrated with hardware components such as sensors, actuators, and control systems to form intelligent systems. This combination of software and hardware is one established approach to building AI systems. 

Recent work argues that modern AI research is becoming increasingly privatised and narrowly focused on deep learning--a specific form of ML--at the expense of other relatively underexplored areas, such as symbolic learning \citep{bianchini2020deep, klinger2022narrowing, jurowetzki2021privatization, whittaker2021steep}. Moreover, the transformative potential of AI appears to be relatively underexploited by adopting firms \citep{bresnahan2023innovation}. It remains uncertain whether this narrowing will influence how AI is defined in the future, potentially shaping which inventions are captured under the AI label. Future developments may also involve new sub-fields that are, at present, largely unexplored.

Taken together, these observations reveal both long-term technological trends and a recent concentration of inventions around a limited set of specific technologies. The choice of how broadly to define AI is central to both research and policy debates, yet this decision is far from straightforward.
Efforts to define AI in more technology-neutral terms are already underway. For instance, in EU legislation regulating AI,\footnote{\url{https://www.europarl.europa.eu/doceo/document/TA-9-2023-0236_EN.html}} an \emph{AI system} is defined as:

\begin{quotation}
\emph{``a machine-based system that is designed to operate with varying levels of autonomy and that may exhibit adaptiveness after deployment, and that, for explicit or implicit objectives, infers, from the input it receives, how to generate outputs such as predictions, content, recommendations, or decisions that can influence physical or virtual environments''}

\end{quotation}
and the specific connection to ML is further emphasised in the text as

\begin{quotation}

\emph{``AI systems often have machine learning capacities that allow them to adapt and perform new tasks autonomously. Machine learning refers to the computational process of optimizing the parameters of a model from data, which is a mathematical construct generating an output based on input data. Machine learning approaches include, for instance, supervised, unsupervised and reinforcement learning, using a variety of methods including deep learning with neural networks.} 

[...] 

\emph{Comparably simpler techniques such as knowledge-based approaches, Bayesian estimation or decision-trees may also lead to legal gaps that need to be addressed by this Regulation, in particular when they are used in combination with machine learning approaches in hybrid systems.}

[...] 

\emph{AI systems can be used as stand-alone software system, integrated into a physical product (embedded), used to serve the functionality of a physical product without being integrated therein (non-embedded) or used as an AI component of a larger system. If this larger system would not function without the AI component in question, then the entire larger system should be considered as one single AI system under this Regulation.''}
\end{quotation}
These definitional efforts deliberately avoid being overly specific, highlighting the difficulty of drawing clear system boundaries.
As technological capabilities advance, definitions of AI for policy and regulation are likewise evolving. For example, since 2023, the EU AI Act has been refined to emphasise the potential outputs of systems classified as AI, and more recently, it has incorporated the requirement that such systems possess the ability to adapt and generalise.

Whilst the United States currently lacks a unified definition of AI, California Assembly Bill 331--introduced in 2023 as part of ongoing efforts to regulate AI--diverges from the EU AI Act by not explicitly addressing the autonomous nature of such systems.\footnote{\url{https://leginfo.legislature.ca.gov/faces/billTextClient.xhtml?bill_id=202320240AB331}} 
The UK similarly lacks a unifying definition of AI, with conceptualisations varying significantly across government departments. 
Whilst the Department for
Science, Innovation
\& Technology defines AI systems narrowly, emphasising only those with ``adaptable'' and ``autonomous'' functionality,\footnote{\url{https://www.gov.uk/government/publications/ai-regulation-a-pro-innovation-approach/white-paper}} the UK's National Cyber Security Centre identifies AI based on outputs--specifically, its ability to perform tasks typically associated with human intelligence.\footnote{\url{https://www.ncsc.gov.uk/section/advice-guidance/}} 
Evidently, key definitions differ not only in the types of technologies considered to constitute AI, but also in whether AI should be defined by its outputs, internal composition, adaptability, or autonomy.

\subsection{Patents as a data source}
\label{subsec:Literature_patent_data}
Attempts of defining AI in legislation are driven by a perceived need for regulation, drawing on empirical research on AI, including economic studies based on patent data \citep[e.g.,][]{cockburn2018impact, webb2019impact, wipo19, fujii2018trends, uspto2020inventing, verendel2023tracking}.
Patents provide a rich data source for research, offering detailed records of inventions and the embedded technological knowledge. Patent offices maintain millions of records, categorised into technological fields via hierarchical classification systems. A single patent may be assigned to one or more codes from hierarchical patent classification systems, each reflecting a distinct aspect of the underlying technology \citep{jaffe2017patent}. 
In AI measurement, these codes help describe the qualitative nature of a patent, for example, whether it is related to specific computing techniques or hardware for data transmission, or related to functions in another sector of the economy. 

Patent documents also include citation links to other patents and academic articles, which help differentiate the patented invention from prior technologies. Inventors must demonstrate the novelty of their invention in relation to existing technologies. These citations are commonly used by innovation scholars, as they reveal the cumulative dependencies between technologies and can serve as an indicator of how much an invention builds on existing knowledge and technology \citep{jaffe2017patent}. 

Several studies on AI have leveraged these characteristics to conduct empirical quantitative assessments of AI's impact, with some framing AI as a potential GPT \citep{cockburn2018impact}. However, it is important to note that there are many alternatives to patent data, including industry-, firm-, and technology-level non-patent data, both quantitative and qualitative, as well as conceptual frameworks \citep[e.g.][]{klinger2018deep, bresnahan2023innovation, trajtenberg2018ai, goldfarb2023could}.

\section{The four different AI classification approaches}
\label{sec:approach}
In our analysis, we sample four sets of AI patents using distinct AI definitions applied to patent data, and compare them based on their GPTness and patterns of concentration. For GPTness, we assess the growth, generality, and complementarity of AI patents classified by (1) keywords, (2) science citations, (3) the WIPO method, and (4) the USPTO method, drawing from all USPTO patents granted between 1990 and 2019. To analyse the concentration of inventive activity, we measure the Hirschmann-Herfindahl Index (HHI) and evaluate dominant firms. The four approaches are implemented as follows:

First, the Keyword method replicates \citet{cockburn2018impact}, using text search terms (e.g., NLP, robotics, neural networks). This approach reflects short-term perspectives on AI, attributing its progress to themes that became dominant since the mid-2000s.

Second, we use scientific citations, including grey literature and conference proceedings \citep{marx2020reliance}, as an AI identifier. This approach conceptualises AI technologies as the result of scientific and academic research \citep{arthur2009nature, jee2023firms}. As discussed below, this method has limitations, as citation practices vary across different technological fields.

Third, the WIPO method emphasizes AI’s technical foundations, such as ML and applied computing. It combines a set of keywords with technology-specific classification codes from computer science \citep{wipo2019technologybackground}.

Fourth, the USPTO method adopts the broadest understanding of AI. It employs a trained ML classifier on patent text and citations to identify innovations related to knowledge processing, speech, hardware, evolutionary computation, NLP, ML, computer vision, and planning and control \citep{giczy2021identifying}. This expansive conceptualisation of AI is reflected in the large volume of AI patents identified through this method, including a significant share of downstream AI applications.

For further background on these definitions and details on their implementation, we refer the reader to \ref{sec:GPTs}-\ref{sec:approaches}.

\section{Results}\label{sec:results}
We compare the four AI samples by their general characteristics (\ref{sec:trends}), GPTness (\ref{sec:GPT_char_of_AI}), and concentration (\ref{sec:Market_power}). 

\subsection{General characteristics of AI}
\label{sec:trends}
Table \ref{tab:ai_traj} shows, for each method, growth rates (panel A),  inventor types (panel B),  main industries (panel C), countries of origin (panel D), public support (panel E), and main technology classes (panel F). 

All methods indicate increasing diversification of AI inventions across industries and countries, suggesting economic diffusion. Furthermore, all approaches indicate that AI inventions are disproportionately generated by commercial enterprises (Panel B) and are predominantly associated with the computer and machinery manufacturing sectors (Panel C).
Although the number of patents identified varies by orders of magnitude across methods, all consistently show a dominance of the US in AI patenting. However, relying solely on USPTO data inherently over-represents US-based inventors.

We find differences in the growth rates of the Science and USPTO approaches, suggesting that these two groups have not experienced the same post-2000 acceleration as the other two approaches (Keywords and WIPO).

\begin{table}[H]
    \centering
        \caption{AI patents by four classification approaches}
    \label{tab:ai_traj}
    \footnotesize
    {
\def\sym#1{\ifmmode^{#1}\else\(^{#1}\)\fi}
    \begin{tabular}{lcccc}
    \hline\hline
    &	Keyword	& Science &	WIPO &	USPTO \\
    \hline
\multicolumn{5}{c}{\textit{A.	Growth}} \\
Growth rate (1990-99)&        0.98&        6.69&        2.38&        3.87\\
Growth rate (2000-09)&        0.39&        1.32&        1.21&        1.16\\
Growth rate (2010-19)&        3.77&        0.91&        3.02&        1.07\\
\hline
\multicolumn{5}{c}{\textit{B.	Inventor type (patent assignee)}}\\
\% Commercial       &        0.86&        0.85&        0.90&        0.91\\
\% Individual       &        0.06&        0.03&        0.05&        0.05\\
\% University       &        0.05&        0.10&        0.04&        0.03\\
\% Other non-profit &        0.03&        0.04&        0.02&        0.02\\
\hline
\multicolumn{5}{c}{\textit{C.	Industry affiliation (patent assignee)}} \\
\% Pharmaceuticals (manufacturing)&        0.01&        0.17&        0.01&        0.01\\
\% Computer (manufacturing)&        0.68&        0.76&        0.79&        0.84\\
\% Machinery \& equipment (manufacturing)&        0.49&        0.28&        0.54&        0.22\\
\% Other manufacturing&        0.14&        0.10&        0.09&        0.09\\
\% Computer programming (service)&        0.06&        0.07&        0.09&        0.12\\
\hline
\multicolumn{5}{c}{\textit{D.	Country of origin (patent applicant)}} \\
\% USA              &        0.62&        0.75&        0.65&        0.72\\
\% Japan            &        0.14&        0.08&        0.15&        0.10\\
\% S. Korea         &        0.04&        0.02&        0.03&        0.02\\
\% Germany          &        0.05&        0.03&        0.03&        0.03\\
\% China            &        0.01&        0.01&        0.02&        0.01\\
\% Canada           &        0.02&        0.02&        0.02&        0.02\\
\hline
\multicolumn{5}{c}{\textit{E.	Public support}} \\
\% Public support (1990-99)&        0.31&        0.39&        0.29&        0.21\\
\% Public support (2000-09)&        0.34&        0.48&        0.31&        0.23\\
\% Public support (2010-17)&        0.33&        0.52&        0.28&        0.22\\
\hline
\multicolumn{5}{c}{\textit{F.	CPC 1-digit codes}} \\
\% Human necessities (A)&        0.15&        0.18&        0.08&        0.08\\
\% Performing operations (B)&        0.31&        0.05&        0.11&        0.04\\
\% Chemistry; Metallurgy (C)&        0.03&        0.16&        0.01&        0.01\\
\% Physics (G)      &        0.66&        0.75&        0.95&        0.79\\
\% Electricity (H)  &        0.25&        0.26&        0.27&        0.33\\
\% General/Cross-sectional (Y)&        0.05&        0.04&        0.03&        0.04\\
\hline
Number of patents   &       54,145&      178,004&      158,652&      595,047\\
\hline\hline
    \end{tabular}
}

    \footnotesize \justifying

Notes: This table compares the scale and scope of AI invention identified by each definition, disaggregated by inventor types (commercial, individual, non-profit, university), industry affiliation (based on NACE Rev. 2 classification), country of origin, reliance on public R\&D support, and technological classification. Note that the data on public support ends in 2017. 

\end{table}

When comparing how inventions are classified using CPC 1-digit codes, all approaches show a similar association with the `Electricity (H)' section. WIPO patents are most strongly concentrated in the `Physics (G)' section, while the Science approach most strongly identifies patents in `Chemistry (C)'. The Keyword and Science approaches yield the most diverse range of AI-related patents across CPC categories. The high share (95\%) of WIPO patents in `Physics (G)' is a direct result of the method’s design: the WIPO approach explicitly filters for patents classified under this CPC section \citep{wipo2019technologybackground}.

\begin{figure}[H]
\centering
    \caption{AI Patents by Year (1990-2019)}
    \label{fig:timeseries}
    \begin{subfigure}[b]{0.49\textwidth}
        \centering
        \includegraphics[width=\textwidth]{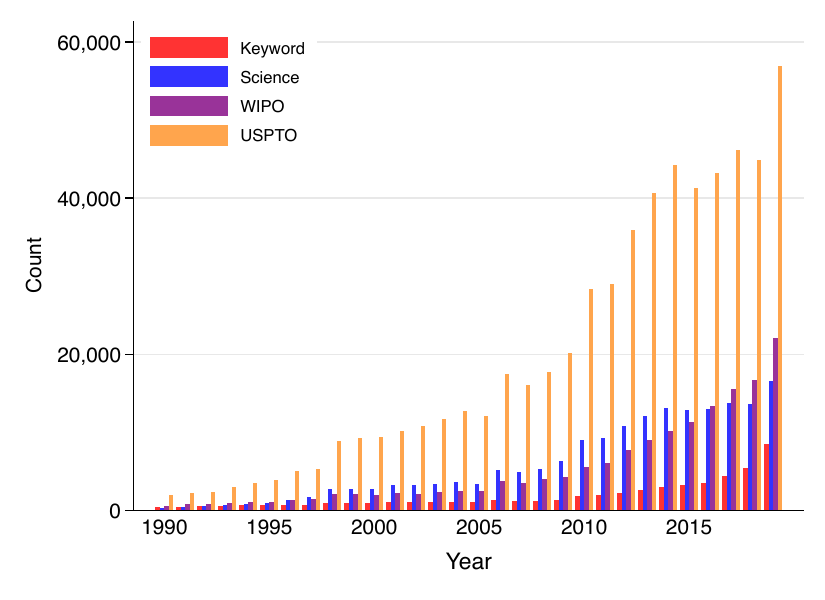}
        \caption{Number of AI patents}
    \end{subfigure}
   \begin{subfigure}[b]{0.49\textwidth}
       \centering
       \includegraphics[width=\textwidth]{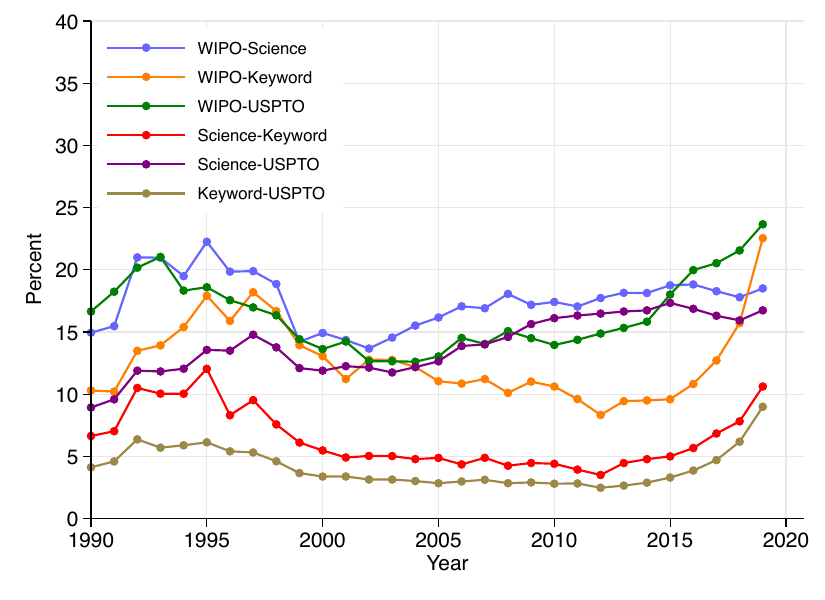}
       \caption{Jaccard Similarities}
        \label{fig:jaccard_similarities}
   \end{subfigure}

     \footnotesize\justifying
    Notes: The right panel shows the number of AI patents over time as identified by the four approaches. The left panel shows the evolution of the Jaccard similarities computed for each year in our dataset.
\end{figure}

Figure \ref{fig:timeseries} illustrates the pace of AI invention. Expressed as a share of all granted US patents, the USPTO approach classifies approximately 16.6\% of patents in 2019 as AI-related (see Figure \ref{app:fig:timeseries} in \ref{app:tables_moved_down_volume_and_timetrends}).
Across all methods, this share has risen from 1–2\% in the 1990s to 3-17\% by 2019.

To quantify the overlap among the four AI patent samples, we compute pairwise Jaccard similarities and track their evolution over time, as shown in Figure \ref{fig:jaccard_similarities}.\footnote{The Jaccard similarity for two sets of patents is given by 
$$J(A,B) = \frac{|\mbox{patents in both A and B}|}{|\mbox{patents in union of A and B}|} = \frac{|A \cap B|}{|A \cup B|}$$
with $J(A,B) \in [0,1]$ where $J(A,B) = 0$ if both sets do not overlap and $J(A,B) = 1$ if both sets are identical. In other terms, the overlap between the sets can range from 0\% to 100\%.} 
All pairs of AI samples exhibit low overlaps, with Jaccard values at or below 20\%. The WIPO approach shows the highest agreement with other methods. Notably, the WIPO–Keyword and Science–USPTO pairs show the most pronounced increases in similarity over time. The Keyword method has the lowest Jaccard similarity with the other approaches, likely due to the relatively small number of patents in this sample.
Overall, only 10,062 patents--or 1.37\% of all unique granted patents--are jointly identified as AI patents by all four methods. This low overlap indicates that quantitative assessments of the scale and diffusion of AI can be highly sensitive to the choice of classification method.

In the following section, we adopt a qualitative perspective to examine how these definitional differences influence the extent to which AI qualifies as a GPT. 

\subsection{GPT characteristics of AI}
\label{sec:GPT_char_of_AI}
We study the GPTness of AI by comparing the four patent samples along three key dimensions: growth, generality, and complementarity. The results are summarised in Table \ref{tab:gpt_summary} and discussed in detail below.

\begin{table}[H]
\centering
\caption{\label{tab:gpt_summary}Measure of GPT characteristics of AI}
{
\def\sym#1{\ifmmode^{#1}\else\(^{#1}\)\fi}
\begin{tabular}{lcccc}
\hline\hline
&   Keyword &     Science &     WIPO &   USPTO \\
\midrule
\multicolumn{5}{c}{\textit{A.	Growth}} \\
Avg. growth rate    &        0.12&        0.15&        0.14&        0.13\\
\hline
\multicolumn{5}{c}{\textit{B. Generality}} \\
Avg. generality index (1 digit)&        0.81&        0.77&        0.76&        0.73\\ [0.25em]
Avg. generality index (3 digit)&        0.94&        0.90&        0.90&        0.87\\ [0.25em] 
Avg. generality index (4 digit)&        0.97&        0.96&        0.95&        0.95\\ [0.25em]
\hline
\multicolumn{5}{c}{\textit{C. Complementarity}} \\
Avg. number of CPC (1 digit)&        1.43&        1.40&        1.36&        1.27\\ [0.25em] 
Avg. number of CPC (3 digit)&        1.67&        1.64&        1.64&        1.43\\ [0.25em] 
Avg. number of CPC (4 digit)&        1.92&        1.97&        2.05&        1.64\\ [0.25em] 
\hline\hline
\end{tabular}
}

\footnotesize \justifying
Notes: This table gives a comparison of the GPT-like characteristics of AI inventions classified by each distinct technique. Note that the generality index is defined as share of citations to patents in different CPC classes at different aggregation levels. Citations within the same class are excluded. 
\end{table}

\subsubsection{Growth}
\begin{figure}[H]
    \centering
    \caption{Growth of Patents by Year}
    \label{fig:growth_rates}
    \begin{subfigure}{0.45\textwidth}
        \centering
        \includegraphics[width=0.95\textwidth]{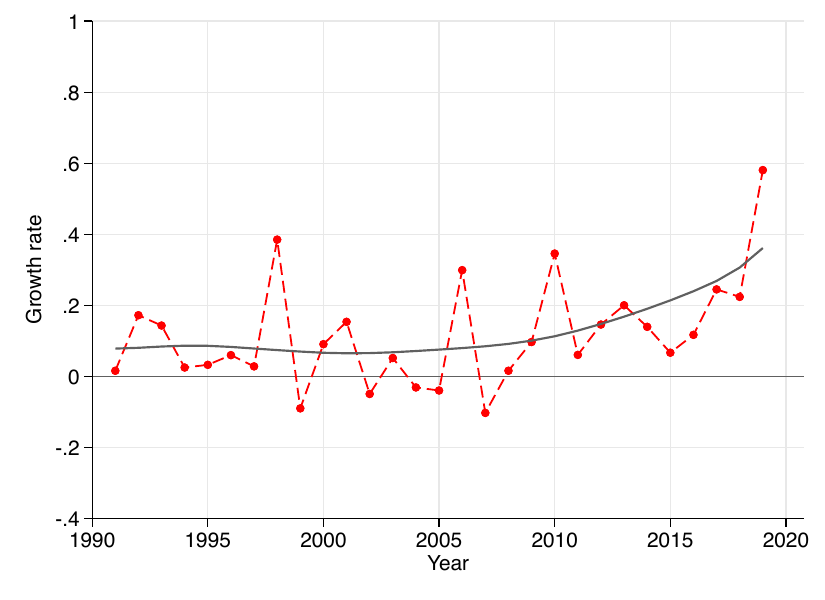}
        \caption{Keyword}
         \label{fig:growth_rates_key}
    \end{subfigure}
    \begin{subfigure}{0.45\textwidth}
        \centering
        \includegraphics[width=0.95\textwidth]{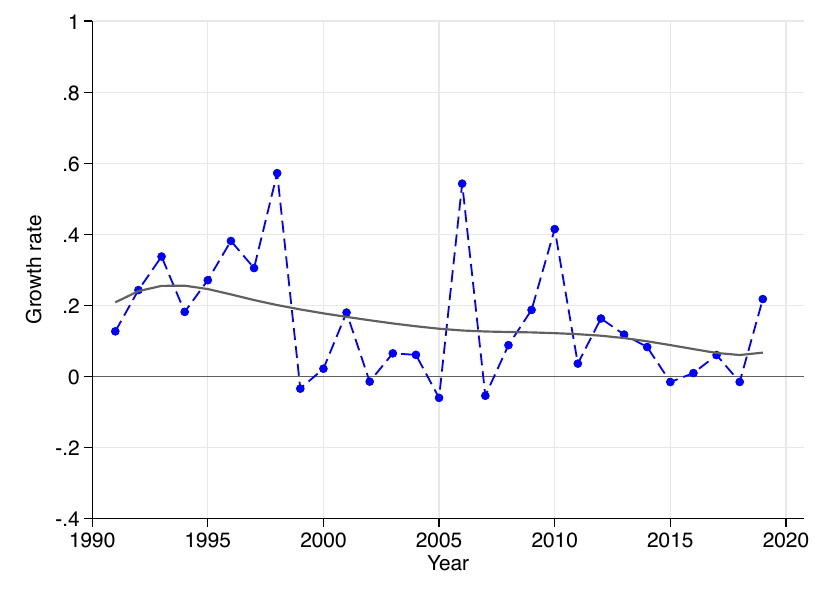}
        \caption{Science}
         \label{fig:growth_rates_sci}
    \end{subfigure}    
    
    \begin{subfigure}{0.45\textwidth}
        \centering
        \includegraphics[width=0.95\textwidth]{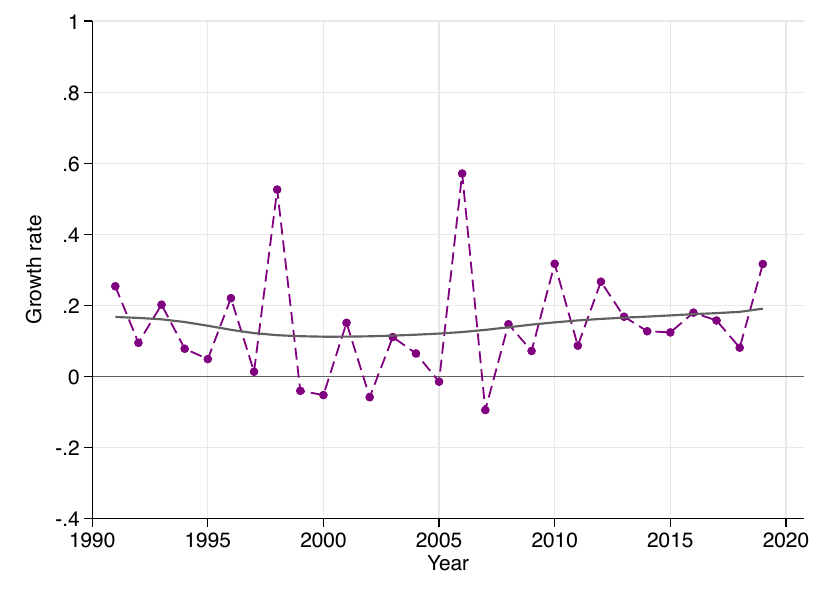}
        \caption{WIPO}
         \label{fig:growth_rates_wip}
    \end{subfigure}
    \begin{subfigure}{0.45\textwidth}
        \centering
        \includegraphics[width=0.95\textwidth]{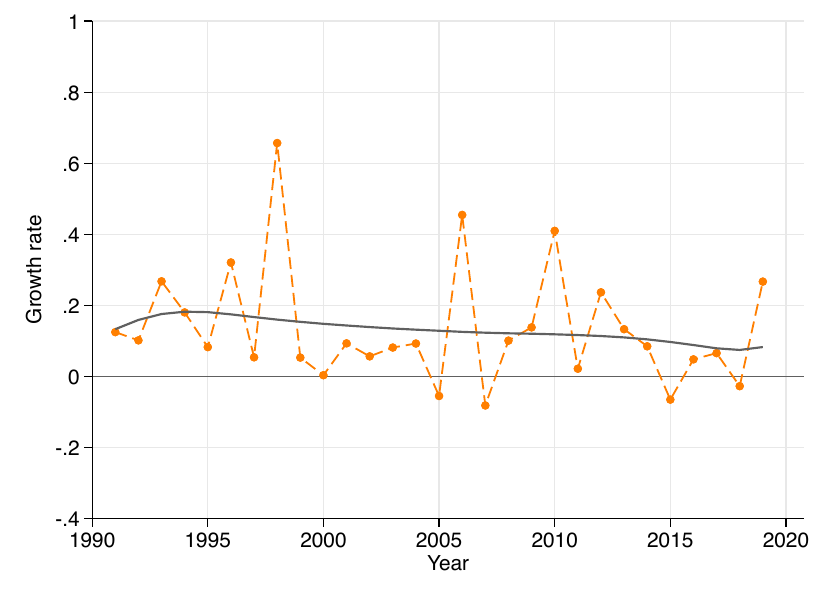}
        \caption{USPTO}
         \label{fig:growth_rates_uspto}
    \end{subfigure} 
    
    \footnotesize\justifying
    Notes: The four AI approaches have different growth patterns over time. The averages for all are positive, but the Keyword and WIPO approaches both have increasing growth rates.
\end{figure}

Panel A of Table \ref{tab:gpt_summary} reports the average annual growth rates of AI patents, while Figure \ref{fig:growth_rates} illustrates how these rates have evolved over time. To show overall trends, we apply local regression (LOWESS) smoothing \citep{cleveland1979robust} to each time series.

We observe the following: First, all methods exhibit positive growth rates, consistent with GPT expectations. Second, each series exhibits a dip in growth during the early 2000s, followed by renewed acceleration in more recent years.\footnote{This slowdown is not unique to AI; patenting activity declined across multiple sectors during this period (see Figure \ref{fig:growth_rates_bench} in \ref{appendix_bench_growth}).} Third, the smaller AI samples identified by the Keyword and WIPO methods show a strong acceleration in the later years, in contrast to the Science and USPTO samples, which exhibit more moderate growth. 
Overall, the results indicate persistent growth across all methods, albeit with varying dynamics and strength depending on the classification approach.

In the Appendix \ref{appendix_add_compare_bench}, we compare our AI patent samples to benchmarks that have been discussed in the literature as potential GPTs--such as nanotechnology, climate technology, biochemistry, and general computing. Overall, the AI samples exhibit higher growth rates than most of these GPT candidates. A Wilcoxon test shows that AI patents grow significantly faster than average, though differences across AI types are mostly not significant (see \ref{appendix:significance_growth}).\footnote{It should be noted that these statistical tests are based on a relatively small number of observations, and the growth trajectories of the four AI samples vary substantially over the three-decade period.} Among the AI classification methods, growth rate differences are generally not statistically significant, with the exception of the USPTO approach, which shows significantly lower growth compared to the others.

\begin{figure}[ht]
\centering
    \caption{Patents Citing AI}
    \label{fig:citing_AI}
    \begin{subfigure}{0.45\textwidth}
        \centering
        \includegraphics[width=0.95\textwidth]{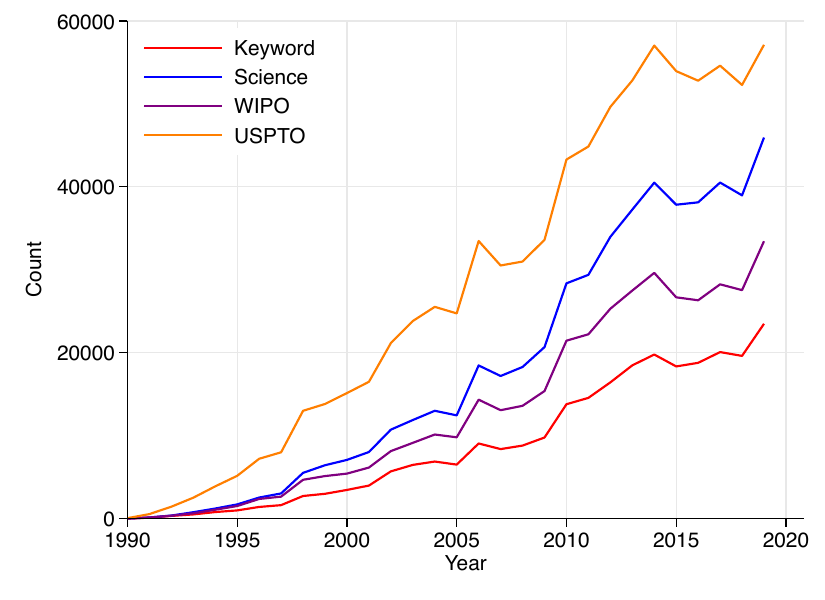}
        \caption{Counts}
         \label{fig:citing_AI_counts}
    \end{subfigure}
    \begin{subfigure}{0.45\textwidth}
        \centering
        \includegraphics[width=0.95\textwidth]{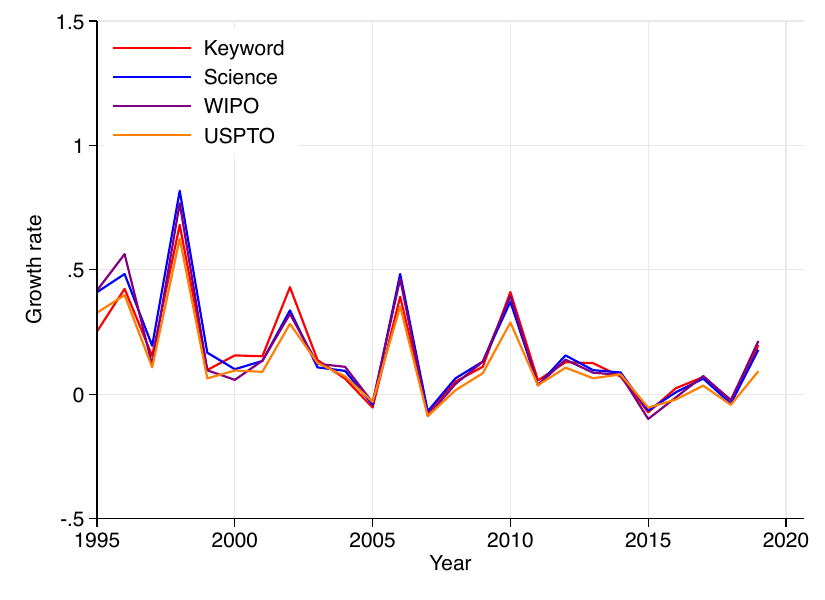}
        \caption{Growth rate (1=100\% growth)}
         \label{fig:citing_AI_growth_rates}
    \end{subfigure}

     \footnotesize\justifying
    Notes: Panel (a) shows the actual number of AI citing patents. Panel (b) shows growth rates plotted from 1995 and onwards. 
\end{figure}

If AI functions as a GPT, it should enable follow-on inventions in non-AI sectors--reflected in the growth of patents that cite AI patents but are not themselves classified as AI. 
Figure \ref{fig:citing_AI} shows rising numbers of AI-citing patents across all methods. The relative sizes of the groups are consistent with earlier patterns (see Figure \ref{fig:growth_rates}). 
The positive downstream growth suggests that each approach generates a growing number of invention spillovers to non-AI sectors. 

indicate that differences between the Keyword, Science, and WIPO approaches are not statistically significant--except for the Keyword sample, which showed slower uptake in the 1990s before accelerating. These three approaches score significantly higher than the USPTO sample. 
Overall, these findings support the view that these classification methods capture AI technologies that have a persistent impact on the wider innovation landscape. 

\FloatBarrier
\subsubsection{Generality}
\label{sec:results_generality}

We use two indicators to evaluate the generality of AI inventions--that is, the extent to which AI patents are cited across diverse technological fields.
First, we calculate a generality index \citep{trajtenberg1997university, hall2006uncovering}, which reflects the dispersion of CPC 1-digit codes among patents citing AI inventions. This measure is conceptually similar to the HHI, but inverted to capture breadth rather than concentration.\footnote{The formula of the generality index $GI_i$ is given by $$GI_i = 1 - \sum_j^{N_j} \left( \frac{\#cites_{ij,t}}{\sum^{N_j}_{j=1} \#cites_{ij,t}} \right)^2$$
where $\#cites_{ij}$ is the number of citations to patents labeled as AI by method $i$ from CPC class $j$, using CPC codes at the 1-digit, 3-digit, and 4-digit level; $\#cites_{ij}$ excludes citations within the same class. $N_j$ is the number of different CPC classes.} Technical details and additional results are provided in \ref{sec:methods_generality} and \ref{app:tables_moved_down_generality}. 

\begin{figure}[H]
\centering
    \caption{Generality Index at the 1-digit CPC-section level}
    \label{fig:generality_index}
    \begin{subfigure}{0.45\textwidth}
        \centering
        \includegraphics[width=0.95\textwidth]{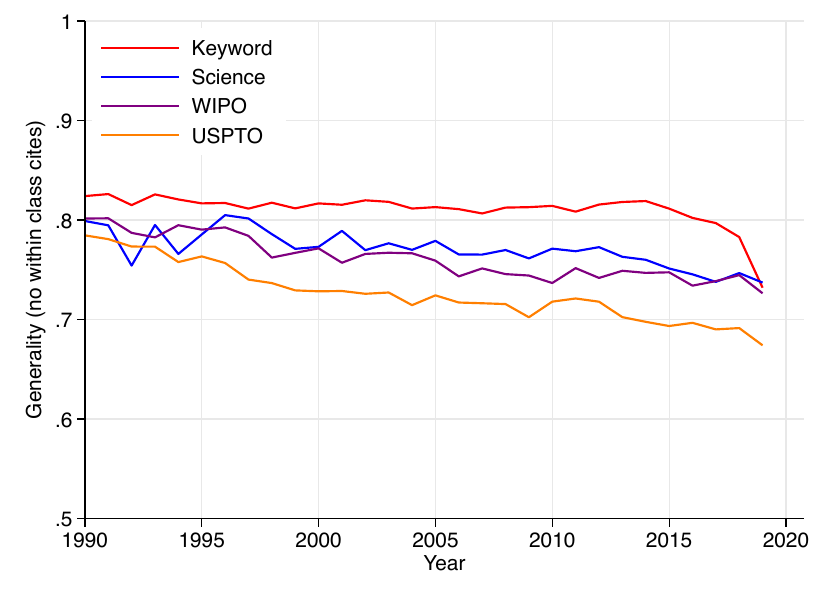}
        \caption{Generality index}
         \label{fig:generality_AI}
    \end{subfigure}
    \begin{subfigure}{0.45\textwidth}
        \centering
        \includegraphics[width=0.95\textwidth]{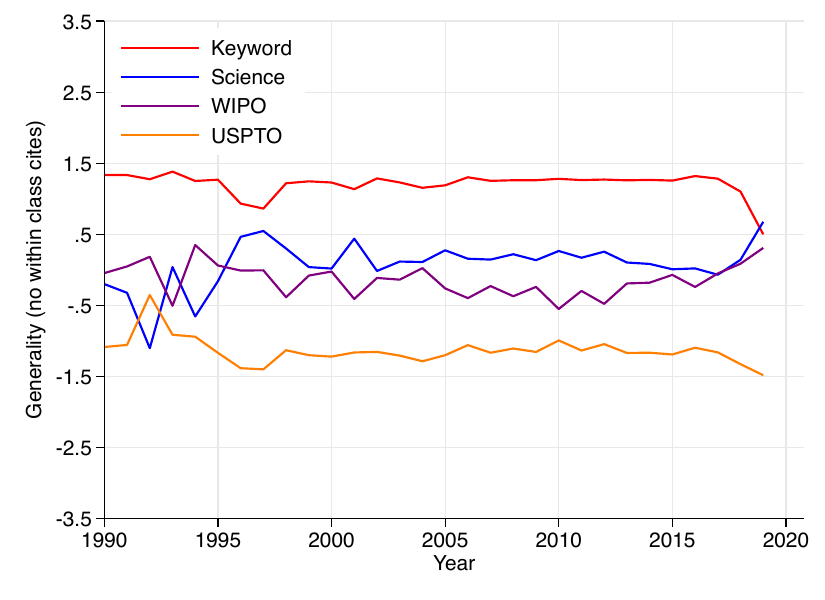}
        \caption{Z-score scaled generality}
         \label{fig:generality_AI_scaled}
    \end{subfigure}

     \footnotesize\justifying
    Notes: The z-scored value equals the level of the generality index minus its average across the four approaches divided by the standard deviation for each year. 

\end{figure}

Figure \ref{fig:generality_index} shows how the generality index evolved over time.
The Keyword sample consistently shows highest generality, with citations broadly distributed across technology fields. The WIPO and Science samples follow, with the Science-based approach scoring slightly higher than WIPO.
The USPTO sample shows the lowest generality at all CPC aggregation levels--somewhat surprisingly, given the large size of this sample.
The decline in the generality of the Keyword and USPTO samples towards the end of the period should be interpreted with caution, as recently granted patents have had less time to be cited.

Second, we assess generality using the mean annual number of unique CPC classes that cite each AI patent. This measure accounts for the overall growth in patenting activity over time and avoids over-representing the generality of more recent patents. As before, we compute this metric at multiple CPC aggregation levels.

\begin{figure}[H]
    \centering
    \caption{Average number of classes citing AI}
    \label{fig:avg_classes_citing_AI}
    
    \begin{subfigure}{0.45\textwidth}
        \centering
        \includegraphics[width=0.95\textwidth]{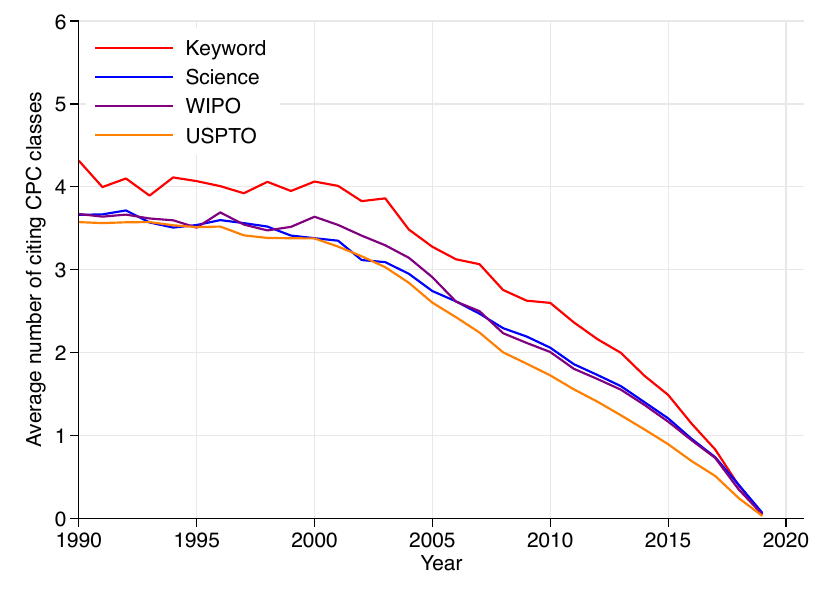}
        \caption{All patents}
         \label{fig:avg_class_cit_AI}
    \end{subfigure}
    \begin{subfigure}{0.45\textwidth}
        \centering
        \includegraphics[width=0.95\textwidth]{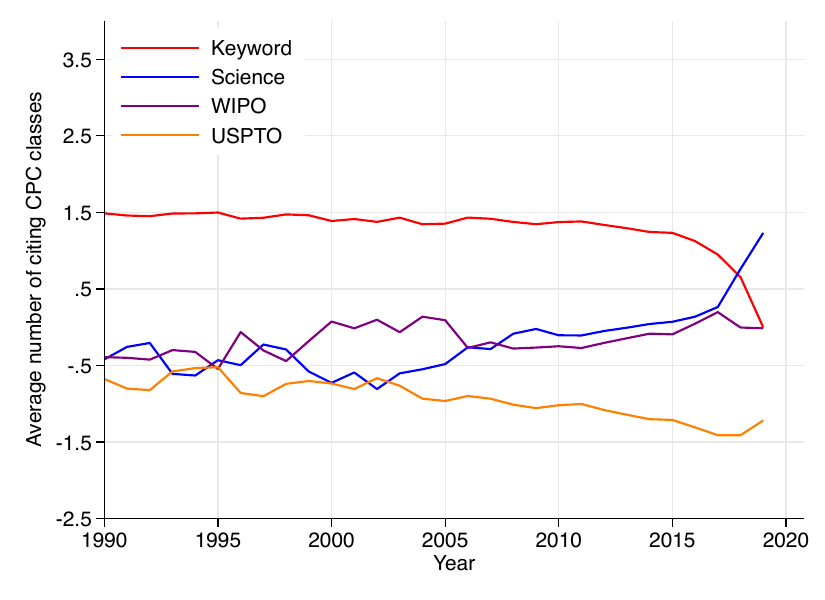}
        \caption{All patents (z-score scaled)}
         \label{fig:avg_class_cit_AI_scaled}
    \end{subfigure}    
    
    \footnotesize\justifying
    The z-scored value equals the level of the generality index calculated at the 1-digit level minus its average across the four approaches divided by the standard deviation for each year. 
\end{figure}

At all CPC aggregation levels, Keyword patents show highest generality (Table \ref{tab:gpt_summary}, Figure \ref{fig:avg_classes_citing_AI}).
At the 1-digit level, the Science sample ranks second, closely followed by WIPO. However, WIPO scores higher than Science at the more granular 3- and 4-digit levels.
The USPTO sample consistently shows the lowest number of unique citing CPC classes across all levels. As before, the decline observed in more recent years can largely be attributed to citation time lags.

Significance tests (\ref{appendix:significance_generality}) confirm that the Keyword sample's generality is statistically higher, and the USPTO sample's lower generality is also significant. Differences between the Science and WIPO methods are negligible at the 1-digit level.

In \ref{app:tables_moved_down_generality}, we additionally report generality based only on AI patents that received at least one citation (Table \ref{app:tab:avg_citations_cited}). This refines the analysis to include only relevant inventions, given that backward citations are often considered a proxy for patent quality \citep{barbieri2025evolving}.
Here too, the Keyword sample shows the highest generality across all CPC levels. Science again ranks slightly higher than WIPO at the 1-digit level, with WIPO scoring higher at the 3- and 4-digit levels. The USPTO sample remains the least general across all metrics.

As before, we benchmark our AI samples against all patents and other GPT candidates. Table \ref{tab:generality_benchmark} shows that the generality index of all patents is higher than our AI samples, which is a natural feature, confirming the usefulness of the generality index in capturing the widespread use of patents. 
Among other groups, biochemistry/genetic engineering, nanotechnology, and climate inventions related patents have high generality scores. Figure Figure \ref{subfig:generality_generality_benchmark} reveals that the generality index follows a relatively stable time series pattern, with some fluctuations toward the end of the period likely due to citation lags. 

Using our second measure of generality--the mean number of unique citing CPC classes per AI patent--we find that all four AI samples exhibit higher generality than the overall patent population across all CPC aggregation levels (Table \ref{tab:generality2_benchmark}). Again, biochemistry/genetic engineering and climate-related technologies rank high by this measure. These results support the validity of our generality metrics, at least in relation to other established GPT candidates.

Additionally, we examine the generality of non-AI patents that cite AI inventions. While differences between the four AI samples are smaller for this group, the qualitative patterns remain consistent (see Appendix \ref{app:add_results_generality}).

\begin{table}[H]
 \centering
 \caption{Average Citation Lags by Approach}
 \label{tab:avg_citation_lags}
 \centering
{
\def\sym#1{\ifmmode^{#1}\else\(^{#1}\)\fi}
 \begin{tabular}{lcccc}
   \hline
   \hline
  Period & Keyword & Science & WIPO & USPTO \\[0.25em] 
   \hline
all periods         &       10.16&        8.90&        9.63&        9.80\\ [0.25em] 
1990-1999           &       14.17&       13.26&       13.77&       13.64\\ [0.25em] 
2000-2009           &        9.92&        9.08&        9.38&        9.34\\ [0.25em] 
2010-2019           &        4.33&        4.15&        4.19&        4.33\\ [0.25em] 
   \hline
    \hline
 \end{tabular}
}

\footnotesize \justifying
 
 Notes: This table shows the average number of years taken until a patent in the sample is cited. The average number of years is lower in recent years, because the data ends in 2019 causing a truncation of the maximal time lag. 
 \end{table}
 
Patent citations to GPTs often occur with long time lags \citep{hall2006uncovering}, as the diffusion of such technologies typically involves an early phase of ``learning and destruction''--a period requiring organisational adjustments and complementary innovations before the GPT can be widely adopted \citep{crafts2021artificial, bresnahan2023innovation}.
In Table \ref{tab:avg_citation_lags}, we report the average citation lags for each AI sample, calculated as the average number of years between a patent's grant year and the grant years of citing patents.
Keyword patents show the longest citation lags and once again score highest for GPTness.  
Taken together, the Keyword sample consistently ranks highest across all generality-related metrics, supporting its identification as the most GPT-like among the four classification approaches.

\FloatBarrier
\subsubsection{Complementarity}
To measure complementarity, we examine the co-classification of AI by multiple CPC codes attached to patents. If AI complements a wide range of other technologies, then AI patents would be co-classified across a variety of technological fields.

\begin{figure}[h]
     \centering
     {
      \caption{Diversity of AI -- Share of technology classes}
        \label{fig:diverse}
     \begin{subfigure}[b]{0.45\textwidth}
         \includegraphics[width=\textwidth]{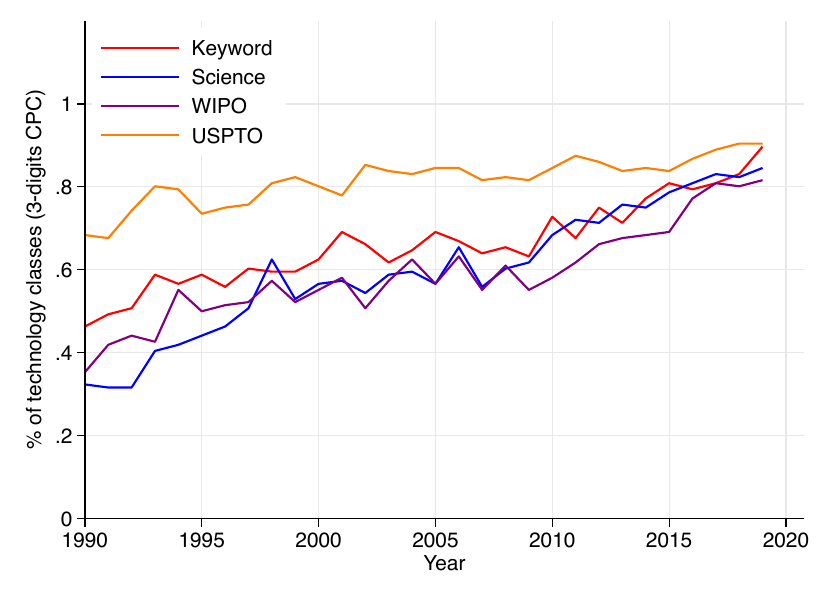}
         \caption{\% of all 3-digit CPC}
         \label{fig:diverse3d}
     \end{subfigure}
     \hfill
     \begin{subfigure}[b]{0.45\textwidth}
         \includegraphics[width=\textwidth]{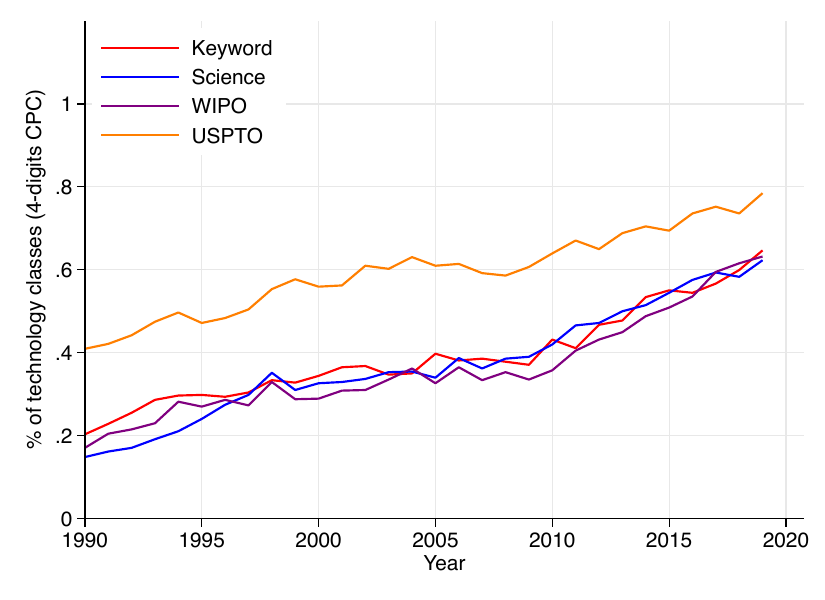}
         \caption{\% of all 4-digit CPC}
         \label{fig:diverse4d}
     \end{subfigure}
     }
     
     \footnotesize \justifying
     Note: Panel (a) shows the percentage of 3-digit CPC codes and panel (b) shows the percentage of 4-digit CPC as a share of all codes in the respective category. Note that the total number of 3-digit and 4-digit CPC codes are 136 and 674, respectively according to the February 2022 version.

\end{figure}   

Figure \ref{fig:diverse} shows the share of 3- and 4-digit CPC classes covered by each AI patent set. The USPTO sample spans the broadest range of technology classes, covering between 70-90\% of all possible CPC codes. This can be explained by the large number of AI patents in the USPTO approach, compared to the others (Table \ref{tab:ai_traj}). 

However, the smaller AI samples took off over time: beginning around 2010, the share of CPC codes associated with the Keyword, Science, and WIPO patents increased rapidly.
Significance tests indicate that the differences among the Keyword, Science, and WIPO approaches are not statistically significant. However, all three score significantly lower than the USPTO sample (see Tables \ref{tab:wilcoxon_complementarity_shareCPC3} and \ref{tab:wilcoxon_complementarity_shareCPC4}).

\begin{figure}[h]
     \centering
     { \caption{Patent-level diversity - Average technology classes}
        \label{fig:diverse_perpatent}
     \begin{subfigure}[b]{0.45\textwidth}
         \centering
         \includegraphics[width=\textwidth]{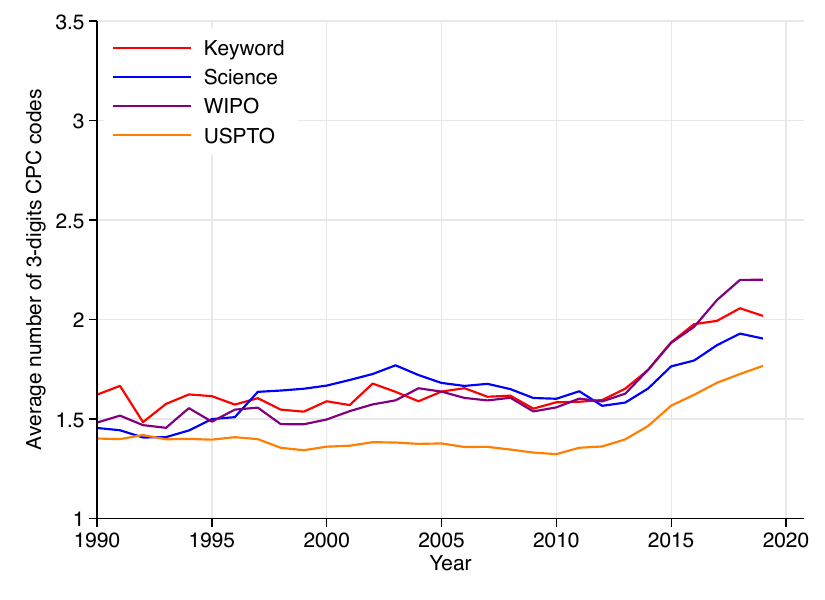}
         \caption{Average number of 3-digit CPC}
         \label{fig:diverse_perpatent_3d}
     \end{subfigure}
     \hfill
     \begin{subfigure}[b]{0.45\textwidth}
         \centering
         \includegraphics[width=\textwidth]{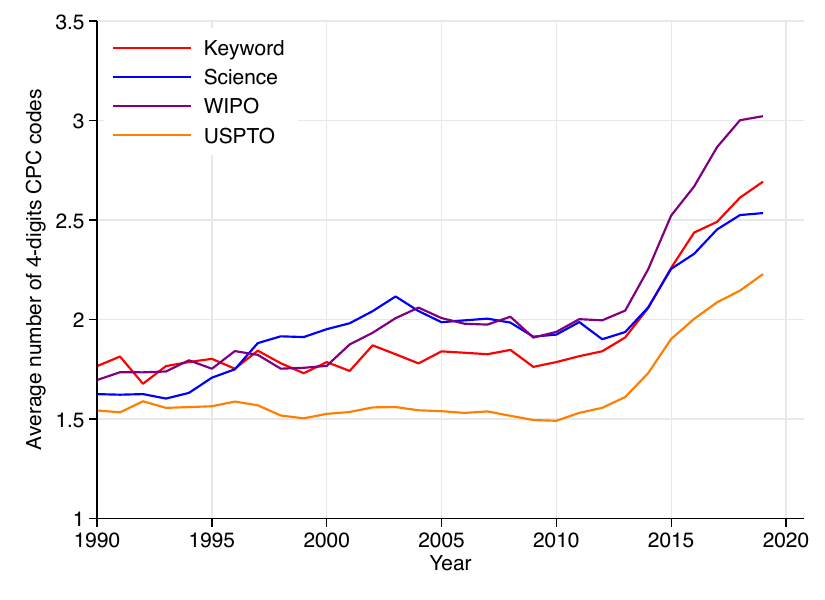}
         \caption{Average number of 4-digit CPC}
         \label{fig:diverse_perpatent_4d}
     \end{subfigure}
     }
\end{figure}   

To account for differences in sample sizes, we compute the annual average number of 1-, 3-, and 4-digit CPC codes per patent (see Table \ref{tab:gpt_summary}, \ref{app:cpc_134sum}, and \ref{app:tables_moved_down_generality}). Across all years, an average WIPO-classified patent is linked to 1.64 3-digit classes and 2.05 4-digit classes. Keyword and Science patents are associated with slightly fewer CPC classes on average, while USPTO patents exhibit the lowest degree of multidisciplinarity by this measure. The superior performance of the WIPO method at the 4-digit level and the lower diversity of the USPTO sample are both statistically significant.

At the 1-digit CPC level, Keyword and Science AI patents exhibit similar values (1.39–1.40), and the difference between them is not statistically significant. By contrast, the other two approaches rank significantly lower, with the USPTO method displaying the least complementarity.
Figure \ref{fig:diverse_perpatent} illustrates the evolution of average co-classifications over time at the 3- and 4-digit levels. All panels indicate a trend toward increasing technological diversity in the second half of the last decade, with the most pronounced growth observed for WIPO patents.  

In Appendix \ref{appendix_bench_growth}, we report results for other benchmark GPT candidates, confirming that our measures effectively reflect the broadening scope of these technologies, as suggested in the literature.\footnote{Due to the heterogeneity of CPC classes, comparing diversity across different GPT candidates is challenging. Our selection of benchmarks is based entirely on CPC codes at the 3- and 4-digit levels. For instance, the Y02 codes used to identify climate technologies naturally span a broad range of technological areas, making them inherently more diverse.}

Summing up, all AI samples exhibit increasing technological diversity over time.
When accounting for differences in sample size, we find that the WIPO method captures the highest diversity at more disaggregated CPC levels, while the Science and Keyword approaches show greater diversity at the broader 1-digit CPC section level.
The high diversity of WIPO patents at finer levels of classification likely reflects the method’s design, which targets patents assigned to various definitions of computing, operational processes, and hardware components of AI (see Section \ref{sec:approaches_wipo}).
In contrast, the Science and Keyword approaches are not constrained by the CPC classes included. 

\FloatBarrier
\subsection{Concentration in AI}
\label{sec:Market_power}

Many policy debates around AI highlight concerns about increasing concentration in the market for AI technology, raising concerns about competition and equitable access to the benefits of AI \citep{petit2017antitrust, babina2022artificial, mazzucato2022governing, jelinek2021policy, schmitt2022mapping}.
In the following, we provide empirical evidence on these concerns across different conceptions of AI, as captured by our four AI patent samples.

The Science approach captures the highest share of non-commercial AI inventions--those filed by individuals, non-profits, and universities. It also shows a strong concentration of AI patenting in the pharmaceutical sector, in contrast to the WIPO and Keyword approaches, which are more centred in the machinery and manufacturing industries.
This reflects the Science sample’s higher share of biotech-related patents, which are often filed by academic inventors and originate from university-based research.

Geographically (Panel E), the Keyword and WIPO samples include a higher share of AI patents filed by foreign inventors, particularly from Japan. Notably, Chinese inventors are absent from the top ranks, despite China’s increasingly prominent role in AI development in recent years \citep{jacobides2021evolutionary}. This absence may be partly due to the time period under study, as China’s emergence at the global frontier of high-quality patenting has primarily occurred over the past decade. However, it may also reflect a methodological bias stemming from the exclusive use of US patent data in this analysis.\footnote{A discussion of the implications of this bias can be found in Section \ref{sec:limitations}.}

Moreover, the four classification methods draw different pictures of the extent to which AI inventions rely on public R\&D support (Table \ref{tab:ai_traj}, Panel F). In the USPTO sample, only about one-fifth of AI patents received public funding, whereas in the Science sample, this share rises to nearly half in the most recent decade.
This pattern aligns with differences in inventor types (Table \ref{tab:ai_traj}, Panel C): AI patents identified by the USPTO method appear primarily commercially driven, while those in the Science sample are more often linked to academic inventors. This is consistent with the fact that public R\&D funding is frequently channelled through universities and research institutions.

\begin{table}[h]
\centering
\caption{Top AI-producing firms}
\label{tab:ai_actors1}
\footnotesize
{
\def\sym#1{\ifmmode^{#1}\else\(^{#1}\)\fi}
\begin{tabular}{lclclclc}
  \hline
  \hline
\multicolumn{2}{c}{Keyword}  & \multicolumn{2}{c}{Science} & \multicolumn{2}{c}{WIPO} & \multicolumn{2}{c}{USPTO} \\[0.25em] 
Company name &	\% &	Company name &	\% &	Company name &	\% &	Company name &	\% \\[0.25em]
  \hline
IBM Corp&0.05&IBM Corp&0.07&IBM Corp&0.07&IBM Corp&0.09 \\[0.25em]
Microsoft&0.02&Microsoft&0.06&Microsoft&0.04&Microsoft&0.04 \\[0.25em]
Samsung&0.02&Google&0.03&Google&0.03&Google&0.02 \\[0.25em]
Fanuc Corp&0.02&Apple&0.02&Canon&0.02&Intel&0.02 \\[0.25em]
Google&0.02&Sony&0.01&Samsung&0.02&Samsung&0.01 \\[0.25em]
Siemens&0.02&Siemens&0.01&Sony&0.02&Hewlett Packard&0.01 \\[0.25em]
Honda Motor&0.01&Hewlett Packard&0.01&Intel&0.01&AT\&T&0.01 \\[0.25em]
Amazon&0.01&Intel&0.01&Amazon&0.01&Sony&0.01 \\[0.25em]
Intel&0.01&AT\&T&0.01&Siemens&0.01&Amazon&0.01 \\[0.25em]
Sony&0.01&Canon&0.01&Fujitsu&0.01&Canon&0.01 \\[0.25em]
\hline
   \hline
\end{tabular}
}
\footnotesize \justifying

Notes: This table reports the top-ten AI producing firms for each AI definition. IBM, Microsoft, and Google are among the top-five AI patenting firms across all four groups. The column share reports the share of commercial patents accounted by a firm within each AI definition. For example, IBM accounts for 4-7\% of all AI patents produced by commercial firms. 
\end{table}

In Table \ref{tab:ai_actors1}, we present the top ten firms by their share of AI inventions for each of our AI samples. \KH{Additional results, including top assignees and concentration by technological field at the 1- and 3-digit CPC level, are reported in \ref{app:concentration_by_field}.}
All methods consistently highlight AI patenting dominated by a small number of leading technology and communication firms. Notably, IBM, Microsoft, and Google occupy the top ranks across all methods. 
\KH{However, their top positions are concentrated within CPC sections G (Physics) and H (Electricity), regardless of the AI classification method. At a more granular level, their dominance is most pronounced in specific domains--namely, Electronic Communication (H04) and Computing (G06). For other technology fields, we observe a wider range of leading firms (see \ref{app:concentration_by_field}).}

\begin{table}[h]
\centering
\caption{Concentration of Firms innovating in AI}
\label{tab:ai_actors2}
{
\def\sym#1{\ifmmode^{#1}\else\(^{#1}\)\fi}
\begin{tabular}{lcccc}
  \hline
  \hline
 & Keyword & Science & WIPO & USPTO \\[0.25em] 
  \hline
  \multicolumn{5}{c}{\textit{A. Concentration Ratio (CR)}}   \\ [0.25em] 
Four-firm CR        &       0.117&       0.174&       0.162&       0.177\\[0.25em]
Eight-firm CR       &       0.175&       0.219&       0.221&       0.228\\[0.25em] 
\hline
\multicolumn{5}{c}{\textit{B. Herfindahl–Hirschman Index (HHI)}}   \\ [0.25em] 
HHI (overall)       &       0.007&       0.012&       0.011&       0.014\\[0.25em]
HHI (1990-1999)     &       0.008&       0.012&       0.013&       0.014\\[0.25em]
HHI (2000-2009)     &       0.006&       0.013&       0.015&       0.016\\[0.25em]
HHI (2010-2019)     &       0.009&       0.013&       0.011&       0.014\\[0.25em]
  \hline
   \hline
\end{tabular}
}

\footnotesize \justifying

Notes: This table shows the measures of concentration of AI-producing commercial firms.  The concentration ratio (CR) measures the market share of top-four (or top-eight) firms. The Herfindahl–Hirschman Index (HHI) is calculated as the sum of squares of the shares of patent produced by each firm within each of the four AI definition patent samples.
\end{table}

To assess concentration beyond the top-ten firms, we calculate standard concentration measures--the concentration ratio (CR) and the Herfindahl-Hirschman Index (HHI)--reported in Table \ref{tab:ai_actors2}. The four-firm (eight-firm) CR captures the share of AI patents attributed to the top four (top eight) assignees, while the HHI is calculated as the sum of squared patent shares of all firms, providing a more comprehensive measure of market concentration.
Across measures, Keyword patents exhibit lowest concentration (HHI and CR values).
\KH{Further disaggregation by 1-digit and 3-digit CPC codes is presented in \ref{app:concentration_by_field}, showing how concentration varies across technological fields.}

Contrary to concerns raised in the literature \citep{petit2017antitrust}, we do not find evidence that AI patenting has become increasingly concentrated among fewer firms over the past three decades. While we observe some fluctuations in concentration levels, there is no consistent upward trend across any of the four AI samples. It is important to emphasize that our analysis is limited to the distribution of patented AI inventions and does not capture other dimensions of concentration.
In particular, broader concerns about AI-driven concentration often relate to final goods markets--where productivity gains may accrue disproportionately to dominant firms \citep{babina2022artificial}--or to the control over strategic inputs such as data, computational infrastructure, and digital platforms \citep{jacobides2021evolutionary, franco2023producing}, which fall outside the scope of our patent-based approach.

\KH{Our analysis of field-specific dominance and concentration of AI inventions across 1- and 3-digit CPC classes reveals no substantial variation in concentration levels across technological fields or classification approaches. While aggregate indicators suggest that overall concentration remains limited, we do observe a set of dominant actors operating within particular niches of AI technology. This pattern is consistent with the conceptualisation of AI as an ecosystem comprising interdependent technologies, applications, and complementary assets \citep{jacobides2021evolutionary}. From the perspective of patented inventions, our findings do not provide evidence of a general trend toward rising concentration across the AI landscape.}

\FloatBarrier
\section{Discussion}
\label{sec:discussion}

Defining AI remains challenging, as it is an emerging technology whose full impact is yet unknown \citep{krafft2020defining, schumpeter2005joseph, barbieri2025evolving}. The evolution of AI is shaped not only by technological progress but also by socio-technical dynamics and regulatory frameworks \citep{geels2005technological}. 
Shaping AI has gained increasing attention in current discussions on AI governance, aimed at mitigating risks and maximising benefits for all \citep{mazzucato2022governing, bresnahan2023innovation}. 
Effective AI policy-making relies on the ability to clearly delineate the domain of AI and to identify areas where regulatory intervention or targeted support may be warranted \citep{krafft2020defining}.

Our analysis of GPTness and concentration provides insights relevant to two key dimensions of AI policy.
First, policy efforts could prioritise forms of AI that generate the greatest public value--those characterised by broad technological spillovers and societal benefits \citep{mazzucato2022governing, bresnahan2023innovation}. Such AI inventions may be identified by their high levels of GPTness, which reflect their capacity to stimulate follow-on innovation across the economy \citep{lipsey2005economic}.

Second, AI policy could aim to mitigate the risks of power concentration that arise when core resources to develop and deploy AI are unequally distributed.
Addressing concentration helps prevent the premature narrowing of innovation pathways toward areas with the greatest short-term private returns, while neglecting other areas with broader societal benefits \citep{klinger2022narrowing, jacobides2021evolutionary, bresnahan2023innovation}.

\subsection{AI as a General-Purpose Technology}

All four AI classification methods show consistent trends: limited patenting in the 1990s (``AI winter'') followed by strong growth post-2000 \citep{stuart2003artificial, klinger2022narrowing}. Each supports the view of AI as a GPT and associate AI with similar industries and core technologies. They also reveal comparable patterns of concentration among a similar set of key firms.
These shared characteristics are consistent and informative for scholars and regulators who are searching for a robust definition of AI, as a basis for research and regulatory action \citep{krafft2020defining}. 

\begin{table}[h]
\small
\centering
\caption{\label{fig_summary} Summary of Findings \\}

\begin{tabularx}{1\textwidth}{lccccll}
\toprule
{} &  Keyword &  USPTO &  WIPO &  Science &                Metric &         Based on \\
\midrule
Growth     &        \ccell{BrickRed}{ } &      \ccell{GreenYellow}{ } &     \ccell{bronze}{ } &        \ccell{bronze}{ } &           Growth rate &           Counts \\
Generality      &        \ccell{BrickRed}{ } &      \ccell{GreenYellow}{ } &     \ccell{bronze}{ } &        \ccell{bronze}{ } &      Generality index &        Citations \\
Complementarity &        \ccell{bronze}{ } &      \ccell{GreenYellow}{ } &     \ccell{BrickRed}{ } &        \ccell{bronze}{ } &  Avg. \# tech. classes &  CPC codes \\
\bottomrule
\end{tabularx}


\footnotesize \justifying

This brief summary of our results shows which patent group generates the strongest average estimate of each GPT characteristic over the last 10 years. Red (yellow) colour indicates the strongest (weakest) performance.
\end{table}
Quantitatively, we observe some differences across approaches: AI patenting identified by the Keyword and WIPO methods shows an acceleration over the past decade, whereas the other approaches suggest a relative slowdown. 
Keywords and WIPO also identify patents that are most GPT-like and capture more from machinery manufacturing. This is reflected in the key firms emerging as top AI inventors\KH{, particularly within computing and electronic communication technologies}. 
The Keyword method, while capturing the narrowest set of inventions ($\sim$54k patents), shows the highest levels of GPTness (Table \ref{fig_summary}) and the lowest levels of concentration in inventive activity.
Framing AI as a GPT using our patent-based metrics aligns with patterns observed in historical GPTs--such as computers, communication technologies, and electrical engineering--in patent data \citep{petralia2020mapping}. 

\subsection{Concentration in AI patenting}

Concentration and GPTness interact, as GPTs have public good characteristics, with high innovation spillovers and lagged private returns for diverse actors, reducing the incentives for short-term profit maximisation \citep{bresnahan1995general}. 
This may help explain the lower concentration levels in AI inventions that are most GPT-like. The low concentration of the Keyword-based AI sample aligns with their high GPTness ranking (Section \ref{sec:GPT_char_of_AI}), reinforcing the interpretation that they capture -like segment of AI. 
The Keywords many patents centred on ML, a technology also highlighted by others as having high GPT potential \citep{goldfarb2023could}. 

The full realisation of GPT benefits is not guaranteed. Commercial actors may prioritise private returns over social value, potentially resulting in premature lock-in to suboptimal trajectories of AI development \citep{klinger2022narrowing, mazzucato2022governing, bresnahan2023innovation}.
Our identification of core technological domains--particularly those captured by Keywords with the strongest GPTness and lowest concentration--may indicate areas in which public support could be justified. However, any such decisions should be informed by methodologically pluralist \KH{and up-to-date} evidence, as each classification approach--including our own--has inherent limitations (see Section \ref{sec:limitations}).

Given their GPTness, Keyword-identified technologies are strong candidates for targeted support. However, when addressing innovation bottlenecks, the challenges may not lie primarily at the level of invention--as captured by patents--but rather at the diffusion stage, where business transformation and AI adoption take place \citep{bresnahan2023innovation}. This aligns with concerns about concentration, as the capacity to adopt and benefit from AI systems remains unevenly distributed across firms \citep{babina2022artificial}.

We observed no evidence of increasing concentration at the invention stage, suggesting that entry barriers in AI innovation have not intensified--contrary to some earlier concerns \citep{tambe2020digital, agrawal2019economic, mazzucato2022governing}. 
This may reflect the nature of digital inventions, which are typically characterised by low capital intensity, short technological cycle time, and high modularity \citep{bresnahan2023innovation}. The modular structure of the AI ecosystem facilitates the entry of new players, particularly when computational resources and hardware components can be sourced externally at reasonable costs \citep{jacobides2021evolutionary, bresnahan2023innovation}.
\KH{Our analysis of concentration across technological subfields further reveals a diverse range of key actors, each specialised in particular fields.}
However, our patent-based analysis does not capture trends in vertical integration across the AI value chain, control over key network resources, or emerging patterns of polarised AI-driven productivity growth--all of which are relevant to competition and antitrust policy \citep{petit2017antitrust, ducuing2020data, franco2023producing, rikap2024varieties}.

\subsection{Research guidance on measuring AI patents}

Our comparison highlights that empirical conclusions in patent-based AI research vary significantly with classification method. For instance, the USPTO and Science approaches suggest a recent slowdown in AI invention activity--a trend not observed with the Keyword or WIPO methods. These discrepancies can lead to diverging interpretations and explanations, such as whether a decline or rise reflects a narrowing and corporatisation of AI research \citep{klinger2022narrowing, jee2023firms}.
We also find minor but meaningful differences in the top-ranked firms identified as AI leaders and their nationalities. These distinctions matter when analysing the performance of national innovation systems, given their influence on emergent pathways of AI specialisation \citep{rikap2024varieties, jacobides2021evolutionary}. 
For example, the Science approach includes Apple among the top-ten AI patentees but excludes Samsung, which consistently ranks among the top five in the other methods.
Such variations can have material consequences when investors, banks, and policymakers rely on patent-based indicators to assess economic performance and inform funding decisions.

Our comparison may inform patent-based AI research in selecting classification methods aligned with specific research objectives. Our findings suggest that \emph{AI as a GPT} might be best studied by using the Keyword or WIPO approaches. The USPTO method appears more appropriate for analysing the broad diffusion of AI, while the Science approach may be best suited for studying the interactions between research and AI development.

The simple Keyword approach--which yields a relatively narrow set of key patents--appears well-suited for research on emerging GPTs. Smaller patent groups may offer a clearer distinction from the broader landscape \citep{kovacs2021categories} and may signal greater potential for future growth. This contrasts with the USPTO method, which classifies 16.6\% of all US patents in 2019 as AI-related, potentially overstating its current breadth.\footnote{During our analysis, we also discovered that the Keyword method reproduced from \citet{cockburn2018impact} may be further simplified. We found that the majority of patents can be identified using a narrower set of four terms (ML, neural network, robot, pattern recognition), rather than the original list of over forty words.} 
However, we cannot rule out that the high levels of GPTness of the Keyword method reflect only a short-term trend of the ML uptake in commercially valuable inventions, as captured by patents, while the full scope of AI remains unleashed \citep{klinger2022narrowing}. 

Measuring emerging technologies is inherently difficult as definitions clarify only after development, diffusion, and final use \citep{schumpeter2005joseph, barbieri2025evolving, lafond2019long}. This paper quantifies areas of consensus and nuanced differences in AI classification. While we can only speculate, this emerging consensus may signal the type of AI that is most likely to be advanced in the future.

\section{Limitations}
\label{sec:limitations}
\KH{Our data ends in 2019 and reflects a view of AI shaped by pre-ChatGPT developments--prior to the rapid rise of large language models (LLMs) and generative AI. As such, it captures how AI was defined and measured before these transformative technologies reshaped the field. Assessing how recent trends affect our four classification approaches and GPT assessment is beyond the scope of this study, given its complexity. 
Patent classification systems co-evolve with technology, especially rapidly changing areas of technology \citep{barbieri2025evolving}. Non-systematic checks of recent amendments in the CPC system reveal a high revision intensity in CPC codes related to those used by the WIPO approach and used in our GPT assessment. Changes in the CPC codes maybe themselves an indicator of the evolving nature of AI.} 

\KH{Keyword-based methods are also time-sensitive. Exploratory queries to recent large language models (ChatGPT and Mistral) seeking keywords to identify current AI patents generated lists still predominantly centred on ML, but also included terms associated with specific AI edge applications, hardware components, and integrated AI systems.\footnote{We conducted exploratory queries in July 2025 using various versions of ChatGPT (4.5, 4o) and LeChat from Mistral. Notably, Mistral highlighted a larger number of hardware-related terms.} This shift aligns with the observed diffusion of AI across a growing number of sectors. Consequently, we would also expect the range of scientific fields contributing to AI innovations to have expanded significantly.}

\KH{Expanding our cross-sectional analysis of AI classification into a longitudinal dimension would be a valuable endeavour. Such an analysis could yield new insights into the evolution of AI trajectories, which may be relevant for policy. Moreover, reclassification dynamics of emerging technologies is relatively understudied, yet understanding this process could help in technological forecasting and policy implementation, with relevance beyond AI \citep{barbieri2025evolving}.}

Beyond these conceptual limitations, our research comes with other well-documented limitations and biases related to patent data. 
First, patents are utilised heterogeneously across industries, firms, and technologies, and some inventions are unsuitable for patent protection. Alternative mechanisms of intellectual property protection may predominate in specific sectors, leading to biases in patent-based analyses \citep{granstrand2009innovation}. Additionally, variations in the frequency of co-classifications and citations across technological fields \citep{jaffe2017patent, hotte2021rise} may influence empirical measures of GPTness. 

Second, patents capture technical inventions but do not reflect key dimensions of AI-driven technological change, such as diffusion, process innovations, input substitution, AI-related services, data resources, and other complementary assets \citep{goldfarb2023could, jacobides2021evolutionary, bresnahan2023innovation}. Nevertheless, our findings align with studies using alternative indicators--for example, \citet{goldfarb2023could}, who draw on occupational skill requirements and conclude that ML qualifies as a GPT.

Third, our analysis is limited to USPTO patents. While US patents are often considered a good proxy for the global technological frontier, this may not hold for AI. National specialisation in AI varies considerably, reflecting differing priorities and institutional contexts \citep{fujii2018trends, jacobides2021evolutionary, rikap2024varieties}. Although foreign inventors can and do file at the USPTO, they may encounter barriers or have reduced incentives compared to filing at home. In addition, differences in examination practices, citation behaviour, and classification standards across patent offices complicate the generalisation of our analyses to the international level \citep{jaffe2017patent}.

Fourth and in addition to our discussion above, the observed levels of GPTness across the different approaches may be specific to the time period analysed (1990–2019). The high GPTness found for the Keywords and WIPO may reflect the recent dominance of a narrow set of technology-related buzzwords (especially ML). These results might differ if reproduced in the future, as the boundaries and focus of AI continue to evolve. This limitation relates to the debate on ``AI narrowing'' \citep{klinger2022narrowing}, which illustrated how recent AI research has become less diverse, concentrated within a few large companies, and focused on a limited set of ML techniques. However, the future breadth of AI development may be shaped by political and regulatory decisions made today.

Lastly, our analysis is restricted to GPTness and concentration as criteria relevant for AI policy, while remaining silent on other objectives such as safety, societal impact, and privacy \citep{roche2023ethics, krafft2020defining}. Efforts to empirically assess the impact of policy on the evolution of AI should draw on a range of AI measurement methods and incorporate evaluation criteria that are aligned with the goals set by policy.

\section{Conclusions}
\label{sec:conclusions}
We conducted a systematic analysis of four distinct approaches to identifying patented AI inventions, each capturing different and only partially overlapping sets of patents. Our findings show that both quantitative and qualitative assessments of AI are sensitive to the chosen classification method. In addition, we examined patterns of concentration in AI innovation. 
Despite methodological differences, we find consistent qualitative overlaps across key dimensions, which are strong compared to benchmark technologies. This suggests the emergence of a shared understanding of what constitutes AI. 

Our analysis offers guidance for policymakers and innovation scholars seeking to identify patented AI inventions. For example, researchers aiming to study the role of AI as a GPT may find that a simple keyword-based method captures a narrow set of patents that best exhibit canonical GPT features: endogenous growth, broad usefulness, and technological complementarity.

Taken together, our results (1) provide robust empirical support for conceptualising AI (particularly ML) as exhibiting key characteristics of a GPT, (2) demonstrate the usefulness of patent data for tracking AI innovation, and (3) underscore the importance of employing multiple classification methods to reduce methodological biases, especially when assessing emerging technologies and issues like market concentration.

\FloatBarrier
\newpage
\section*{Acknowledgements}
The authors thankfully acknowledge valuable feedback received from the participants of the OMPTEC-FoW and INET complexity group meetings. 
K.H. and T.T. thankfully acknowledge support from OMPTEC and Citi. V. V. acknowledges support from Intelligence Advanced Research Projects Activity (IARPA), via contract no. 2019-1902010003, for developing capabilities that this work builds upon. L. B. gratefully acknowledges support from the General Sir John Monash Foundation. 
\printbibliography

\newpage

\appendix
\renewcommand{\appendixname}{Appendix}
\renewcommand{\thesection}{\Alph{section}} \setcounter{section}{0}
\renewcommand{\thefigure}{\Alph{section}.\arabic{figure}} \setcounter{figure}{0}
\renewcommand{\thetable}{\Alph{section}.\arabic{table}} \setcounter{table}{0}
\renewcommand{\theequation}{\Alph{section}.\arabic{table}} \setcounter{equation}{0}

\section{Measuring GPTs}
\label{sec:GPTs}
Currently, there exists a number of alternative metrics to capture GPT characteristics. Given the lack of consensus, many believe GPTs should be better identified as sophisticated networks of technologies sharing ``underlying principles and mutual dependencies'' \citep{petralia2020mapping}.

Historically, patent growth rates have been used to capture the endogenous elaboration of technologies similar to GPTs \citep{ moser2004electricity, jovanovic2005general, petralia2020mapping}. \citet{petralia2020mapping} uses patent growth rates, co-classifications, and a text-mining algorithm to successfully reproduce the canonical GPTs contained within the broad USPTO categories of electricity and computer communication. However, the author finds great heterogeneity within these pools of patents, which contain both dynamic and stagnant inventions. Moreover, the author notes that the identification of more diffuse and diverse GPTs, such as AI, may require ``bottom-up'' classification approaches using lower levels of aggregation that can scan multiple technological classes for common principles.

\citet{hall2006uncovering} attempt to capture GPTs by measuring the patent growth rates and unbiased generality measures for the most-cited US patents and the patents which cite them. The authors also find great heterogeneity between patents, which underscores the need for multiple metrics to satisfactorily capture GPTs.

In the next section, we motivate our selection of patent measures for GPT characteristics and connect each with empirical facts about AI's dissemination and the three canonical GPT features.

\subsection{Growth}
For more than a decade, AI methods have become more powerful and complex as a result of new technical methods, increased data availability, and improved hardware. Consequently, AI invention has shifted away from specific application-based methods to more generalised learning-orientated systems \citep{cockburn20194}. With this refinement, the performance of many sub-fields of AI, such as image and text recognition, have seen remarkable improvements in performance \citep{brynjolfsson2021productivity}. This is reflected in the exponential growth of patenting activity referencing terms such as ML and deep learning (see Appendix, Figure \ref{keywords_study_learning}). 

Based on these observations, we measure improvements in AI via the growth rates of each group of patents and changes to their share of all patents, from 1990 to 2020 \citep{hall2006uncovering, petralia2020mapping}. We also look at the growth of the patents that cite such technologies: the ``GPT hypothesis'' in previous work has been that inventions that build on GPT-like technologies should spawn more new inventions \citep{hall2006uncovering}.

Let $N_{i,t}$ denote the number of patents in a group $i \in \{ \text{keyword}, \allowbreak  \text{science}, \text{WIPO}, \text{USPTO} \}$ at time $t$, indexed by year. We compute the growth rate as 

\begin{align}
\frac{N_{i,t}-N_{i,t-1}}{N_{i,t-1}}.
\label{eq:growth}
\end{align}


\subsection{Generality}
\label{sec:methods_generality}
AI has already begun to pervade a myriad of industries as it expands beyond computer science into such diverse fields as structural biology, transport, and imaging \citep{cockburn20194}. In the early 1990s, AI methods remained largely confined to computer science. However, over the past decade, the majority of patents referencing these technologies have appeared in secondary domains \citep{cockburn20194}. Based on the work of \citet{trajtenberg1997university}, we capture this stylised fact through the ‘generality’ of patents, measuring the dissemination of AI across different technology fields. 

To do so, we build on patent citation data and assume that a forward citation link entails information about the use of a patent in a subsequent invention \citep{jaffe2017patent}. To operationalise wide usefulness, we rely on a modified version of the generality metric by \citet{trajtenberg1997university} and \citet{hall2006uncovering} given by
\begin{align}
    1 - \sum_j^{N_j} \left( \frac{\#cites_{ij,t}}{\sum^{N_j}_{j=1} \#cites_{ij,t}} \right)^2
    \label{eq:generality}
\end{align}
where $\#cites_{ij}$ is the sum of citations to patents labelled as AI by classification approach $i$ from technology class $j$, whereby we use the CPC 1-digit level as class. The number of citations $\#cites_{ij}$ excludes citations within the same class: $N_j$ is the number of different CPC classes. Our approach differs to that of \citet{trajtenberg1997university} as we apply the method to each group of AI patents belonging to a variety of CPC sections. 
For the main analysis, we focus on 1-digit CPC sections, as these are more technologically distant than 3-digit or 4-digit classes and subclasses, whose results we also report. 

Our generality measure is calculated for the entire group of patents in $i$ with $N_i$ unique patents. To address concerns that this metric may be affected by differences between group sizes, we additionally calculate patent-level metrics given by the average number of citing classes, i.e. 
\begin{align}
    \frac{1}{N_{i,t}} \sum^{N_{i,t}}_{p=1} \sum^{N_d}_{j=1} \mathds{1}(\#ncites_{p,j,t} \geq 0)
    \label{eq:avg_cites}
\end{align}
where $\mathds{1}(\#ncites_{p,j,t} \geq 0) = 1$ if patent $p$ in $i$ is cited by at least one patent in technology class $j$ out of the total number of classes $N_d$ at level with $d \subset \{1,3,4\}$ s in the code. $N_i$ is the number of patents in approach $i$. Again, we exclude within-class citations and present results at both the 1-digit CPC section level ($d=1$) and higher orders of disaggregation ($d=3$ or $d=4$).

\subsection{Complementarity}

Thirdly, GPTs augment existing products and processes in a range of novel contexts to generate productive complementarities throughout the economy \citep{bresnahan1995general, petralia2020mapping}. AI technologies have been shown to complement and rely on secondary inventions, related to areas such as cloud computing and big data, which increase access to larger and more affordable data-sets \citep{brynjolfsson2019artificial}. Furthermore, because diverse AI systems share similar underlying structures and can share information, advances in one application of ML, such as machine vision, can spur inventions in other fields, such as autonomous vehicles. 

Following the approach of  \citet{petralia2020mapping}, we measure the extent to which AI patents enhance and supplement other inventions through the diversity of their technology class co-occurrences. For our analysis, we calculate the share of 3- and 4-digit CPC codes ($d=3,4$) assigned to the patents in each group of AI patents. Specifically, we calculate the following \emph{diversity measure} over time is

\begin{equation}\label{eq:eq_diversity}
    \frac{\#CPCs_{i,d,t}}{N_{d}}
\end{equation}

where $i$ denotes each of the four patent classification approaches, $d$ is the classification level and $t$ is year. $N_d$ refers to the number of CPC codes found in use for a particular group of patents, where the codes include $d$ digits. Note that there are 136 and 674 CPC codes, respectively at the 3- and 4-digit level (according to the February 2022 version of CPC codes).

As the above measure could be biased by patent volume, we also calculate the average number of distinct 1-, 3-, and 4-digit CPC codes per patent per year. The \emph{diversity per patent} over time is

\begin{align}
    \frac{1}{N_{i,t}} \sum_p \#CPCs_{p,i,d,t}
    \label{eq:diversity_per}
\end{align}
where $d$ represent the technology class represented by 1-, 3-, or 4-digit CPC codes. The time series graphs for the latter measures depict how an average patent's complementarity across technology sections evolves over time.

\FloatBarrier

\section{Measuring AI}
\label{sec:approaches}
In our analysis, we compare four methodologically and conceptually distinct approaches to identifying AI inventions in patents based on (1) keyword search, (2) science citations, (3) the WIPO, and the (4) USPTO method. Here, we introduce these classification approaches in detail. 


\subsection{Data source}
\label{sec:data_method}
We apply our methods to all patents granted by the USPTO from 1990-2019. For the analysis, we create four groups of AI patents for each classification method and complement each with supplementary information. 

From PATSTAT (Spring 2021 edition, \citet{patstat_data}) we sourced patent grant dates and from the USPTO we downloaded the Master classification file (April 2021 version) which contains CPC classifications of patents.\footnote{\url{https://bulkdata.uspto.gov/data/patent/classification/cpc/}} 
We added further data on patent-to-patent citations and patent titles from GooglePats obtained in an earlier project \citep{hotte2021rise}. 
For our analysis, we supplemented the citation data with citation year and the technology classes of both the citing and cited patent. In doing so, we obtained networks which represent citations from technology fields at different levels of aggregation to our four sets of AI patents. 
We also made use of the Reliance on Science database \citep{marx2020reliance} for citation data between patents and science. 

\subsection{Keyword search}
\label{sec:approaches_keyword}
Our first classification technique is a straight-forward approach based on keyword search, in which researchers use their discretion to develop a set of terms that reflect the very recent developments in AI. In this paper, we use the set of keywords provided in the appendix of \cite{cockburn2018impact}.\footnote{While we use the keywords from \cite{cockburn2018impact}, we do not fully replicate their approach. They use two subsets of patents: (1) patents classified by the USPC code 706 (Artificial Intelligence) and 901 (Robots); and (2) patents identified by searching titles for the selected keywords. Here we use patents identified by keyword search only, but we extend our search to match keywords also from abstract, claims, and description. We do not use the USPC classification codes since the WIPO method takes a more comprehensive approach combining keywords with IPC or CPC classifications. Also, with our extensive keyword search, we miss only a few patents which are in the first group (i.e., USPC 706 and 901) but not in the second group of \cite{cockburn2018impact}.} The keywords used in this paper focus on three sub-fields of AI: symbolic systems, learning algorithms, and robotics (see Table \ref{tab:keywords} for the full list of keywords). According to the authors, the symbolic systems represent ``complex concepts through logical manipulation of symbolic representations'' and include ``natural language processing'' and ``pattern recognition'. Learning algorithms include core analytic techniques such as neural networks, deep learning, and ML. The last category, robotics, is related to automation or applications of AI (e.g. computer vision and sensory networks). 

We search for these keywords in patent titles, abstracts, claims, and descriptions using USPTO data. 
We match the resulting list with patents granted by the USPTO between 1990 and 2019. The main advantage of the keyword approach is its simplicity and ease of implementation. Moreover, carefully chosen keywords can capture recent changes in the AI field. However, the success of this approach  depends on the judgement and familiarity of the researcher to the field of AI. Missing important keywords could lead to under-representation of a sub-field. Our approach yields 67,187 patents. 

\subsection{Science citations}
\label{sec:approaches_science}
This classification approach harnesses the scientific basis of patents. In particular, we classify a patent as an  AI patent if it makes at least one citation to a scientific paper in the scientific field of ``Computer Science; Artificial Intelligence'' (short, AI paper) as categorised by the Web of Science (WoS). 
Scientific citations are added to patent documents for multiple reasons such as describing the technological content of the invention or distinguishing the legal claim from other publicly available knowledge \citep[see ][]{narin1995linkage, meyer2000does,  tijssen2001global,ahmadpoor2017dual, marx2019reliance_working_paper}. A citation link to an AI paper indicates that the patent is technologically related to AI because it builds on scientific advancements in this field. A limitation of this approach is that it only identifies AI patents within the subset of patents that make citations to science.


For this method, we use data from the Reliance on Science (RoS) database \citep{marx2019reliance_working_paper, marx2020reliance, marx2021v30} which comprises a mapping from patents to scientific articles indexed in Microsoft Academic Graph (MAG) \citep{sinha2015overview}. Scientific articles are tagged by the WoS fields indicating the field of science into which an article is grouped.\footnote{Note that this assignment was made at the paper level using a probabilistic mapping which is different from the journal-based categorisation of Clarivate Analytics (Web of Science).} 

The citation links in the RoS database cover citations made by both the patent applicant and examiner, as well as citations indicated at both the front page and body of the patent document. \citet{marx2019reliance_working_paper} identified citations through a sequential probabilistic text recognition technique. Each citation link is tagged with a confidence score indicating the reliability of the matching approach. 
In the RoS data, roughly one third (34\%) of all US patents granted in 2019 can be attributed with at least one citation to science. 

In our study, we identified AI papers by their WoS categories and extracted all patents with at least one citation link to an AI paper. We kept only citation links with a reliability score greater than three, which corresponds to a precision rate of 99.5\% and a recall of 93\%. This approach yields 178,004 AI patents.

\subsection{World Intellectual Property Organisation (WIPO) method}
\label{sec:approaches_wipo}

The WIPO methodology for classifying AI patents was published in 2019 and validated by a team of patent experts \citep{wipo2019technologybackground,wipo19}. The aim behind the methodology is to capture three aspects of AI invention: (1) core AI techniques (deep learning, other learning methods, various type of logic, clustering, etc.); (2) functional applications of AI that can be used to simulate human-like cognitive capacities (such as vision, language, or decision-making); and (3) end-user application fields (such as automation in business, health, or military).
    
This methodology is based on both a keyword search of patent texts and the use of patent classification codes (CPC and IPC). In this technique, some patents are classified based on only a subset of the technological codes, or keywords, whilst others are identified by a combination of both.
    
The list of keywords used in this approach covers core AI methods as well as computing and mathematical concepts used in such technologies. These keywords are matched to the text in the patent titles, abstracts, and claims. 

This approach identifies 158,652 patents. 

\subsection{United States Patent and Trademark Office (USPTO) classification}
\label{sec:approaches_uspto}
The USPTO approach uses a supervised ML classifier to identify  AI patents \citep[see ][]{giczy2021identifying}. This ML model is trained to classify eight components of AI technologies, namely: ML, evolutionary computation, natural language processing, speech, vision, knowledge processing, planning/control, and AI hardware. The ML model is trained on the abstracts and claims of a seed (positive set) and an anti-seed (negative set). The seeds are chosen carefully for each respective component by taking an intersection of CPC, IPC, and USPC codes, as well as Derwent's World Patents Index\textsuperscript{TM}. The seeds are expanded based on patent families, CPC codes, and citations to identify all patents linked to the seed set. The anti-seed set is selected randomly from all remaining patents. For training, each text is pre-processed and embedded via the Word2Vec algorithm. The ML models also encode backward and forward citations in a citation vector. The predictions from the ML model are further validated using a small subset of patents that are manually examined.

Published in August 2021, the resulting dataset contains 13.2 million USPTO patents and pre-grant publications issued or published between 1976 and 2020. For consistency with our other approaches, we only consider patents granted between 1990 and 2019 and exclude pre-grant publications. The remaining data yields 595,047 patents.


\section{Keywords in detail}
\subsection{Words used in the Keyword approach}

\begin{table}[H]
\centering
      \small
    \caption{List of Keywords from \cite{cockburn2018impact}}  
    \label{tab:keywords}
    \begin{tabular}{p{4.5cm}|p{5cm}|p{4.5cm}}
    \hline\hline
Symbols &	Learning &	Robotics \\
\hline
natural language processing&	machine learning&	computer vision\\
image grammars&	neural networks&	robot\\
pattern recognition&	reinforcement learning&	robots\\
image matching&	logic theorist&	robot systems\\
symbolic reasoning&	bayesian belief networks&	robotics\\
symbolic error analysis&	unsupervised learning&	robotic\\
pattern analysis&	deep learning&	collaborative systems\\
symbol processing&	knowledge representation and reasoning&	humanoid robotics\\
physical symbol system&	crowdsourcing and human computation&	sensor network\\
natural languages&	neuromorphic computing&	sensor networks\\
pattern analysis&	decision making&	sensor data fusion\\
image alignment&	machine intelligence&	systems and control theory\\
optimal search&	neural network&	layered control systems\\
symbolic reasoning&	&	\\
symbolic error analysis&	&	\\
\hline\hline
\end{tabular}
\end{table}

\subsection{AI keywords in patent texts}
We split all the patent texts into three time periods (1990-1999, 2000-2009, 2010-2019) and search through the texts for keywords. Then, in each period (and for each category) we count the unique number of matching documents and what percentages of the AI patents match according to this keyword. Figures \ref{keywords_study_symbolic}, \ref{keywords_study_learning}, and \ref{keywords_study_robotics} below illustrate both counts and shares.

\newpage
\begin{figure}[H]
\centering
\caption{Symbolic keywords: in full texts\label{keywords_study_symbolic}}
\fbox{\includegraphics[width=13cm]{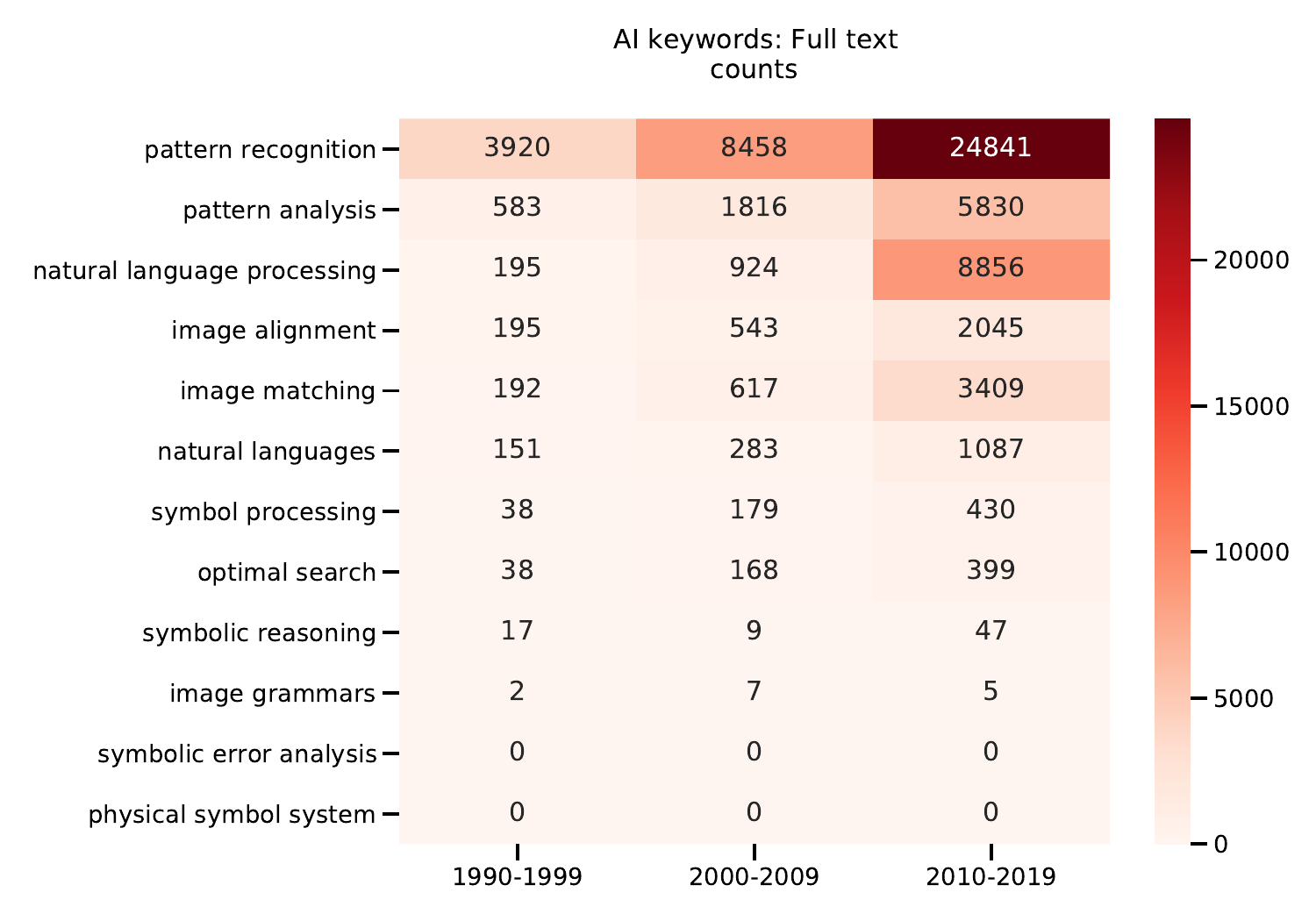}}
\\
\fbox{\includegraphics[width=13cm]{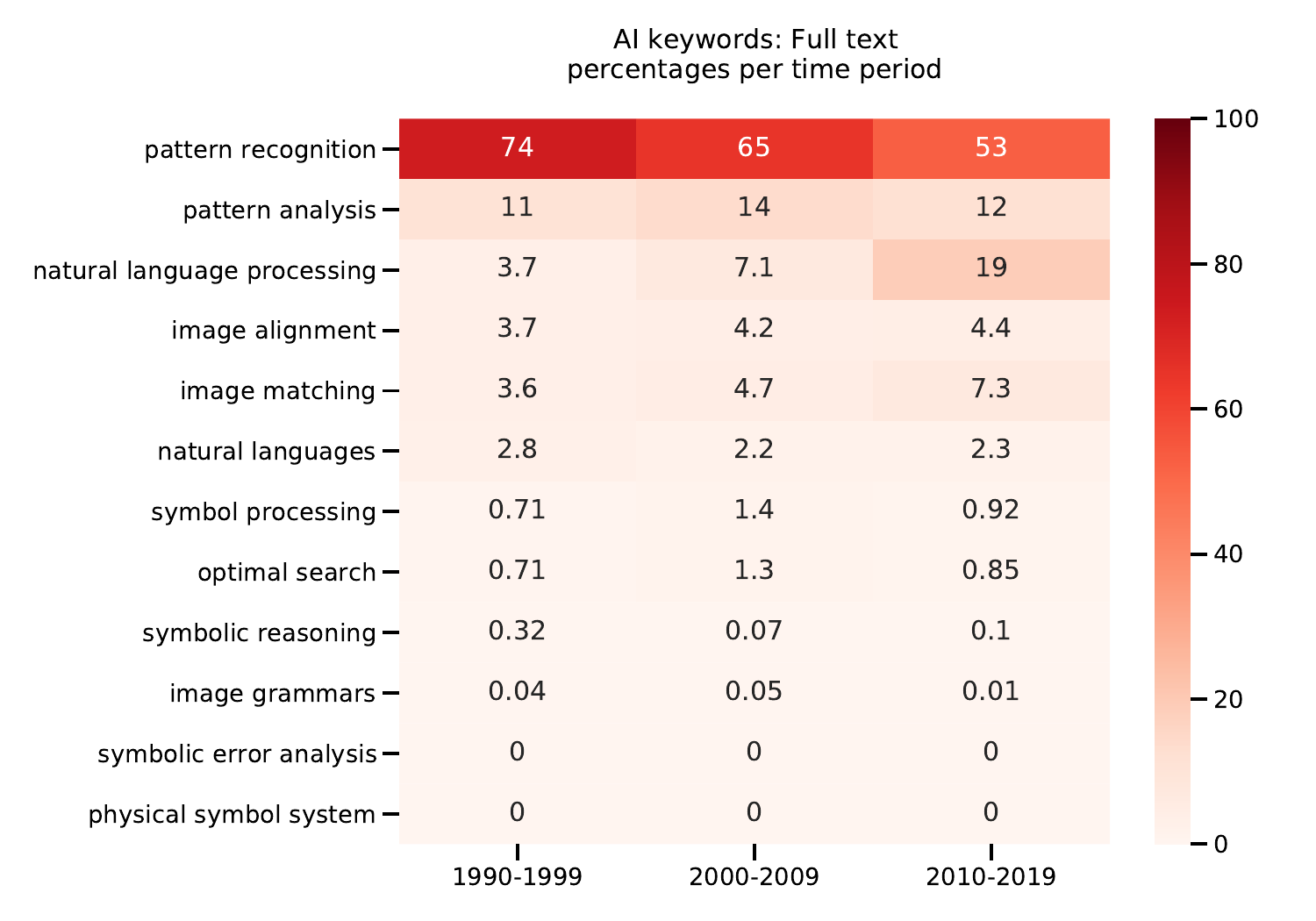}}
\end{figure}

\newpage
\begin{figure}[H]
\centering
\caption{Learning keywords: in full Texts\label{keywords_study_learning}}
\fbox{\includegraphics[width=13cm]{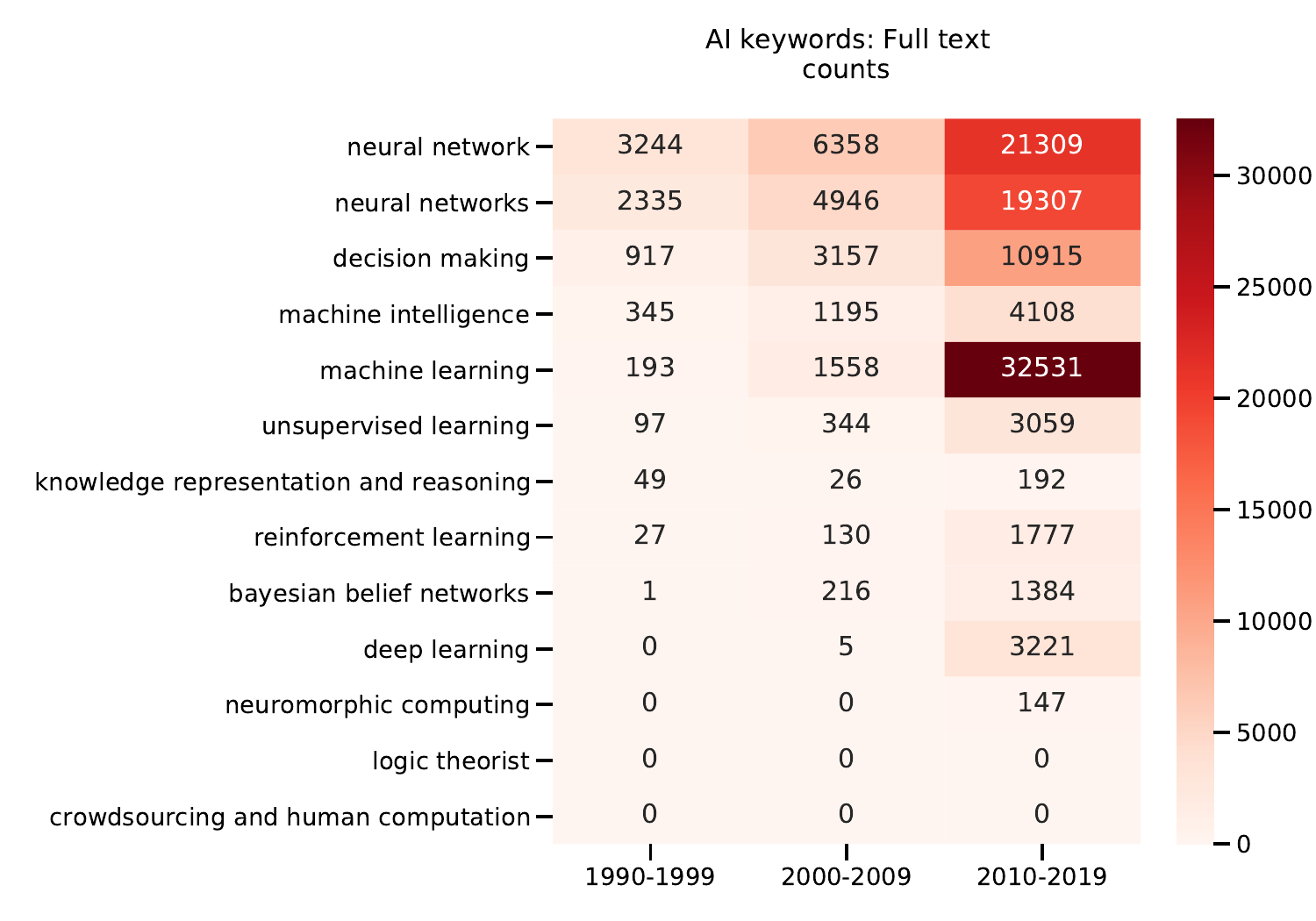}}
\\
\fbox{\includegraphics[width=13cm]{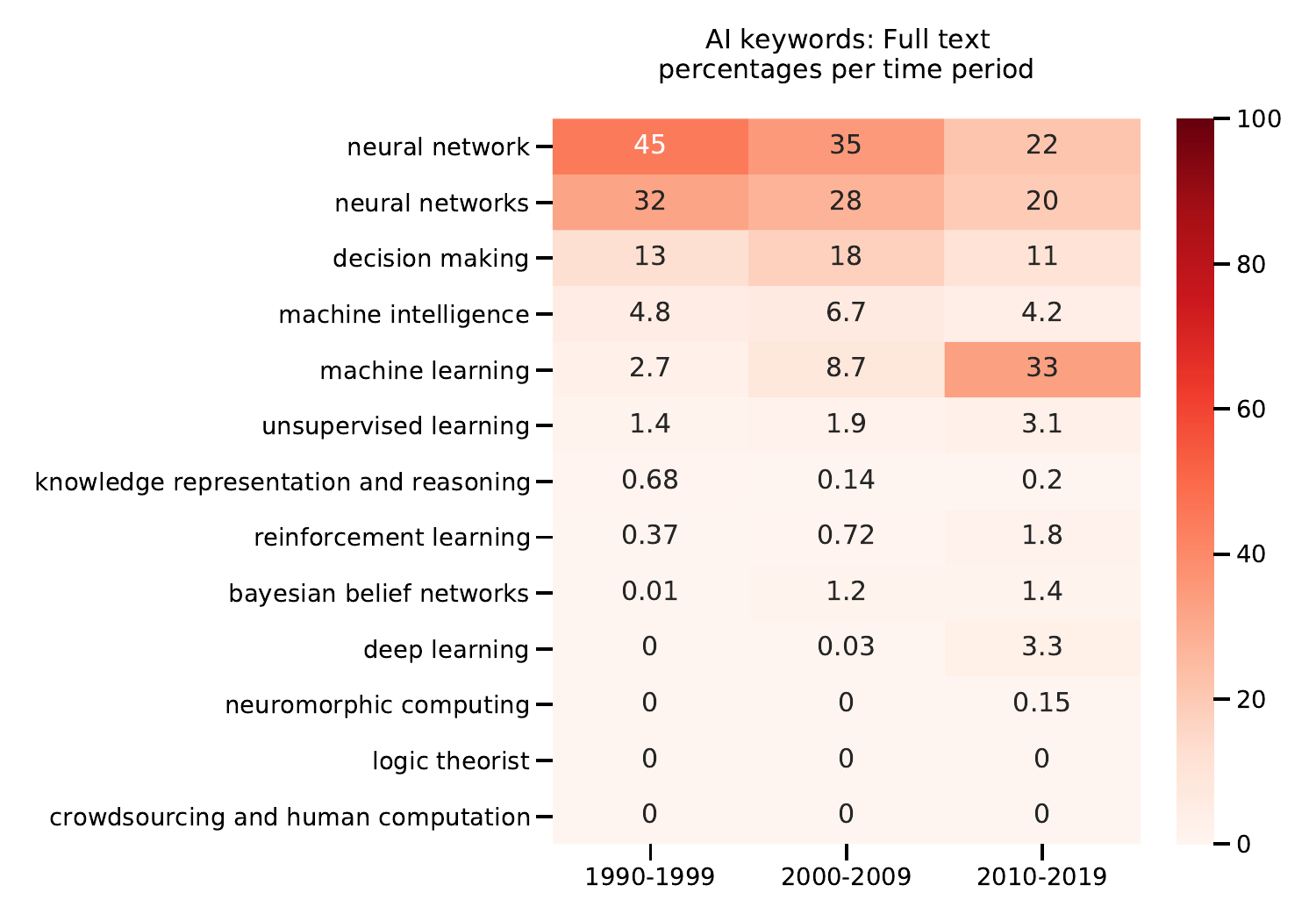}}
\end{figure}

\newpage
\begin{figure}[H]
\centering
\caption{Robotics keywords: in full Texts\label{keywords_study_robotics}}
\fbox{\includegraphics[width=13cm]{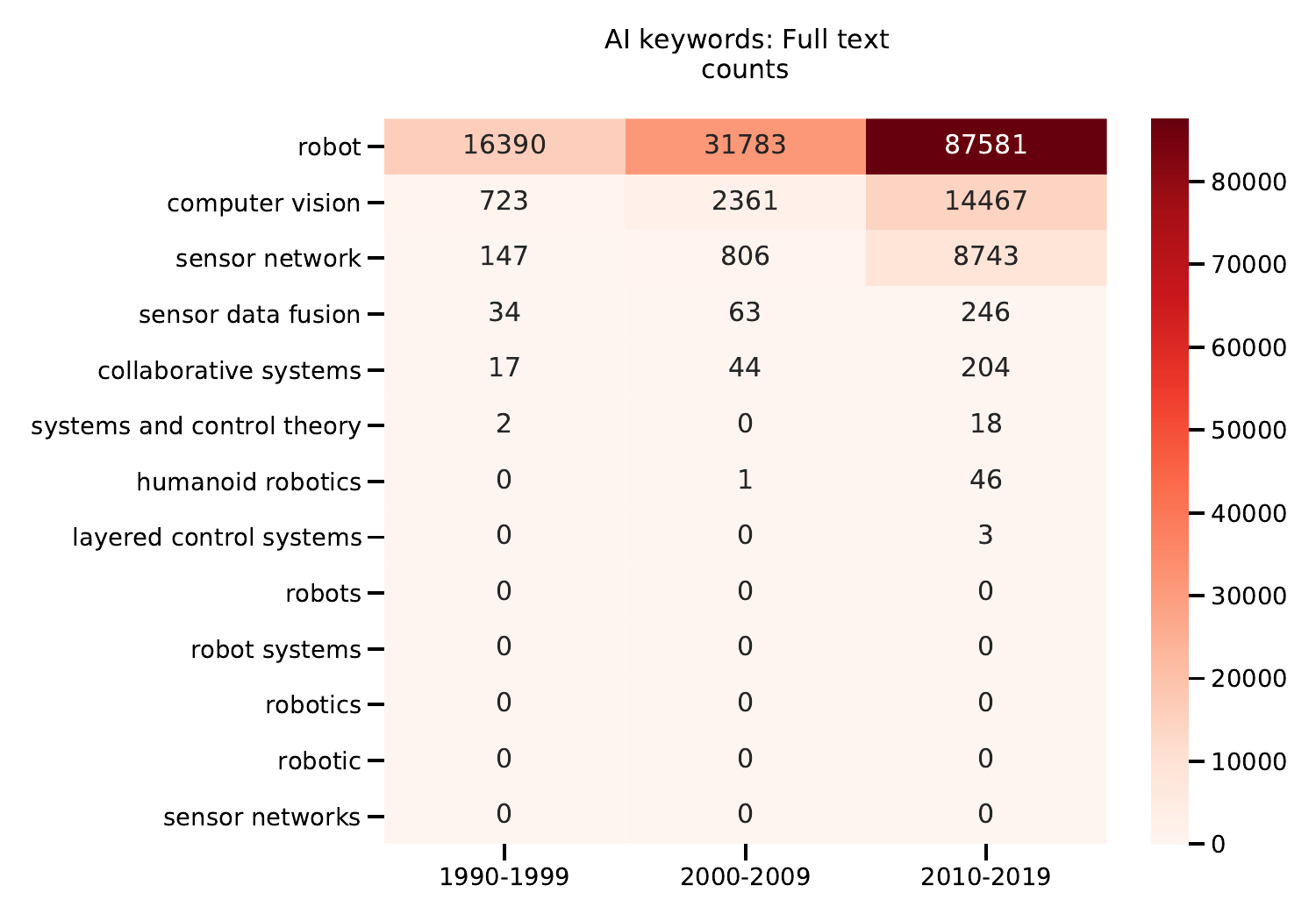}}
\\
\fbox{\includegraphics[width=13cm]{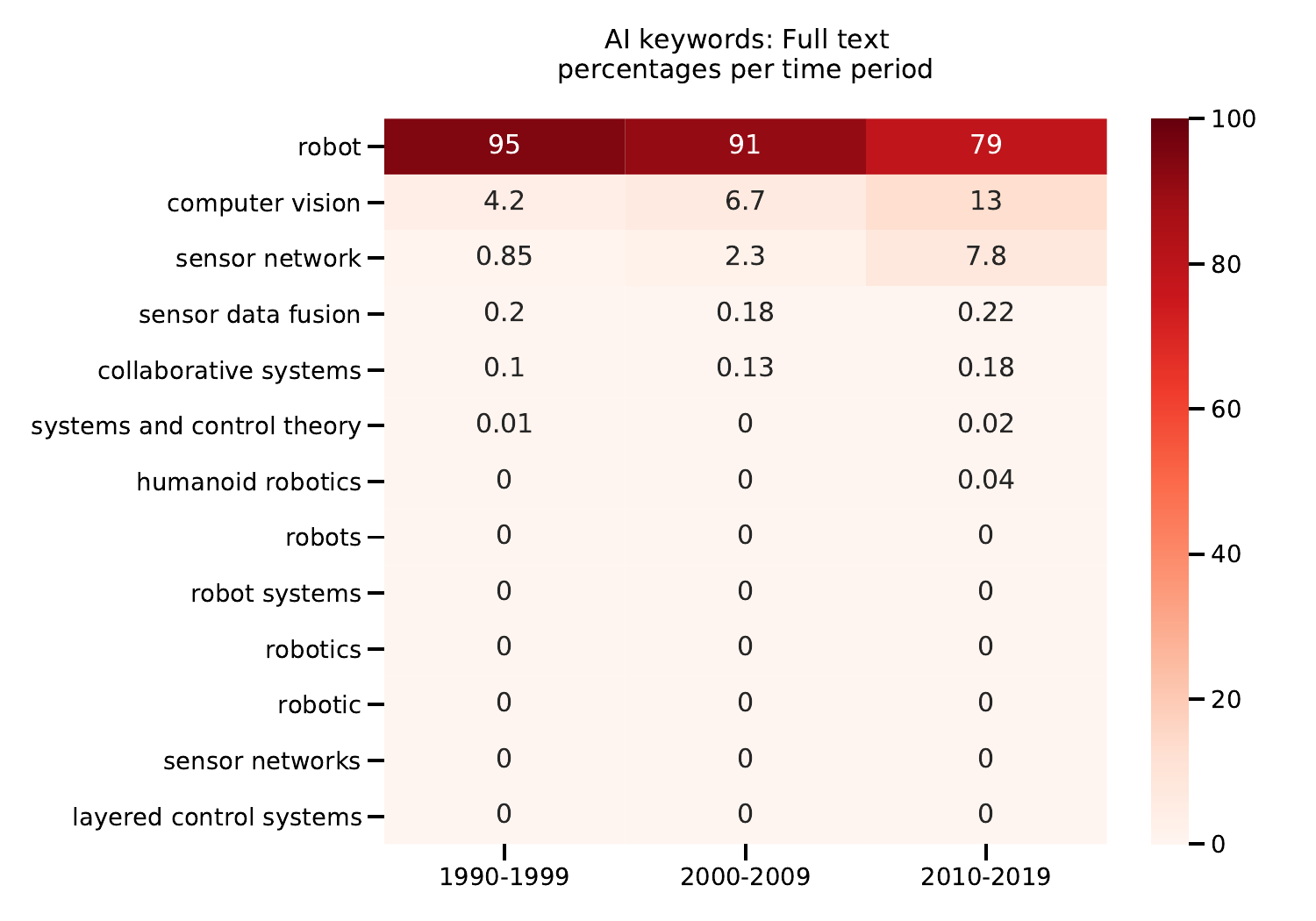}}
\end{figure}

\FloatBarrier

\section{Supplementary results}\label{appendix_additional}
\subsection{Comparison to benchmarks}\label{appendix_add_compare_bench}

The following figures reproduce time series of growth rates, counts, and shares for additional groups of patents. The benchmarks were identified in 
previous discussions of GPT technologies in the literature (nanotechnology, biochemistry, green technologies, computing). Climate patents were also included as a group of technologies where one can expect wide diversity, as climate inventions can be expected to cover many sectors of the economy.

\subsubsection{Growth}\label{appendix_bench_growth}





\begin{figure}[H]
    \centering
    \caption{Growth Rates of Benchmark Patents by Year}
    \label{fig:growth_rates_bench}
    \begin{subfigure}{0.3\textwidth}
        \centering
        \includegraphics[width=0.95\textwidth]{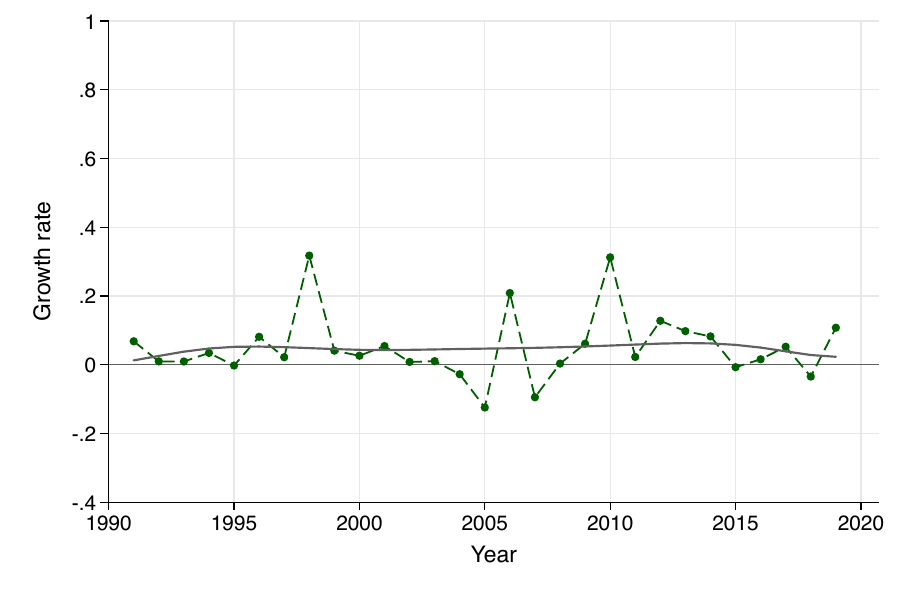}
        \caption{All}
         \label{fig:growth_rates_all}
    \end{subfigure}
    \begin{subfigure}{0.3\textwidth}
        \centering
        \includegraphics[width=0.95\textwidth]{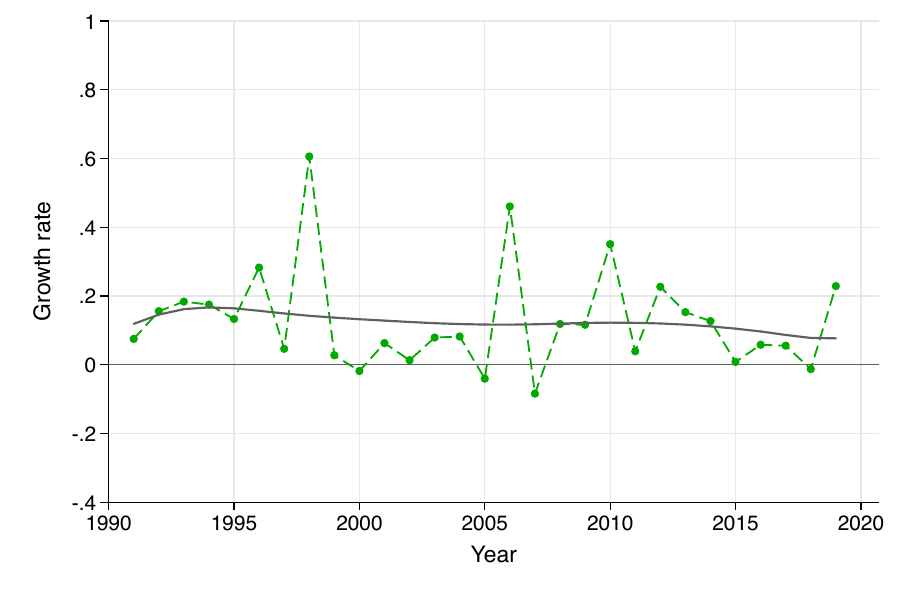}
        \caption{G06}
         \label{fig:growth_rates_g06}
    \end{subfigure}    
    \begin{subfigure}{0.3\textwidth}
        \centering
        \includegraphics[width=0.95\textwidth]{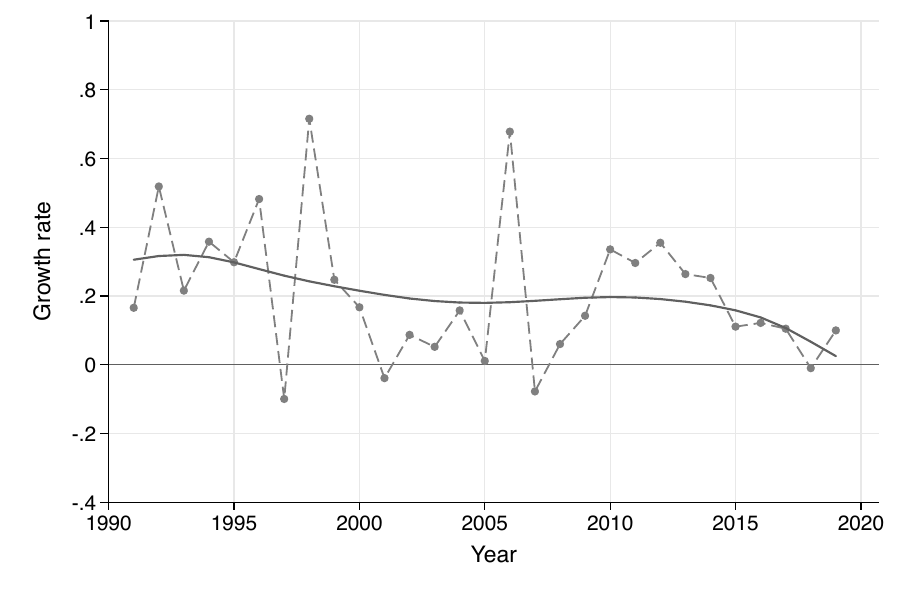}
        \caption{H04W}
         \label{fig:growth_rates_h04w}
    \end{subfigure}
    
    \begin{subfigure}{0.3\textwidth}
        \centering
        \includegraphics[width=0.95\textwidth]{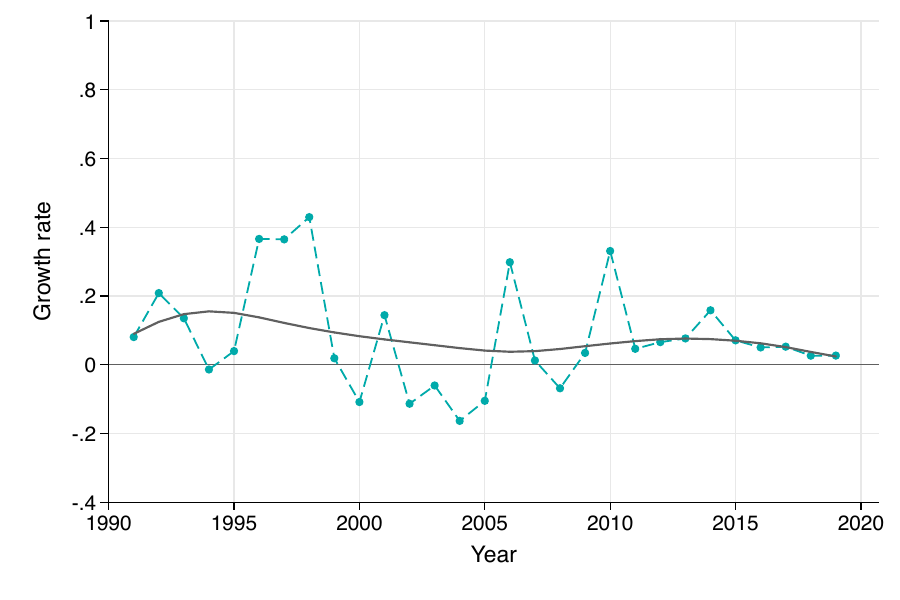}
        \caption{C12}
         \label{fig:growth_rates_c12}
    \end{subfigure}
    \begin{subfigure}{0.3\textwidth}
        \centering
        \includegraphics[width=0.95\textwidth]{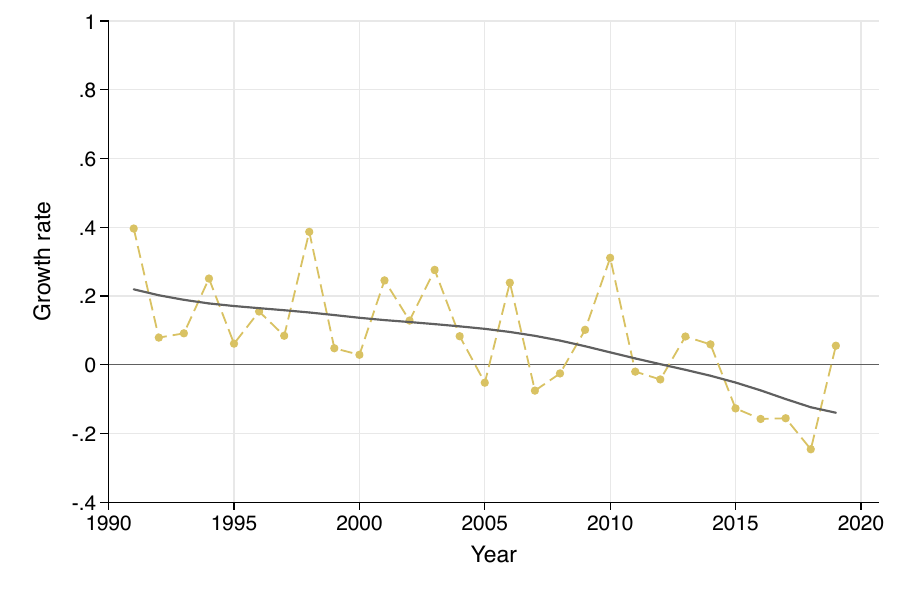}
        \caption{B82}
         \label{fig:growth_rates_b82}
    \end{subfigure} 
    \begin{subfigure}{0.3\textwidth}
        \centering
        \includegraphics[width=0.95\textwidth]{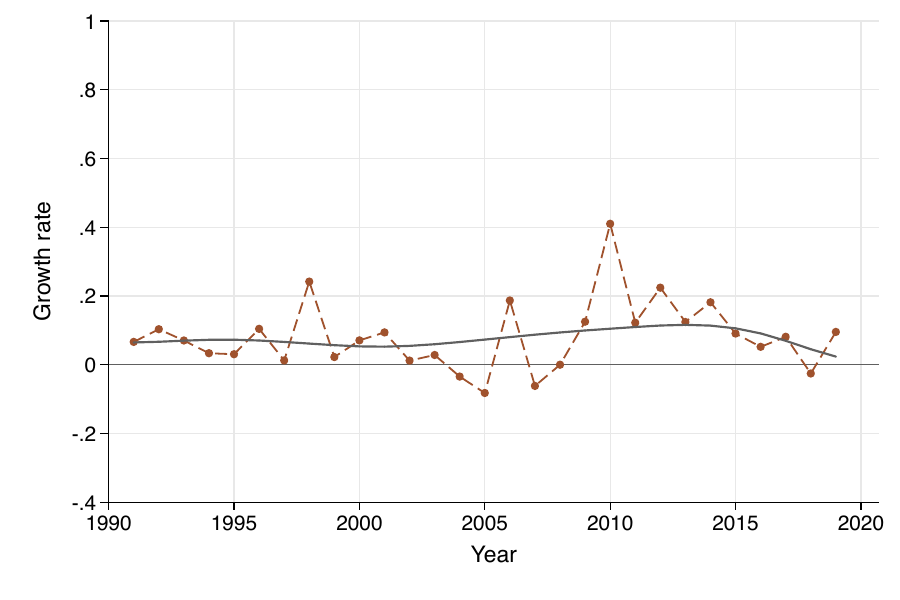}
        \caption{Y02}
         \label{fig:growth_rates_y02}
    \end{subfigure}
    
    \footnotesize\justifying
    Note: ``All'' refers to all patents, G06, H04W, B82, C12, and Y02 refers to computing, wireless communications, biochemistry/genetic engineering, nanotechnology, and climate invention-related patents, respectively.
\end{figure}

\FloatBarrier

\subsubsection{Generality}\label{appendix_bench_general}


\begin{table}[H]
\centering
\caption{Average Generality Index (1990-2019): Benchmark Categories} 
\label{tab:generality_benchmark}
\begin{tabular}{lcccccc}
  \hline
  \hline
& All & G06 & H04W & C12 & B82 & Y02  \\ [0.25em]
  \hline
  1 digit & 0.82 & 0.62 & 0.62 & 0.74 & 0.79 & 0.85  \\ [0.25em]
  3 digit &  0.95 & 0.82 & 0.82 & 0.85 & 0.92 & 0.95 \\ [0.25em]
  4 digit & 0.82 & 0.62 & 0.62 & 0.74 & 0.79 & 0.85 \\ [0.25em]

  \hline\hline
     \end{tabular}
    \footnotesize \justifying
    
   Notes: The generality index is defined as share of citations to patents in different CPC classes at different aggregation levels (see \ref{sec:methods_generality}). Citations within the same class are excluded. ``All'' refers to all patents, G06, H04W, B82, C12, and Y02 refers to computing, wireless communications, biochemistry/genetic engineering, nanotechnology, and climate invention-related patents, respectively.
\end{table}

\begin{figure}[H]
    \centering
    \caption{Generality Index at the 1-digit CPC-section Level}
    \begin{subfigure}{0.45\textwidth}
        \includegraphics[width=\textwidth]{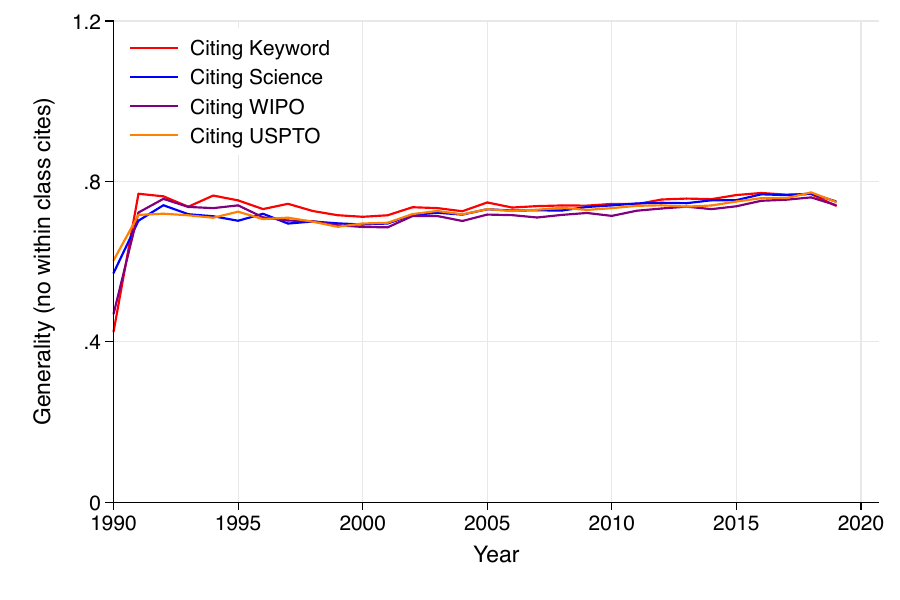}
        \caption{AI citing patents}
        \label{subfig:generality_generality_AI_citing}
    \end{subfigure}
    \begin{subfigure}{0.45\textwidth}
        \includegraphics[width=\textwidth]{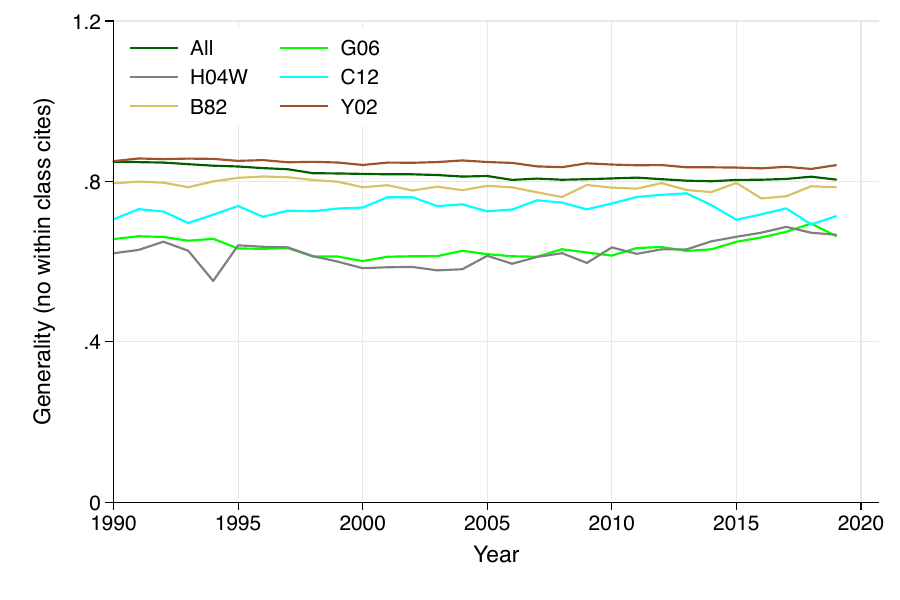}
        \caption{Benchmark}
        \label{subfig:generality_generality_benchmark}
    \end{subfigure}
\end{figure}


\begin{table}[H]
\centering
\caption{Average CPC Classes Making Citations: Benchmark Categories} 
\label{tab:generality2_benchmark}
\begin{tabular}{lcccccc}
  \hline
  \hline
& All & G06 & H04W & C12 & B82 & Y02  \\ [0.25em]
  \hline
  1 digit & 1.27 & 1.00 & 0.68 & 1.46 & 2.36 & 1.99  \\ [0.25em]
  3 digit & 2.48 & 1.99 & 1.19 & 2.52 & 4.39 & 3.32  \\ [0.25em]
  4 digit & 3.97 & 3.31 & 2.80 & 4.18 & 6.42 & 5.10  \\ [0.25em]

  \hline\hline
     \end{tabular}
    \footnotesize \justifying
    
    Notes: The table shows number of different CPC classes making a citation to an average patent of the respective group. Citations within the same class are excluded. ``All'' refers to all patents, G06, H04W, B82, C12, and Y02 refers to computing, wireless communications, biochemistry/genetic engineering, nanotechnology, and climate invention-related patents, respectively. 
\end{table}

\begin{figure}[H]
    \centering
    \caption{Average Number of CPC Classes Citing AI}
    \begin{subfigure}{0.45\textwidth}
        \includegraphics[width=\textwidth]{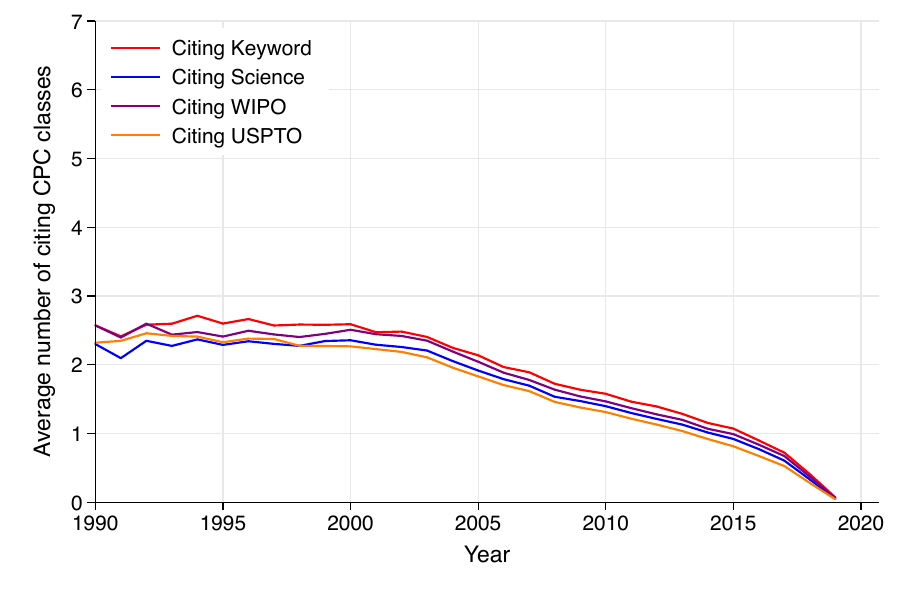}
        \caption{AI citing patents}
        \label{subfig:generality_avg_cites_AI_citing_all}
    \end{subfigure}
    \begin{subfigure}{0.45\textwidth}
        \includegraphics[width=\textwidth]{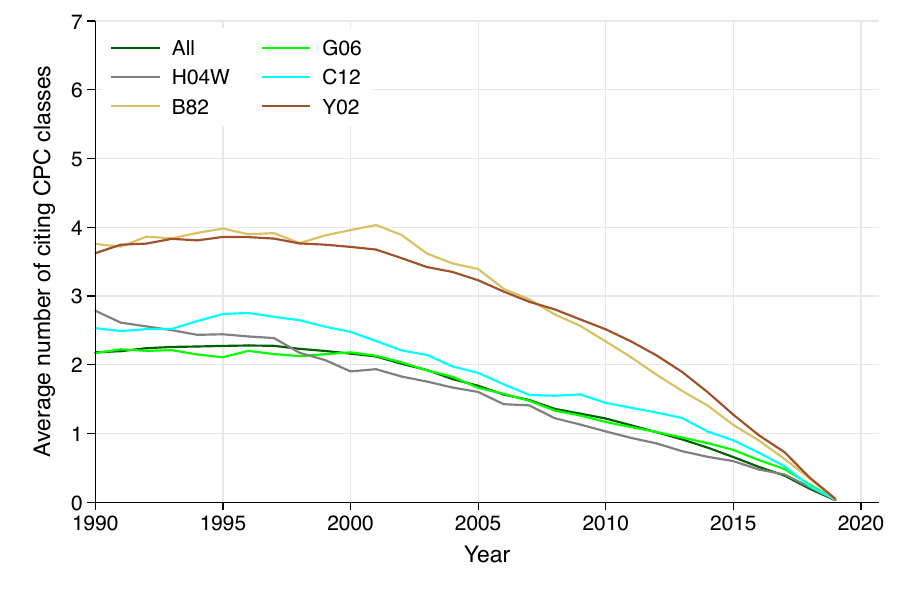}
        \caption{Benchmark}
        \label{subfig:generality_avg_cites_benchmark_all}
    \end{subfigure}
    
    \footnotesize \justifying
     Note: ``All'' refers to all patents, G06, H04W, B82, C12, and Y02 refers to computing, wireless communications, biochemistry/genetic engineering, nanotechnology, and climate invention-related patents, respectively.
\end{figure}


\begin{table}[H]
\centering
\caption{Average Number of Citing CPC Classes -- Cited Patents} 
\label{tab:generality3_benchmark}
\begin{tabular}{lccccccc}
  \hline
  \hline
& All & G06 & H04W & C12 & B82 & Y02  \\ [0.25em]
  \hline
  1 digit & 2.39 & 2.03 & 2.00 & 2.78 & 3.35 & 3.29  \\ [0.25em]
  3 digit & 4.28 & 3.78 & 3.35 & 4.60 & 6.22 & 5.48   \\ [0.25em]
  4 digit & 6.44 & 5.96 & 5.63 & 7.47 & 9.08 & 8.42  \\ [0.25em]

  \hline\hline
     \end{tabular}
    \footnotesize \justifying
    
    Notes: The table reports numbers of different CPC classes making a citation to an average patent of the respective group that receives at least one citation. Citations within the same class are excluded. ``All'' refers to all patents, G06, H04W, B82, C12, and Y02 refers to computing, wireless communications, biochemistry/genetic engineering, nanotechnology, and climate invention-related patents, respectively.
\end{table}

\begin{figure}[H]
    \centering
    \caption{Average Number of CPC Classes Citing AI: Subset of Cited Patents}
    \begin{subfigure}{0.45\textwidth}
        \includegraphics[width=\textwidth]{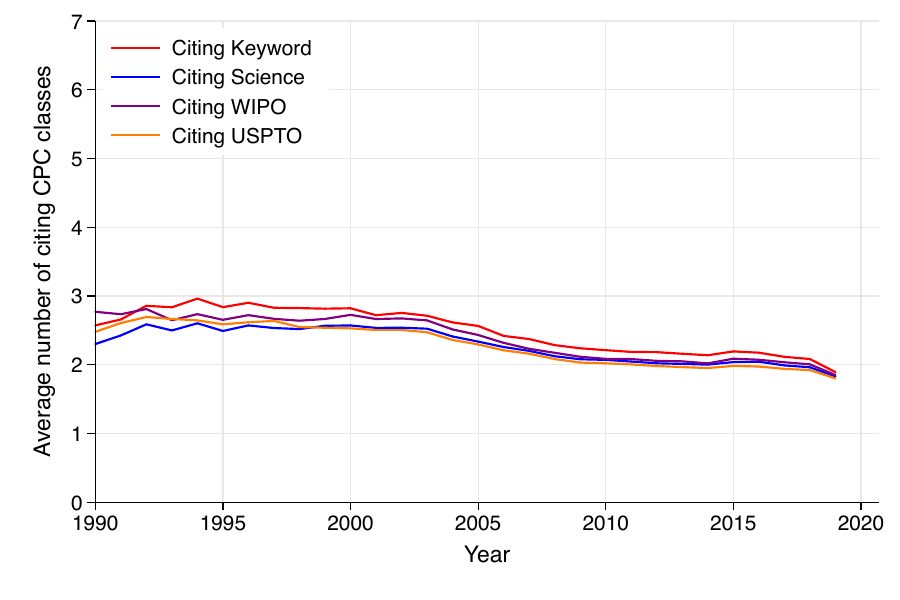}
        \caption{AI citing patents}
        \label{subfig:generality_avg_cites_AI_citing_cited}
    \end{subfigure}
    \begin{subfigure}{0.45\textwidth}
        \includegraphics[width=\textwidth]{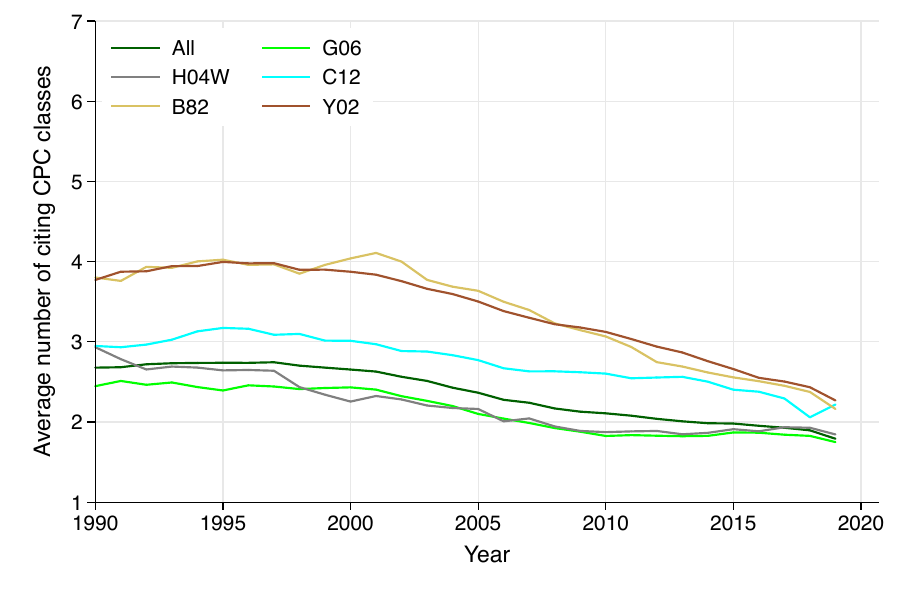}
        \caption{Benchmark}
        \label{subfig:generality_avg_cites_benchmark_cited}
    \end{subfigure}
    
    \footnotesize \justifying
     Note: ``All'' refers to all patents, G06, H04W, B82, C12, and Y02 refers to computing, wireless communications, biochemistry/genetic engineering, nanotechnology, and climate invention-related patents, respectively.
\end{figure}

\begin{table}[H]
\centering

\caption{Average Citation Lags by Patents in Benchmark Categories}
\label{tab:avg_citation_lags_bench}

\begin{center}
\begin{tabular}{lrrrrrr} 
  \hline \hline
 Period & All & G06 & H04W & C12 & B82 & Y02 \\ 
  \hline
1990-1999 & 13.57 & 12.78 & 12.47 & 14.58 & 12.14 & 13.49  \\ 
2000-2009 & 9.19 & 9.00 & 8.75 & 10.15 & 8.58 & 8.92  \\ 
2010-2019 & 4.29 & 4.19 & 3.83 & 4.29 & 4.33 & 4.08  \\ 
   \hline
   \hline
\end{tabular}
\end{center}

\footnotesize \justifying

Notes: This table shows the average number of years it takes until a patent in the sample is cited. The average number of years is lower during the more recent decade as the maximal time lag is truncated since our data ends in 2019. ``All'' refers to all patents, G06, H04W, B82, C12, and Y02 refers to computing, wireless communications, biochemistry/genetic engineering, nanotechnology, and climate invention-related patents, respectively.
\end{table}

\subsubsection{Complementarity}\label{appendix_bench_complement}

\begin{figure}[H]
     \centering
     {
      \caption{Share of technology classes: diversity of benchmark categories}
        \label{fig:diverse_bench}
     \begin{subfigure}[b]{0.45\textwidth}
         \includegraphics[width=\textwidth]{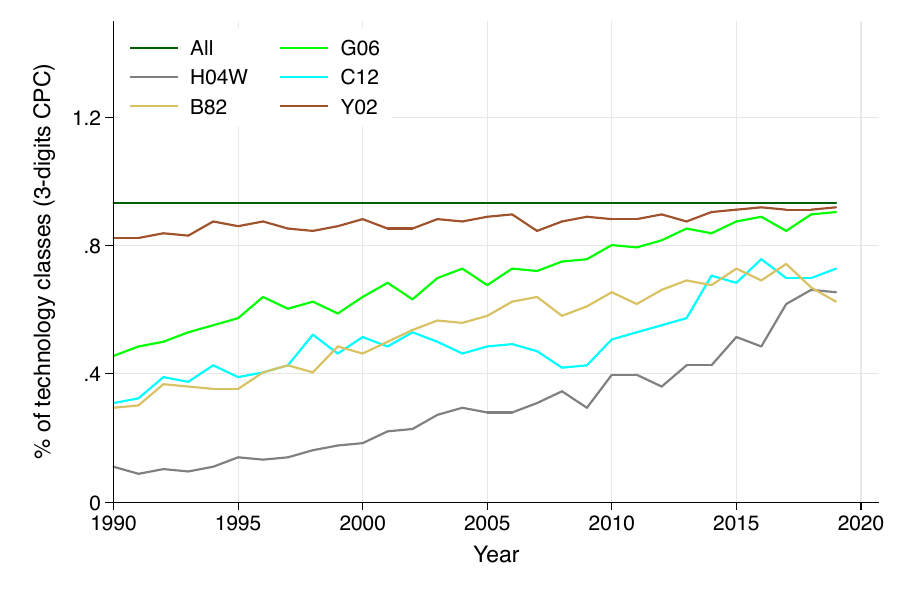}
         \caption{\% of all 3-digit CPC}
         \label{fig:diverse3d_bench}
     \end{subfigure}
     \hfill
     \begin{subfigure}[b]{0.45\textwidth}
         \includegraphics[width=\textwidth]{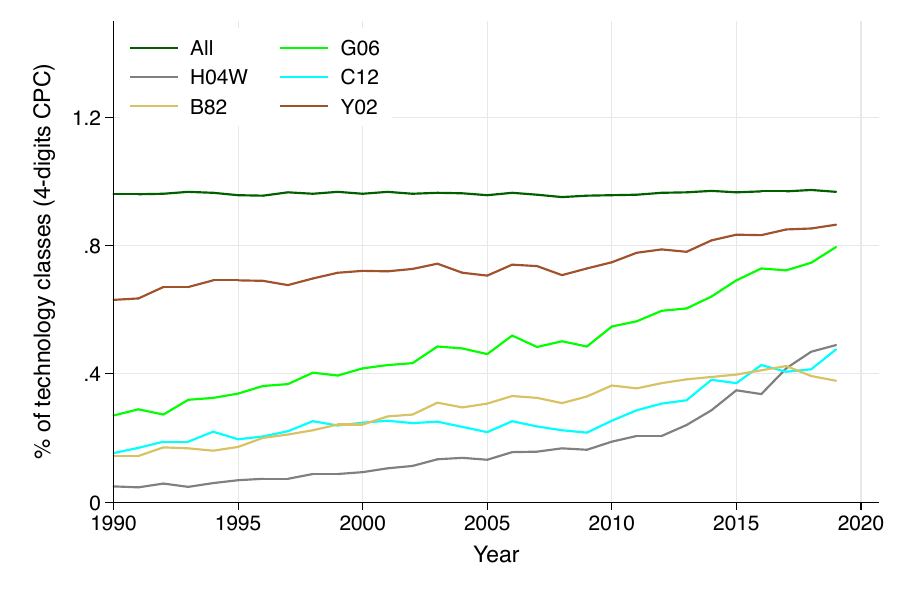}
         \caption{\% of all 4-digit CPC}
         \label{fig:diverse4d_bench}
     \end{subfigure}
     }
     
     \footnotesize \justifying
     Note: Panel (a) shows the percentage of 3-digit CPC codes and panel (b) shows the percentage of 4-digit CPC as a share of all codes in the respective category. Note that the total number of 3-digit and 4-digit CPC codes are 136 and 674, respectively. ``All'' refers to all patents, G06, H04W, B82, C12, and Y02 refers to computing, wireless communications, biochemistry/genetic engineering, nanotechnology, and climate invention-related patents, respectively.
\end{figure}   

\begin{table}[H]
\centering
\caption{Yearly Average Number of 3- and 4-digits CPC Codes per Patent}\label{tab:cpc_134sum_bench}
        \begin{tabular}{l*{6}{c}}
\hline\hline
            &\multicolumn{1}{c}{All}&\multicolumn{1}{c}{G06}&\multicolumn{1}{c}{H04W}&\multicolumn{1}{c}{C12}&\multicolumn{1}{c}{B82}&\multicolumn{1}{c}{Y02}\\
\hline
1 digit     &        1.36&        1.36&        1.32&        1.80&        2.48&        2.47\\
3 digit    &        1.54&        1.62&        1.43&        2.32&        3.03&        2.85\\
4 digit    &        1.80&        1.81&        2.26&        2.93&        3.50&        3.39\\

\hline\hline
\end{tabular}

\footnotesize \justifying

Notes: The table shows the average of annual average number of technology classes by 1-, 3- or 4-digit CPC per patent. ``All'' refers to all patents, G06, H04W, B82, C12, and Y02 refers to computing, wireless communications, biochemistry/genetic engineering, nanotechnology, and climate invention-related patents, respectively.
\end{table}

\begin{figure}[H]
     \centering
      \caption{Average Technology Classes: Patent-Level Diversity of Benchmark Categories}
        \label{fig:diverse_perpatent_bench}
     \begin{subfigure}[b]{0.45\textwidth}
         \centering
         \includegraphics[width=\textwidth]{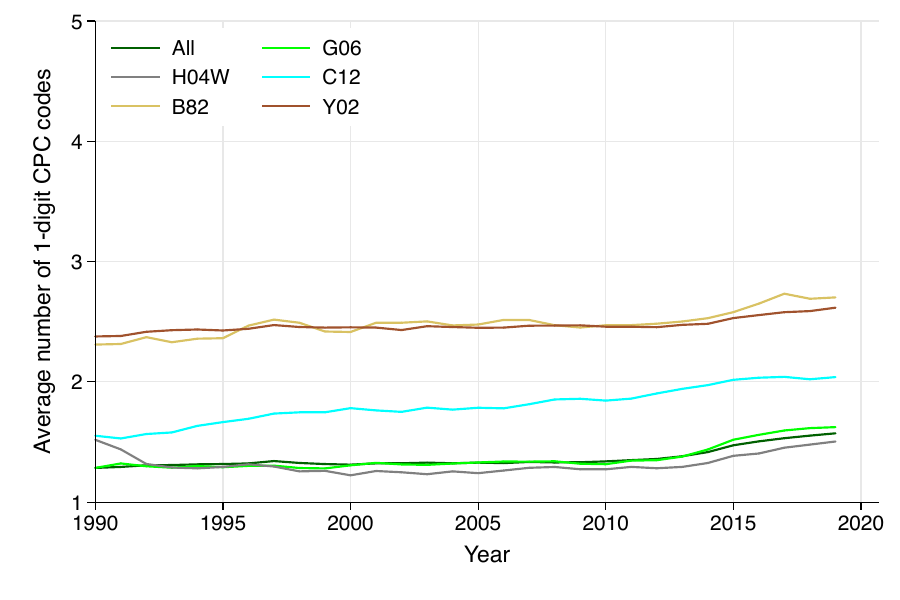}
         \caption{Average Number of 1-digit CPC}
         \label{fig:diverse_perpatent_1d_bench}
     \end{subfigure}
     \hfill
     \begin{subfigure}[b]{0.45\textwidth}
         \centering
         \includegraphics[width=\textwidth]{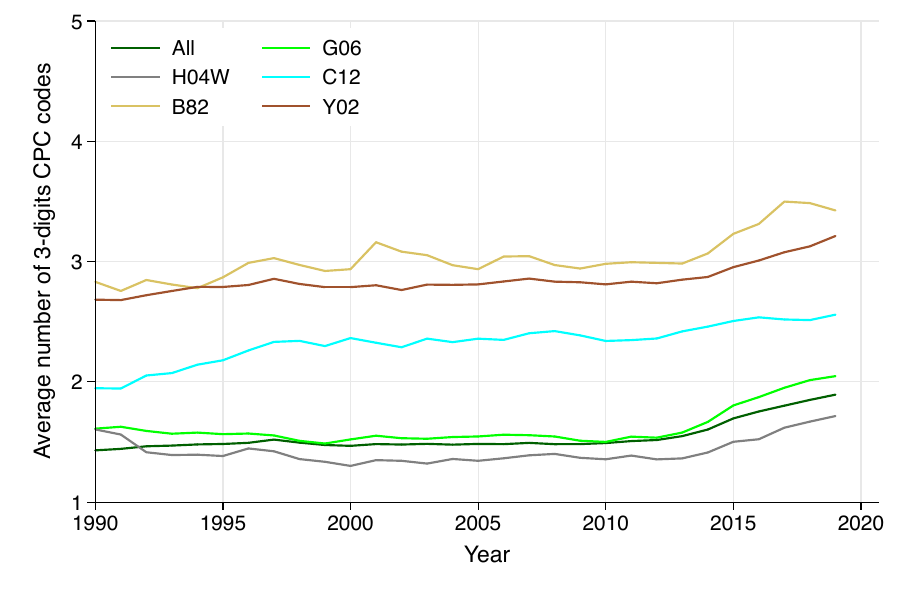}
         \caption{Average Number of 3-digit CPC}
         \label{fig:diverse_perpatent_3d_bench}
     \end{subfigure}
     \begin{subfigure}[c]{0.45\textwidth}
         \centering
         \includegraphics[width=\textwidth]{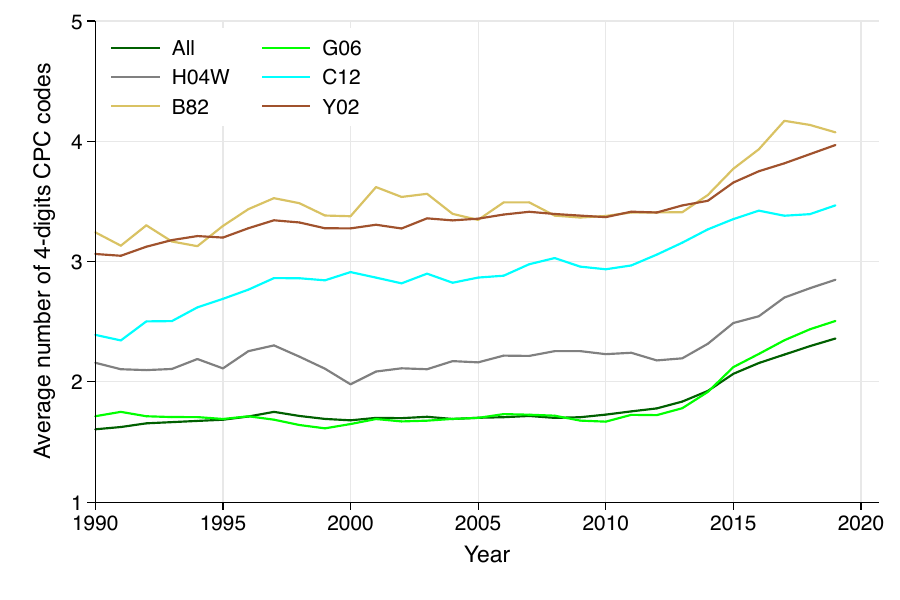}
         \caption{Average Number of 4-digit CPC}
         \label{fig:diverse_perpatent_4d_bench}
     \end{subfigure}
     
         \footnotesize \justifying
     Note: ``All'' refers to all patents, G06, H04W, B82, C12, and Y02 refers to computing, wireless communications, biochemistry/genetic engineering, nanotechnology, and climate-related patents, respectively.
\end{figure}  

\newpage
\subsection{Significance tests}
Here, we provide the results of a series of pair-wise Wilcoxon signed rank tests showing whether the differences between the means reported in Table \ref{tab:gpt_summary}, \ref{app:tab:generality}, \ref{app:tab:avg_citations_all},  \ref{app:tab:avg_citations_cited}, \ref{app:cpc_134sum} are significant. 

\subsubsection{Growth}
\label{appendix:significance_growth}


\begin{table}[H]
\centering
\caption{Growth Rates} 
\label{tab:wilcox_growth}
\begingroup\scriptsize
\begin{tabular}{l|lccccccccc}
  \hline
\hline
period & pair & Keyword & Science & WIPO & USPTO & All & G06 & H04W & C12 & B82 \\ 
  \hline
1990-2019 & Science & 1.00 &  &  &  &  &  &  &  &  \\ 
  1990-2019 & WIPO & 1.00 & 1.00 &  &  &  &  &  &  &  \\ 
  1990-2019 & USPTO & 1.00 & 1.00 & 1.00 &  &  &  &  &  &  \\ 
  1990-2019 & All & 0.02 & 0.00 & 0.00 & 0.00 &  &  &  &  &  \\ 
  1990-2019 & G06 & 1.00 & 1.00 & 1.00 & 1.00 & 0.00 &  &  &  &  \\ 
  1990-2019 & H04W & 0.27 & 0.62 & 0.74 & 0.17 & 0.00 & 0.04 &  &  &  \\ 
  1990-2019 & C12 & 1.00 & 0.08 & 0.40 & 0.76 & 1.00 & 0.75 & 0.01 &  &  \\ 
  1990-2019 & B82 & 1.00 & 0.29 & 1.00 & 0.69 & 1.00 & 0.86 & 0.04 & 1.00 &  \\ 
  1990-2019 & Y02 & 1.00 & 0.60 & 0.32 & 0.67 & 0.03 & 0.86 & 0.00 & 1.00 & 1.00 \\ 
   \hline
1990-1999 & Science & 0.18 &  &  &  &  &  &  &  &  \\ 
  1990-1999 & WIPO & 1.00 & 0.85 &  &  &  &  &  &  &  \\ 
  1990-1999 & USPTO & 0.62 & 1.00 & 1.00 &  &  &  &  &  &  \\ 
  1990-1999 & All & 1.00 & 0.45 & 1.00 & 0.18 &  &  &  &  &  \\ 
  1990-1999 & G06 & 0.32 & 1.00 & 1.00 & 1.00 & 0.45 &  &  &  &  \\ 
  1990-1999 & H04W & 0.45 & 1.00 & 1.00 & 1.00 & 0.45 & 0.85 &  &  &  \\ 
  1990-1999 & C12 & 1.00 & 1.00 & 1.00 & 1.00 & 1.00 & 1.00 & 1.00 &  &  \\ 
  1990-1999 & B82 & 1.00 & 1.00 & 1.00 & 1.00 & 0.18 & 1.00 & 1.00 & 1.00 &  \\ 
  1990-1999 & Y02 & 1.00 & 0.32 & 1.00 & 0.32 & 1.00 & 0.18 & 0.45 & 1.00 & 0.45 \\ 
   \hline
2000-2009 & Science & 1.00 &  &  &  &  &  &  &  &  \\ 
  2000-2009 & WIPO & 1.00 & 1.00 &  &  &  &  &  &  &  \\ 
  2000-2009 & USPTO & 1.00 & 1.00 & 1.00 &  &  &  &  &  &  \\ 
  2000-2009 & All & 1.00 & 0.43 & 1.00 & 0.26 &  &  &  &  &  \\ 
  2000-2009 & G06 & 1.00 & 1.00 & 1.00 & 1.00 & 0.56 &  &  &  &  \\ 
  2000-2009 & H04W & 1.00 & 1.00 & 1.00 & 1.00 & 1.00 & 1.00 &  &  &  \\ 
  2000-2009 & C12 & 1.00 & 0.43 & 1.00 & 0.78 & 1.00 & 1.00 & 1.00 &  &  \\ 
  2000-2009 & B82 & 1.00 & 1.00 & 1.00 & 1.00 & 0.43 & 1.00 & 1.00 & 1.00 &  \\ 
  2000-2009 & Y02 & 1.00 & 1.00 & 1.00 & 1.00 & 1.00 & 1.00 & 1.00 & 1.00 & 1.00 \\ 
   \hline
2010-2019 & Science & 0.55 &  &  &  &  &  &  &  &  \\ 
  2010-2019 & WIPO & 1.00 & 0.71 &  &  &  &  &  &  &  \\ 
  2010-2019 & USPTO & 1.00 & 1.00 & 0.85 &  &  &  &  &  &  \\ 
  2010-2019 & All & 0.09 & 0.71 & 0.09 & 1.00 &  &  &  &  &  \\ 
  2010-2019 & G06 & 1.00 & 1.00 & 0.30 & 1.00 & 0.09 &  &  &  &  \\ 
  2010-2019 & H04W & 1.00 & 1.00 & 1.00 & 1.00 & 0.13 & 1.00 &  &  &  \\ 
  2010-2019 & C12 & 0.55 & 1.00 & 0.30 & 1.00 & 1.00 & 1.00 & 0.19 &  &  \\ 
  2010-2019 & B82 & 0.09 & 0.09 & 0.09 & 0.09 & 0.09 & 0.09 & 0.09 & 0.40 &  \\ 
  2010-2019 & Y02 & 1.00 & 1.00 & 1.00 & 1.00 & 0.19 & 1.00 & 0.85 & 0.71 & 0.09 \\ 
   \hline
\hline
\end{tabular}
\endgroup
\caption*{Notes: Entries show the p-value of a two-sided paired Wilcoxon signed rank test for the hypothesis that the compared pair ranks equal.} 
\end{table}
\begin{table}[H]
\centering
\caption{Summary Statistics for Growth Rates} 
\label{summ_growth}
\begingroup\scriptsize

\begin{tabular}{l|cccccccccc}
  \hline
\hline
 & Keyword & Science & WIPO & USPTO & All & G06 & H04W & C12 & B82 & Y02 \\ 
  \hline
Mean  
1990-2019 & 0.12 & 0.15 & 0.14 & 0.13 & 0.05 & 0.13 & 0.21 & 0.08 & 0.08 & 0.08 \\ 
  Median  
1990-2019 & 0.09 & 0.12 & 0.12 & 0.09 & 0.03 & 0.08 & 0.17 & 0.05 & 0.08 & 0.07 \\ 
  St.dev.  
1990-2019 & 0.15 & 0.17 & 0.16 & 0.16 & 0.10 & 0.15 & 0.20 & 0.15 & 0.16 & 0.10 \\ 
   \hline
Mean  
1990-1999 & 0.09 & 0.26 & 0.16 & 0.20 & 0.06 & 0.19 & 0.32 & 0.18 & 0.17 & 0.08 \\ 
  Median  
1990-1999 & 0.03 & 0.27 & 0.09 & 0.12 & 0.03 & 0.16 & 0.30 & 0.13 & 0.09 & 0.07 \\ 
  St.dev.  
1990-1999 & 0.14 & 0.17 & 0.17 & 0.19 & 0.10 & 0.18 & 0.23 & 0.17 & 0.14 & 0.07 \\ 
   \hline
Mean  
2000-2009 & 0.05 & 0.10 & 0.09 & 0.09 & 0.01 & 0.08 & 0.12 & -0.01 & 0.09 & 0.03 \\ 
  Median  
2000-2009 & 0.03 & 0.06 & 0.07 & 0.09 & 0.01 & 0.07 & 0.07 & -0.06 & 0.09 & 0.02 \\ 
  St.dev.  
2000-2009 & 0.12 & 0.18 & 0.19 & 0.15 & 0.09 & 0.15 & 0.21 & 0.14 & 0.13 & 0.09 \\ 
   \hline
Mean  
2010-2019 & 0.21 & 0.11 & 0.18 & 0.12 & 0.08 & 0.12 & 0.19 & 0.09 & -0.02 & 0.14 \\ 
  Median  
2010-2019 & 0.17 & 0.07 & 0.16 & 0.08 & 0.07 & 0.09 & 0.19 & 0.06 & -0.03 & 0.11 \\ 
  St.dev.  
2010-2019 & 0.16 & 0.13 & 0.09 & 0.15 & 0.10 & 0.12 & 0.12 & 0.09 & 0.16 & 0.12 \\ 
   \hline
\hline
\end{tabular}

\endgroup
\end{table}

\begin{table}[H]
\centering
\caption{Growth Rates (Citing AI)} 
\label{tab:appendix_growth_citing_AI_significance}
\begingroup\tiny

\begin{tabular}{l|lcccccccc}
  \hline
\hline
period & pair & Keyword & Science & WIPO & USPTO & G06 & H04W & C12 & B82 \\ 
  \hline
1990-2019 & Science & 1.00 &  &  &  &  &  &  &  \\ 
  1990-2019 & WIPO & 1.00 & 1.00 &  &  &  &  &  &  \\ 
  1990-2019 & USPTO & 0.02 & 0.00 & 0.01 &  &  &  &  &  \\ 
  1990-2019 & G06 & 0.14 & 0.00 & 0.00 & 1.00 &  &  &  &  \\ 
  1990-2019 & H04W & 0.91 & 0.89 & 1.00 & 0.01 & 0.00 &  &  &  \\ 
  1990-2019 & C12 & 0.89 & 0.62 & 0.96 & 1.00 & 1.00 & 0.10 &  &  \\ 
  1990-2019 & B82 & 1.00 & 1.00 & 1.00 & 1.00 & 1.00 & 1.00 & 1.00 &  \\ 
  1990-2019 & Y02 & 0.07 & 0.12 & 0.96 & 1.00 & 1.00 & 0.05 & 1.00 & 1.00 \\ 
   \hline
1990-1999 & Science & 1.00 &  &  &  &  &  &  &  \\ 
  1990-1999 & WIPO & 1.00 & 1.00 &  &  &  &  &  &  \\ 
  1990-1999 & USPTO & 1.00 & 0.14 & 0.26 &  &  &  &  &  \\ 
  1990-1999 & G06 & 1.00 & 1.00 & 0.14 & 1.00 &  &  &  &  \\ 
  1990-1999 & H04W & 1.00 & 1.00 & 1.00 & 1.00 & 0.14 &  &  &  \\ 
  1990-1999 & C12 & 1.00 & 1.00 & 1.00 & 1.00 & 1.00 & 0.88 &  &  \\ 
  1990-1999 & B82 & 1.00 & 1.00 & 1.00 & 1.00 & 1.00 & 1.00 & 1.00 &  \\ 
  1990-1999 & Y02 & 1.00 & 1.00 & 1.00 & 1.00 & 1.00 & 1.00 & 1.00 & 1.00 \\ 
   \hline
2000-2009 & Science & 1.00 &  &  &  &  &  &  &  \\ 
  2000-2009 & WIPO & 1.00 & 1.00 &  &  &  &  &  &  \\ 
  2000-2009 & USPTO & 0.68 & 0.68 & 1.00 &  &  &  &  &  \\ 
  2000-2009 & G06 & 1.00 & 1.00 & 1.00 & 1.00 &  &  &  &  \\ 
  2000-2009 & H04W & 1.00 & 1.00 & 1.00 & 1.00 & 0.90 &  &  &  \\ 
  2000-2009 & C12 & 1.00 & 1.00 & 1.00 & 1.00 & 1.00 & 1.00 &  &  \\ 
  2000-2009 & B82 & 1.00 & 1.00 & 1.00 & 1.00 & 1.00 & 1.00 & 1.00 &  \\ 
  2000-2009 & Y02 & 0.07 & 1.00 & 1.00 & 1.00 & 1.00 & 1.00 & 1.00 & 1.00 \\ 
   \hline
2010-2019 & Science & 1.00 &  &  &  &  &  &  &  \\ 
  2010-2019 & WIPO & 1.00 & 1.00 &  &  &  &  &  &  \\ 
  2010-2019 & USPTO & 0.33 & 0.62 & 1.00 &  &  &  &  &  \\ 
  2010-2019 & G06 & 0.14 & 0.21 & 0.62 & 1.00 &  &  &  &  \\ 
  2010-2019 & H04W & 1.00 & 1.00 & 1.00 & 1.00 & 0.45 &  &  &  \\ 
  2010-2019 & C12 & 1.00 & 1.00 & 1.00 & 1.00 & 1.00 & 1.00 &  &  \\ 
  2010-2019 & B82 & 1.00 & 1.00 & 1.00 & 1.00 & 1.00 & 1.00 & 1.00 &  \\ 
  2010-2019 & Y02 & 1.00 & 1.00 & 1.00 & 1.00 & 1.00 & 1.00 & 1.00 & 1.00 \\ 
   \hline
\hline
\end{tabular}

\endgroup
\caption*{Notes: Table excludes those patents that themselves are AI by the respective classification approach. Entries show the p-value of a two-sided paired Wilcoxon signed rank test for the hypothesis that the compared pair ranks equal.} 
\end{table}
\begin{table}[H]
\centering
\caption{Summary Statistics for Growth Rates (Citing AI)} 
\label{summ_citing_ai}
\begingroup\tiny
\begin{tabular}{l|ccccccccc}
  \hline
\hline
 & Keyword & Science & WIPO & USPTO & G06 & H04W & C12 & B82 & Y02 \\ 
  \hline
Mean  
1990-2019 & 0.95 & 0.75 & 1.74 & 0.53 & 0.96 & 1.29 & 0.70 & 1.67 & 0.70 \\ 
  Median  
1990-2019 & 0.14 & 0.13 & 0.12 & 0.09 & 0.11 & 0.19 & 0.08 & 0.12 & 0.10 \\ 
  St.dev.  
1990-2019 & 3.80 & 2.56 & 8.01 & 1.71 & 4.08 & 5.34 & 2.58 & 7.77 & 2.65 \\ 
   \hline
Mean  
1990-1999 & 2.81 & 2.15 & 5.38 & 1.52 & 2.91 & 3.89 & 2.07 & 5.14 & 2.07 \\ 
  Median  
1990-1999 & 0.57 & 0.58 & 0.68 & 0.54 & 0.44 & 0.73 & 0.43 & 0.39 & 0.40 \\ 
  St.dev.  
1990-1999 & 6.69 & 4.43 & 14.24 & 2.93 & 7.21 & 9.43 & 4.50 & 13.83 & 4.63 \\ 
   \hline
Mean  
2000-2009 & 0.14 & 0.13 & 0.13 & 0.10 & 0.11 & 0.17 & 0.11 & 0.16 & 0.10 \\ 
  Median  
2000-2009 & 0.12 & 0.10 & 0.12 & 0.09 & 0.09 & 0.12 & 0.04 & 0.13 & 0.09 \\ 
  St.dev.  
2000-2009 & 0.17 & 0.16 & 0.16 & 0.13 & 0.13 & 0.19 & 0.22 & 0.17 & 0.15 \\ 
   \hline
Mean  
2010-2019 & 0.10 & 0.09 & 0.09 & 0.06 & 0.05 & 0.09 & 0.07 & 0.07 & 0.07 \\ 
  Median  
2010-2019 & 0.07 & 0.07 & 0.08 & 0.05 & 0.05 & 0.09 & 0.06 & 0.06 & 0.07 \\ 
  St.dev.  
2010-2019 & 0.13 & 0.13 & 0.14 & 0.10 & 0.10 & 0.13 & 0.15 & 0.13 & 0.10 \\ 
   \hline
\hline
\end{tabular}

\endgroup
\caption*{Notes: Table excludes those patents that themselves are AI by the respective classification approach.}
\end{table}

\FloatBarrier

\subsubsection{Generality}
\label{appendix:significance_generality}


\begin{table}[H]
\centering
\caption{Generality Index at 1-Digit Level. } \label{tab:gen_sig_1}

\begingroup\scriptsize
\begin{tabular}{l|lccccccccc}
  \hline
\hline
Period & 1-digit & Keyword & Science & WIPO & USPTO & All & G06 & H04W & C12 & B82 \\ 
  \hline
1990-2019 & Science & 0.00 &  &  &  &  &  &  &  &  \\ 
  1990-2019 & WIPO & 0.00 & 0.00 &  &  &  &  &  &  &  \\ 
  1990-2019 & USPTO & 0.00 & 0.00 & 0.00 &  &  &  &  &  &  \\ 
  1990-2019 & All & 0.00 & 0.00 & 0.00 & 0.00 &  &  &  &  &  \\ 
  1990-2019 & G06 & 0.00 & 0.00 & 0.00 & 0.00 & 0.00 &  &  &  &  \\ 
  1990-2019 & H04W & 0.00 & 0.00 & 0.00 & 0.00 & 0.00 & 0.00 &  &  &  \\ 
  1990-2019 & C12 & 0.00 & 0.00 & 0.00 & 0.00 & 0.00 & 0.00 & 0.00 &  &  \\
\hline
\hline
\end{tabular}

\endgroup
\caption*{Notes: Entries show the p-value of a two-sided paired Wilcoxon signed rank test for the hypothesis that the compared pair ranks equal.} 
\label{tab:wilcoxon_generality_index_1d}
\end{table}
\begin{table}[H]
\centering
\caption{Summary Statistics for Generality Index at 1-Digit Level.}\label{tab:sig_gen_sum1}
\begingroup\scriptsize

\begin{tabular}{l|cccccccccc}
  \hline
\hline
 & Keyword & Science & WIPO & USPTO & All & G06 & H04W & C12 & B82 & Y02 \\ 
  \hline
Mean  
1990-2019 & 0.81 & 0.77 & 0.76 & 0.73 & 0.83 & 0.7 & 0.68 & 0.78 & 0.82 & 0.81 \\ 
  Median  
1990-2019 & 0.81 & 0.77 & 0.76 & 0.72 & 0.83 & 0.7 & 0.68 & 0.79 & 0.82 & 0.81 \\ 
  St.dev.  
1990-2019 & 0.02 & 0.02 & 0.02 & 0.03 & 0.02 & 0.02 & 0.02 & 0.01 & 0.01 & 0.01 \\ 
   \hline
Mean  
1990-1999 & 0.82 & 0.79 & 0.79 & 0.76 & 0.84 & 0.73 & 0.68 & 0.78 & 0.83 & 0.81 \\ 
  Median  
1990-1999 & 0.82 & 0.79 & 0.79 & 0.76 & 0.84 & 0.73 & 0.68 & 0.78 & 0.83 & 0.81 \\ 
  St.dev.  
1990-1999 & 0.01 & 0.02 & 0.01 & 0.02 & 0 & 0.01 & 0.01 & 0 & 0 & 0 \\ 
   \hline
Mean  
2000-2009 & 0.81 & 0.77 & 0.76 & 0.72 & 0.83 & 0.7 & 0.66 & 0.79 & 0.82 & 0.81 \\ 
  Median  
2000-2009 & 0.81 & 0.77 & 0.76 & 0.72 & 0.83 & 0.7 & 0.66 & 0.79 & 0.82 & 0.81 \\ 
  St.dev.  
2000-2009 & 0 & 0.01 & 0.01 & 0.01 & 0 & 0.01 & 0.01 & 0.01 & 0 & 0.01 \\ 
   \hline
Mean  
2010-2019 & 0.8 & 0.76 & 0.74 & 0.7 & 0.81 & 0.68 & 0.7 & 0.78 & 0.81 & 0.81 \\ 
  Median  
2010-2019 & 0.81 & 0.76 & 0.74 & 0.7 & 0.82 & 0.68 & 0.7 & 0.78 & 0.81 & 0.81 \\ 
  St.dev.  
2010-2019 & 0.03 & 0.01 & 0.01 & 0.01 & 0.03 & 0.01 & 0.02 & 0.02 & 0.02 & 0.01 \\ 
   \hline
\hline
\end{tabular}

\endgroup
\end{table}

\begin{table}[ht]
\centering
\caption{Generality Index at 3-Digit Level}\label{tab:gen_sig_3}
\begingroup\scriptsize
\begin{tabular}{l|lccccccccc}
  \hline
\hline
Period & 3-digit & Keyword & Science & WIPO & USPTO & All & G06 & H04W & C12 & B82 \\ 
  \hline
1990-2019 & Science & 0.00 &  &  &  &  &  &  &  &  \\ 
  1990-2019 & WIPO & 0.00 & 0.74 &  &  &  &  &  &  &  \\ 
  1990-2019 & USPTO & 0.00 & 0.00 & 0.00 &  &  &  &  &  &  \\ 
  1990-2019 & All & 0.00 & 0.00 & 0.00 & 0.00 &  &  &  &  &  \\ 
  1990-2019 & G06 & 0.00 & 0.00 & 0.00 & 0.00 & 0.00 &  &  &  &  \\ 
  1990-2019 & H04W & 0.00 & 0.00 & 0.00 & 0.22 & 0.00 & 0.00 &  &  &  \\ 
  1990-2019 & C12 & 0.00 & 0.00 & 0.00 & 0.05 & 0.00 & 0.00 & 0.01 &  &  \\
\hline
\hline
\end{tabular}

\endgroup
\caption*{Notes: Entries show the p-value of a two-sided paired Wilcoxon signed rank test for the hypothesis that the compared pair ranks equal.} 
\label{tab:wilcoxon_generality_index_3d}
\end{table}
\begin{table}[H]
\centering
\caption{Summary Statistics for Generality Index at 3-Digit Level}\label{tab:sig_gen_sum3}
\begingroup\scriptsize

\begin{tabular}{l|cccccccccc}
  \hline
\hline
 & Keyword & Science & WIPO & USPTO & All & G06 & H04W & C12 & B82 & Y02 \\ 
  \hline
Mean  
1990-2019 & 0.94 & 0.9 & 0.9 & 0.87 & 0.95 & 0.86 & 0.87 & 0.88 & 0.93 & 0.94 \\ 
  Median  
1990-2019 & 0.94 & 0.9 & 0.9 & 0.87 & 0.95 & 0.86 & 0.86 & 0.88 & 0.93 & 0.94 \\ 
  St.dev.  
1990-2019 & 0.01 & 0.01 & 0.01 & 0.02 & 0.01 & 0.01 & 0.02 & 0.01 & 0.02 & 0 \\ 
   \hline
Mean  
1990-1999 & 0.94 & 0.91 & 0.91 & 0.89 & 0.96 & 0.87 & 0.88 & 0.88 & 0.93 & 0.94 \\ 
  Median  
1990-1999 & 0.94 & 0.91 & 0.92 & 0.9 & 0.96 & 0.88 & 0.88 & 0.88 & 0.93 & 0.94 \\ 
  St.dev.  
1990-1999 & 0 & 0.01 & 0.01 & 0.02 & 0 & 0.01 & 0.01 & 0.01 & 0 & 0 \\ 
   \hline
Mean  
2000-2009 & 0.94 & 0.9 & 0.89 & 0.86 & 0.95 & 0.85 & 0.85 & 0.88 & 0.93 & 0.94 \\ 
  Median  
2000-2009 & 0.94 & 0.9 & 0.89 & 0.86 & 0.95 & 0.85 & 0.85 & 0.88 & 0.93 & 0.94 \\ 
  St.dev.  
2000-2009 & 0 & 0.01 & 0.01 & 0.01 & 0 & 0 & 0.01 & 0 & 0 & 0 \\ 
   \hline
Mean  
2010-2019 & 0.93 & 0.89 & 0.9 & 0.86 & 0.94 & 0.86 & 0.87 & 0.88 & 0.92 & 0.94 \\ 
  Median  
2010-2019 & 0.94 & 0.89 & 0.9 & 0.86 & 0.95 & 0.86 & 0.88 & 0.88 & 0.93 & 0.94 \\ 
  St.dev.  
2010-2019 & 0.02 & 0 & 0.01 & 0.01 & 0.02 & 0 & 0.02 & 0.01 & 0.03 & 0 \\ 
   \hline
\hline
\end{tabular}

\endgroup
\end{table}
\begin{table}[H]
\centering
\caption{Generality Index at 4-Digit Level}\label{tab:wilcoxon_generality_index_4d}

\begingroup\scriptsize

\begin{tabular}{l|lccccccccc}
  \hline
\hline
Period & 4-digit & Keyword & Science & WIPO & USPTO & All & G06 & H04W & C12 & B82 \\ 
  \hline
1990-2019 & Science & 0.00 &  &  &  &  &  &  &  &  \\ 
  1990-2019 & WIPO & 0.00 & 0.05 &  &  &  &  &  &  &  \\ 
  1990-2019 & USPTO & 0.00 & 0.00 & 0.00 &  &  &  &  &  &  \\ 
  1990-2019 & All & 0.00 & 0.00 & 0.00 & 0.00 &  &  &  &  &  \\ 
  1990-2019 & G06 & 0.00 & 0.00 & 0.00 & 0.00 & 0.00 &  &  &  &  \\ 
  1990-2019 & H04W & 0.00 & 0.00 & 0.00 & 0.00 & 0.00 & 0.00 &  &  &  \\ 
  1990-2019 & C12 & 0.00 & 0.00 & 0.00 & 0.69 & 0.00 & 0.00 & 0.00 &  &  \\
\hline
\hline
\end{tabular}

\endgroup
\caption*{Notes: Entries show the p-value of a two-sided paired Wilcoxon signed rank test for the hypothesis that the compared pair ranks equal.} 
\end{table}
\begin{table}[H]
\centering
\caption{Summary Statistics for Generality Index at 4-Digit Level.}\label{tab:sig_sum_4}
\begingroup\scriptsize

\begin{tabular}{l|cccccccccc}
  \hline
\hline
 & Keyword & Science & WIPO & USPTO & All & G06 & H04W & C12 & B82 & Y02 \\ 
  \hline
Mean  
1990-2019 & 0.97 & 0.96 & 0.95 & 0.95 & 0.98 & 0.94 & 0.91 & 0.95 & 0.97 & 0.98 \\ 
  Median  
1990-2019 & 0.97 & 0.96 & 0.95 & 0.94 & 0.98 & 0.94 & 0.9 & 0.95 & 0.97 & 0.99 \\ 
  St.dev.  
1990-2019 & 0.01 & 0 & 0.01 & 0.01 & 0.01 & 0.01 & 0.02 & 0.01 & 0.01 & 0 \\ 
   \hline
Mean  
1990-1999 & 0.97 & 0.96 & 0.96 & 0.96 & 0.99 & 0.94 & 0.91 & 0.95 & 0.97 & 0.99 \\ 
  Median  
1990-1999 & 0.98 & 0.96 & 0.96 & 0.96 & 0.99 & 0.94 & 0.91 & 0.95 & 0.97 & 0.99 \\ 
  St.dev.  
1990-1999 & 0 & 0.01 & 0.01 & 0.01 & 0 & 0.01 & 0.01 & 0 & 0 & 0 \\ 
   \hline
Mean  
2000-2009 & 0.97 & 0.95 & 0.95 & 0.94 & 0.98 & 0.93 & 0.9 & 0.95 & 0.97 & 0.98 \\ 
  Median  
2000-2009 & 0.97 & 0.95 & 0.95 & 0.94 & 0.98 & 0.93 & 0.9 & 0.95 & 0.97 & 0.98 \\ 
  St.dev.  
2000-2009 & 0 & 0 & 0 & 0 & 0 & 0 & 0 & 0 & 0 & 0 \\ 
   \hline
Mean  
2010-2019 & 0.97 & 0.96 & 0.95 & 0.94 & 0.98 & 0.94 & 0.92 & 0.94 & 0.96 & 0.98 \\ 
  Median  
2010-2019 & 0.97 & 0.96 & 0.96 & 0.94 & 0.98 & 0.94 & 0.92 & 0.94 & 0.97 & 0.98 \\ 
  St.dev.  
2010-2019 & 0.02 & 0 & 0.01 & 0 & 0.01 & 0 & 0.02 & 0.01 & 0.02 & 0 \\ 
   \hline
\hline
\end{tabular}

\endgroup
\end{table}

\begin{table}[H]
\centering
\caption{Average Number of Citing Classes (All) at 1-Digit Level}\label{tab:sig_gen_cite1}
\begingroup\scriptsize

\begin{tabular}{l|lccccccccc}
  \hline
\hline
Period & 1-digit & Keyword & Science & WIPO & USPTO & All & G06 & H04W & C12 & B82 \\ 
  \hline
1990-2019 & Science & 0.00 &  &  &  &  &  &  &  &  \\ 
  1990-2019 & WIPO & 0.00 & 1.00 &  &  &  &  &  &  &  \\ 
  1990-2019 & USPTO & 0.00 & 0.00 & 0.00 &  &  &  &  &  &  \\ 
  1990-2019 & All & 0.00 & 0.00 & 0.00 & 0.00 &  &  &  &  &  \\ 
  1990-2019 & G06 & 0.00 & 0.00 & 0.00 & 0.00 & 0.74 &  &  &  &  \\ 
  1990-2019 & H04W & 0.00 & 0.00 & 0.00 & 0.00 & 1.00 & 1.00 &  &  &  \\ 
  1990-2019 & C12 & 0.00 & 0.00 & 0.00 & 0.00 & 1.00 & 1.00 & 1.00 &  &  \\
\hline
\hline
\end{tabular}

\endgroup
\caption*{Notes: Entries show the p-value of a two-sided paired Wilcoxon signed rank test for the hypothesis that the compared pair ranks equal.} 
\end{table}
\begin{table}[H]
\centering
\caption{Summary Statistics for Average Number of Citing Classes (All) at 1-Digit Level}\label{tab:sig_sum_all1}
\begingroup\scriptsize
\begin{tabular}{l|cccccccccc}
  \hline
\hline
 & Keyword & Science & WIPO & USPTO & All & G06 & H04W & C12 & B82 & Y02 \\ 
  \hline
Mean  
1990-2019 & 2.98 & 2.53 & 2.57 & 2.38 & 2.14 & 2.17 & 2.14 & 2.15 & 3.08 & 2.77 \\ 
  Median  
1990-2019 & 3.38 & 2.85 & 3.03 & 2.72 & 2.45 & 2.51 & 2.43 & 2.26 & 3.79 & 3.3 \\ 
  St.dev.  
1990-2019 & 1.24 & 1.1 & 1.15 & 1.17 & 1.06 & 1.03 & 1.06 & 1.08 & 1.5 & 1.24 \\ 
   \hline
Mean  
1990-1999 & 4.04 & 3.58 & 3.59 & 3.5 & 3.16 & 3.11 & 3.22 & 3.28 & 4.31 & 3.81 \\ 
  Median  
1990-1999 & 4.03 & 3.57 & 3.61 & 3.53 & 3.17 & 3.11 & 3.26 & 3.29 & 4.33 & 3.81 \\ 
  St.dev.  
1990-1999 & 0.12 & 0.09 & 0.08 & 0.08 & 0.04 & 0.04 & 0.15 & 0.1 & 0.09 & 0.07 \\ 
   \hline
Mean  
2000-2009 & 3.41 & 2.82 & 2.94 & 2.68 & 2.43 & 2.49 & 2.36 & 2.28 & 3.73 & 3.24 \\ 
  Median  
2000-2009 & 3.38 & 2.85 & 3.03 & 2.72 & 2.45 & 2.51 & 2.43 & 2.26 & 3.79 & 3.3 \\ 
  St.dev.  
2000-2009 & 0.52 & 0.42 & 0.55 & 0.54 & 0.44 & 0.46 & 0.36 & 0.43 & 0.6 & 0.38 \\ 
   \hline
Mean  
2010-2019 & 1.48 & 1.2 & 1.17 & 0.94 & 0.84 & 0.9 & 0.85 & 0.88 & 1.2 & 1.25 \\ 
  Median  
2010-2019 & 1.61 & 1.31 & 1.27 & 0.99 & 0.85 & 0.97 & 0.88 & 0.95 & 1.18 & 1.27 \\ 
  St.dev.  
2010-2019 & 0.86 & 0.65 & 0.64 & 0.57 & 0.58 & 0.55 & 0.54 & 0.58 & 0.89 & 0.89 \\ 
   \hline
\hline
\end{tabular}

\endgroup
\end{table}
\begin{table}[H]
\centering
\caption{Average Number of Citing Classes (All) at 3-Digit Level}\label{tab:sig_gen_cite3}
\begingroup\scriptsize

\begin{tabular}{l|lccccccccc}
  \hline
\hline
Period & 3-digit & Keyword & Science & WIPO & USPTO & All & G06 & H04W & C12 & B82 \\ 
  \hline
1990-2019 & Science & 0.00 &  &  &  &  &  &  &  &  \\ 
  1990-2019 & WIPO & 0.00 & 0.00 &  &  &  &  &  &  &  \\ 
  1990-2019 & USPTO & 0.00 & 0.01 & 0.00 &  &  &  &  &  &  \\ 
  1990-2019 & All & 0.00 & 0.00 & 0.00 & 0.00 &  &  &  &  &  \\ 
  1990-2019 & G06 & 0.00 & 0.00 & 0.00 & 0.00 & 0.00 &  &  &  &  \\ 
  1990-2019 & H04W & 0.00 & 0.00 & 0.00 & 0.00 & 0.65 & 0.91 &  &  &  \\ 
  1990-2019 & C12 & 0.00 & 0.00 & 0.00 & 0.00 & 1.00 & 0.00 & 0.00 &  &  \\
\hline
\hline
\end{tabular}

\endgroup
\caption*{Notes: Entries show the p-value of a two-sided paired Wilcoxon signed rank test for the hypothesis that the compared pair ranks equal.} 
\end{table}
\begin{table}[H]
\centering
\caption{Summary Statistics for Average Number of Citing Classes (All) at 3-Digit Level}\label{tab:sig_sum_all3}
\begingroup\scriptsize

\begin{tabular}{l|cccccccccc}
  \hline
\hline
 & Keyword & Science & WIPO & USPTO & All & G06 & H04W & C12 & B82 & Y02 \\ 
  \hline
Mean  
1990-2019 & 6 & 4.82 & 5.07 & 4.57 & 3.59 & 4 & 3.92 & 3.56 & 5.88 & 5.12 \\ 
  Median  
1990-2019 & 6.73 & 5.31 & 5.92 & 4.95 & 4 & 4.41 & 4.11 & 3.64 & 7.29 & 6.06 \\ 
  St.dev.  
1990-2019 & 2.9 & 2.47 & 2.62 & 2.67 & 1.95 & 2.22 & 2.4 & 2 & 3.23 & 2.59 \\ 
   \hline
Mean  
1990-1999 & 8.65 & 7.37 & 7.56 & 7.36 & 5.54 & 6.24 & 6.65 & 5.75 & 8.54 & 7.49 \\ 
  Median  
1990-1999 & 8.58 & 7.27 & 7.54 & 7.44 & 5.57 & 6.27 & 6.78 & 5.74 & 8.64 & 7.57 \\ 
  St.dev.  
1990-1999 & 0.4 & 0.44 & 0.26 & 0.24 & 0.1 & 0.14 & 0.68 & 0.29 & 0.39 & 0.23 \\ 
   \hline
Mean  
2000-2009 & 6.88 & 5.22 & 5.78 & 4.97 & 4.01 & 4.44 & 3.96 & 3.69 & 7.29 & 5.94 \\ 
  Median  
2000-2009 & 6.73 & 5.31 & 5.92 & 4.95 & 4 & 4.41 & 4.11 & 3.64 & 7.29 & 6.06 \\ 
  St.dev.  
2000-2009 & 1.4 & 1.05 & 1.47 & 1.42 & 0.91 & 1.19 & 0.9 & 0.86 & 1.74 & 0.99 \\ 
   \hline
Mean  
2010-2019 & 2.46 & 1.86 & 1.88 & 1.38 & 1.21 & 1.32 & 1.16 & 1.24 & 1.82 & 1.91 \\ 
  Median  
2010-2019 & 2.6 & 1.96 & 2.01 & 1.41 & 1.19 & 1.39 & 1.18 & 1.3 & 1.69 & 1.83 \\ 
  St.dev.  
2010-2019 & 1.54 & 1.07 & 1.1 & 0.88 & 0.87 & 0.83 & 0.76 & 0.85 & 1.44 & 1.44 \\ 
   \hline
\hline
\end{tabular}

\endgroup
\end{table}
\begin{table}[H]
\centering
\caption{Average Number of Citing Classes (All) at 4-Digit Level}\label{tab:sig_gen_cite4}
\begingroup\scriptsize

\begin{tabular}{l|lccccccccc}
  \hline
\hline
Period & 4-digit & Keyword & Science & WIPO & USPTO & All & G06 & H04W & C12 & B82 \\ 
  \hline
1990-2019 & Science & 0.00 &  &  &  &  &  &  &  &  \\ 
  1990-2019 & WIPO & 0.00 & 0.00 &  &  &  &  &  &  &  \\ 
  1990-2019 & USPTO & 0.00 & 0.10 & 0.00 &  &  &  &  &  &  \\ 
  1990-2019 & All & 0.00 & 0.00 & 0.00 & 0.00 &  &  &  &  &  \\ 
  1990-2019 & G06 & 0.00 & 0.00 & 0.00 & 0.00 & 0.00 &  &  &  &  \\ 
  1990-2019 & H04W & 0.00 & 0.88 & 0.07 & 0.88 & 0.00 & 0.00 &  &  &  \\ 
  1990-2019 & C12 & 0.00 & 0.00 & 0.00 & 0.00 & 0.02 & 0.00 & 0.00 &  &  \\
\hline
\hline
\end{tabular}

\endgroup
\caption*{Notes: Entries show the p-value of a two-sided paired Wilcoxon signed rank test for the hypothesis that the compared pair ranks equal.} 
\end{table}
\begin{table}[H]
\centering
\caption{Summary Statistics for Average Number of Citing Classes (All) at 4-Digit Level}\label{tab:sig_sum_all4}
\begingroup\scriptsize

\begin{tabular}{l|cccccccccc}
  \hline
\hline
 & Keyword & Science & WIPO & USPTO & All & G06 & H04W & C12 & B82 & Y02 \\ 
  \hline
Mean  
1990-2019 & 9.26 & 7.82 & 8.42 & 7.5 & 5.26 & 6.15 & 7.62 & 5.72 & 8.81 & 7.87 \\ 
  Median  
1990-2019 & 10.62 & 8.63 & 9.9 & 8.04 & 5.84 & 6.74 & 7.93 & 5.66 & 10.77 & 9.37 \\ 
  St.dev.  
1990-2019 & 4.55 & 4.13 & 4.47 & 4.55 & 2.93 & 3.51 & 4.85 & 3.38 & 5.16 & 4.12 \\ 
   \hline
Mean  
1990-1999 & 13.32 & 12.11 & 12.56 & 12.27 & 8.2 & 9.67 & 13.16 & 9.52 & 13.2 & 11.65 \\ 
  Median  
1990-1999 & 13.44 & 11.9 & 12.6 & 12.44 & 8.24 & 9.67 & 13.47 & 9.58 & 13.29 & 11.8 \\ 
  St.dev.  
1990-1999 & 0.57 & 0.68 & 0.41 & 0.31 & 0.2 & 0.16 & 1.33 & 0.59 & 0.89 & 0.5 \\ 
   \hline
Mean  
2000-2009 & 10.78 & 8.48 & 9.72 & 8.15 & 5.87 & 6.85 & 7.64 & 5.78 & 10.84 & 9.21 \\ 
  Median  
2000-2009 & 10.62 & 8.63 & 9.9 & 8.04 & 5.84 & 6.74 & 7.93 & 5.66 & 10.77 & 9.37 \\ 
  St.dev.  
2000-2009 & 2.28 & 1.85 & 2.69 & 2.56 & 1.43 & 1.99 & 1.91 & 1.48 & 3.07 & 1.71 \\ 
   \hline
Mean  
2010-2019 & 3.68 & 2.88 & 2.98 & 2.09 & 1.7 & 1.93 & 2.06 & 1.86 & 2.39 & 2.76 \\ 
  Median  
2010-2019 & 3.81 & 3 & 3.17 & 2.12 & 1.67 & 2.04 & 2.07 & 1.92 & 2.18 & 2.56 \\ 
  St.dev.  
2010-2019 & 2.4 & 1.7 & 1.78 & 1.37 & 1.24 & 1.23 & 1.39 & 1.3 & 1.93 & 2.14 \\ 
   \hline
\hline
\end{tabular}

\endgroup
\end{table}

\begin{table}[H]
\centering
\caption{Average Number of Citing Classes (Cited) at 1-Digit Level}\label{tab:sig_gen_citec1}
\begingroup\scriptsize
\begin{tabular}{l|lccccccccc}
  \hline
\hline
Period & 1-digit & Keyword & Science & WIPO & USPTO & All & G06 & H04W & C12 & B82 \\ 
  \hline
1990-2019 & Science & 0.00 &  &  &  &  &  &  &  &  \\ 
  1990-2019 & WIPO & 0.00 & 0.26 &  &  &  &  &  &  &  \\ 
  1990-2019 & USPTO & 0.00 & 0.00 & 0.00 &  &  &  &  &  &  \\ 
  1990-2019 & All & 0.00 & 0.00 & 0.00 & 0.42 &  &  &  &  &  \\ 
  1990-2019 & G06 & 0.00 & 0.00 & 0.00 & 0.00 & 0.00 &  &  &  &  \\ 
  1990-2019 & H04W & 0.00 & 0.00 & 0.00 & 0.00 & 0.00 & 0.00 &  &  &  \\ 
  1990-2019 & C12 & 0.00 & 0.42 & 0.32 & 0.01 & 0.00 & 0.00 & 0.00 &  &  \\
\hline
\hline
\end{tabular}

\endgroup
\caption*{Notes: Entries show the p-value of a two-sided paired Wilcoxon signed rank test for the hypothesis that the compared pair ranks equal.} 
\end{table}
\begin{table}[H]
\centering
\caption{Summary Statistics for Average Number of Citing Classes (Cited) at 1-Digit Level}\label{tab:sig_sum_gen_cite_c1}  
\begingroup\scriptsize
\begin{tabular}{l|cccccccccc}
  \hline
\hline
 & Keyword & Science & WIPO & USPTO & All & G06 & H04W & C12 & B82 & Y02 \\ 
  \hline
Mean  
1990-2019 & 3.44 & 2.99 & 2.96 & 2.83 & 2.81 & 2.67 & 2.53 & 3.05 & 3.74 & 3.4 \\ 
  Median  
1990-2019 & 3.67 & 3.17 & 3.19 & 2.96 & 2.92 & 2.79 & 2.54 & 3.07 & 4.05 & 3.57 \\ 
  St.dev.  
1990-2019 & 0.78 & 0.67 & 0.71 & 0.74 & 0.6 & 0.58 & 0.6 & 0.5 & 0.76 & 0.56 \\ 
   \hline
Mean  
1990-1999 & 4.14 & 3.66 & 3.66 & 3.62 & 3.42 & 3.25 & 3.24 & 3.59 & 4.38 & 3.95 \\ 
  Median  
1990-1999 & 4.12 & 3.66 & 3.65 & 3.63 & 3.44 & 3.24 & 3.28 & 3.6 & 4.4 & 3.95 \\ 
  St.dev.  
1990-1999 & 0.1 & 0.07 & 0.06 & 0.07 & 0.03 & 0.04 & 0.14 & 0.08 & 0.09 & 0.07 \\ 
   \hline
Mean  
2000-2009 & 3.68 & 3.14 & 3.12 & 2.94 & 2.91 & 2.78 & 2.49 & 3.09 & 4.04 & 3.55 \\ 
  Median  
2000-2009 & 3.67 & 3.17 & 3.19 & 2.96 & 2.92 & 2.79 & 2.54 & 3.07 & 4.05 & 3.57 \\ 
  St.dev.  
2000-2009 & 0.41 & 0.35 & 0.45 & 0.42 & 0.3 & 0.35 & 0.27 & 0.2 & 0.39 & 0.26 \\ 
   \hline
Mean  
2010-2019 & 2.5 & 2.16 & 2.11 & 1.95 & 2.08 & 1.97 & 1.86 & 2.48 & 2.78 & 2.71 \\ 
  Median  
2010-2019 & 2.61 & 2.16 & 2.12 & 1.94 & 2.06 & 1.97 & 1.83 & 2.52 & 2.73 & 2.68 \\ 
  St.dev.  
2010-2019 & 0.43 & 0.22 & 0.19 & 0.18 & 0.22 & 0.15 & 0.12 & 0.25 & 0.33 & 0.25 \\ 
   \hline
\hline
\end{tabular}

\endgroup
\end{table}
\begin{table}[H]
\centering
\caption{Average Number of Citing Classes (Cited) at 3-Digit Level}\label{tab:sig_gen_citec3}
\begingroup\scriptsize
\begin{tabular}{l|lccccccccc}
  \hline
\hline
Period & 3-digit & Keyword & Science & WIPO & USPTO & All & G06 & H04W & C12 & B82 \\ 
  \hline
1990-2019 & Science & 0.00 &  &  &  &  &  &  &  &  \\ 
  1990-2019 & WIPO & 0.00 & 0.08 &  &  &  &  &  &  &  \\ 
  1990-2019 & USPTO & 0.00 & 0.00 & 0.00 &  &  &  &  &  &  \\ 
  1990-2019 & All & 0.00 & 0.00 & 0.00 & 0.01 &  &  &  &  &  \\ 
  1990-2019 & G06 & 0.00 & 0.00 & 0.00 & 0.00 & 0.08 &  &  &  &  \\ 
  1990-2019 & H04W & 0.00 & 0.00 & 0.00 & 0.00 & 0.54 & 0.08 &  &  &  \\ 
  1990-2019 & C12 & 0.00 & 0.00 & 0.00 & 0.08 & 0.00 & 0.31 & 0.05 &  &  \\
\hline
\hline
\end{tabular}

\endgroup
\caption*{Notes: Entries show the p-value of a two-sided paired Wilcoxon signed rank test for the hypothesis that the compared pair ranks equal.} 
\end{table}
\begin{table}[H]
\centering
\caption{Summary Statistics for Average Number of Citing Classes (Cited) at 3-Digit Level}\label{tab:sig_sum_gen_cite_c3} 
\begingroup\scriptsize

\begin{tabular}{l|cccccccccc}
  \hline
\hline
 & Keyword & Science & WIPO & USPTO & All & G06 & H04W & C12 & B82 & Y02 \\ 
  \hline
Mean  
1990-2019 & 6.76 & 5.54 & 5.71 & 5.27 & 4.57 & 4.76 & 4.47 & 4.9 & 6.85 & 6.1 \\ 
  Median  
1990-2019 & 7.3 & 5.91 & 6.23 & 5.37 & 4.78 & 4.89 & 4.29 & 4.96 & 7.79 & 6.56 \\ 
  St.dev.  
1990-2019 & 2.25 & 1.89 & 2 & 2.14 & 1.37 & 1.65 & 1.83 & 1.28 & 2.27 & 1.67 \\ 
   \hline
Mean  
1990-1999 & 8.86 & 7.55 & 7.69 & 7.6 & 6.01 & 6.51 & 6.69 & 6.29 & 8.67 & 7.76 \\ 
  Median  
1990-1999 & 8.82 & 7.43 & 7.67 & 7.65 & 6.04 & 6.52 & 6.81 & 6.24 & 8.76 & 7.82 \\ 
  St.dev.  
1990-1999 & 0.36 & 0.39 & 0.25 & 0.23 & 0.09 & 0.14 & 0.67 & 0.26 & 0.4 & 0.23 \\ 
   \hline
Mean  
2000-2009 & 7.4 & 5.78 & 6.12 & 5.42 & 4.79 & 4.94 & 4.18 & 4.98 & 7.88 & 6.49 \\ 
  Median  
2000-2009 & 7.3 & 5.91 & 6.23 & 5.37 & 4.78 & 4.89 & 4.29 & 4.96 & 7.79 & 6.56 \\ 
  St.dev.  
2000-2009 & 1.22 & 0.96 & 1.3 & 1.25 & 0.73 & 1.03 & 0.79 & 0.55 & 1.4 & 0.8 \\ 
   \hline
Mean  
2010-2019 & 4.03 & 3.28 & 3.32 & 2.79 & 2.93 & 2.82 & 2.55 & 3.42 & 4.02 & 4.06 \\ 
  Median  
2010-2019 & 4.22 & 3.24 & 3.35 & 2.76 & 2.87 & 2.84 & 2.49 & 3.44 & 3.89 & 3.98 \\ 
  St.dev.  
2010-2019 & 1.04 & 0.49 & 0.49 & 0.43 & 0.43 & 0.34 & 0.26 & 0.53 & 0.85 & 0.62 \\ 
   \hline
\hline
\end{tabular}

\endgroup
\end{table}
\begin{table}[H]
\centering
\caption{Average Number of Citing Classes (Cited) at 4-Digit Level}\label{tab:sig_gen_citec4}
\begingroup\scriptsize
\begin{tabular}{l|lccccccccc}
  \hline
\hline
Period & 4-digit & Keyword & Science & WIPO & USPTO & All & G06 & H04W & C12 & B82 \\ 
  \hline
1990-2019 & Science & 0.00 &  &  &  &  &  &  &  &  \\ 
  1990-2019 & WIPO & 0.00 & 0.00 &  &  &  &  &  &  &  \\ 
  1990-2019 & USPTO & 0.00 & 0.04 & 0.00 &  &  &  &  &  &  \\ 
  1990-2019 & All & 0.00 & 0.00 & 0.00 & 0.00 &  &  &  &  &  \\ 
  1990-2019 & G06 & 0.00 & 0.00 & 0.00 & 0.00 & 0.00 &  &  &  &  \\ 
  1990-2019 & H04W & 0.00 & 0.37 & 0.06 & 0.90 & 0.00 & 0.00 &  &  &  \\ 
  1990-2019 & C12 & 0.00 & 0.00 & 0.00 & 0.07 & 0.00 & 0.01 & 0.19 &  &  \\
\hline
\hline
\end{tabular}

\endgroup
\caption*{Notes: Entries show the p-value of a two-sided paired Wilcoxon signed rank test for the hypothesis that the compared pair ranks equal.} 
\end{table}
\begin{table}[H]
\centering
\caption{Summary Statistics for Average Number of Citing Classes (Cited) at 4-Digit Level}\label{tab:sig_sum_gen_cite_c4} 
\begingroup\scriptsize
\begin{tabular}{l|cccccccccc}
  \hline
\hline
 & Keyword & Science & WIPO & USPTO & All & G06 & H04W & C12 & B82 & Y02 \\ 
  \hline
Mean  
1990-2019 & 10.4 & 8.94 & 9.43 & 8.56 & 6.66 & 7.26 & 8.56 & 7.75 & 10.12 & 9.29 \\ 
  Median  
1990-2019 & 11.53 & 9.61 & 10.42 & 8.72 & 6.96 & 7.49 & 8.29 & 7.7 & 11.51 & 10.14 \\ 
  St.dev.  
1990-2019 & 3.62 & 3.28 & 3.53 & 3.79 & 2.15 & 2.71 & 3.89 & 2.38 & 3.93 & 2.87 \\ 
   \hline
Mean  
1990-1999 & 13.64 & 12.4 & 12.79 & 12.66 & 8.89 & 10.1 & 13.25 & 10.42 & 13.4 & 12.07 \\ 
  Median  
1990-1999 & 13.76 & 12.2 & 12.82 & 12.78 & 8.94 & 10.09 & 13.52 & 10.41 & 13.53 & 12.26 \\ 
  St.dev.  
1990-1999 & 0.54 & 0.61 & 0.42 & 0.28 & 0.21 & 0.17 & 1.31 & 0.55 & 0.91 & 0.51 \\ 
   \hline
Mean  
2000-2009 & 11.6 & 9.4 & 10.29 & 8.86 & 7 & 7.6 & 8.05 & 7.77 & 11.68 & 10.05 \\ 
  Median  
2000-2009 & 11.53 & 9.61 & 10.42 & 8.72 & 6.96 & 7.49 & 8.29 & 7.7 & 11.51 & 10.14 \\ 
  St.dev.  
2000-2009 & 2 & 1.71 & 2.43 & 2.31 & 1.19 & 1.77 & 1.7 & 1.04 & 2.61 & 1.43 \\ 
   \hline
Mean  
2010-2019 & 5.97 & 5.03 & 5.21 & 4.17 & 4.1 & 4.09 & 4.4 & 5.05 & 5.28 & 5.75 \\ 
  Median  
2010-2019 & 6.19 & 4.97 & 5.29 & 4.17 & 4.03 & 4.16 & 4.29 & 5.08 & 5.02 & 5.52 \\ 
  St.dev.  
2010-2019 & 1.69 & 0.88 & 0.9 & 0.76 & 0.65 & 0.56 & 0.45 & 0.91 & 1.13 & 1.06 \\ 
   \hline
\hline
\end{tabular}

\endgroup
\end{table}

\FloatBarrier

\subsubsection{Complementarity}
\label{appendix:significance_complementarity}
    

\begin{table}[H]
\centering
\caption{Share of CPC Classes at 3-Digit Level}\label{tab:wilcoxon_complementarity_shareCPC3}
\begingroup\scriptsize
\begin{tabular}{l|lccccccccc}
  \hline
\hline
Period & 3-digit & Keyword & Science & WIPO & USPTO & All & G06 & H04W & C12 & B82 \\ 
  \hline
1990-2019 & Science & 0.29 &  &  &  &  &  &  &  &  \\ 
  1990-2019 & WIPO & 0.15 & 0.29 &  &  &  &  &  &  &  \\ 
  1990-2019 & USPTO & 0.00 & 0.00 & 0.00 &  &  &  &  &  &  \\ 
  1990-2019 & All & 0.00 & 0.00 & 0.00 & 0.00 &  &  &  &  &  \\ 
  1990-2019 & G06 & 0.00 & 0.00 & 0.00 & 0.00 & 0.00 &  &  &  &  \\ 
  1990-2019 & H04W & 0.00 & 0.00 & 0.00 & 0.00 & 0.00 & 0.00 &  &  &  \\ 
  1990-2019 & C12 & 0.00 & 0.00 & 0.00 & 0.00 & 0.00 & 0.00 & 0.00 &  &  \\ 
  1990-2019 & B82 & 0.00 & 0.00 & 0.01 & 0.00 & 0.00 & 0.00 & 0.00 & 0.29 &  \\ 
  1990-2019 & Y02 & 0.00 & 0.00 & 0.00 & 0.00 & 0.00 & 0.00 & 0.00 & 0.00 & 0.00 \\ 
   \hline
1990-1999 & Science & 0.32 &  &  &  &  &  &  &  &  \\ 
  1990-1999 & WIPO & 0.32 & 0.14 &  &  &  &  &  &  &  \\ 
  1990-1999 & USPTO & 0.09 & 0.09 & 0.13 &  &  &  &  &  &  \\ 
  1990-1999 & All & 0.13 & 0.13 & 0.13 & 0.09 &  &  &  &  &  \\ 
  1990-1999 & G06 & 0.10 & 0.13 & 0.13 & 0.13 & 0.09 &  &  &  &  \\ 
  1990-1999 & H04W & 0.13 & 0.09 & 0.09 & 0.09 & 0.13 & 0.09 &  &  &  \\ 
  1990-1999 & C12 & 0.10 & 0.32 & 0.09 & 0.09 & 0.13 & 0.09 & 0.13 &  &  \\ 
  1990-1999 & B82 & 0.13 & 0.13 & 0.09 & 0.09 & 0.13 & 0.09 & 0.09 & 0.17 &  \\ 
  1990-1999 & Y02 & 0.09 & 0.09 & 0.13 & 0.13 & 0.13 & 0.09 & 0.09 & 0.13 & 0.09 \\ 
   \hline
2000-2009 & Science & 1.00 &  &  &  &  &  &  &  &  \\ 
  2000-2009 & WIPO & 1.00 & 0.77 &  &  &  &  &  &  &  \\ 
  2000-2009 & USPTO & 0.18 & 0.09 & 0.18 &  &  &  &  &  &  \\ 
  2000-2009 & All & 0.18 & 0.18 & 0.18 & 0.18 &  &  &  &  &  \\ 
  2000-2009 & G06 & 0.18 & 0.18 & 0.18 & 0.18 & 0.18 &  &  &  &  \\ 
  2000-2009 & H04W & 0.18 & 0.09 & 0.09 & 0.18 & 0.18 & 0.09 &  &  &  \\ 
  2000-2009 & C12 & 0.09 & 0.09 & 0.12 & 0.09 & 0.18 & 0.09 & 0.09 &  &  \\ 
  2000-2009 & B82 & 0.63 & 0.63 & 1.00 & 0.09 & 0.18 & 0.18 & 0.18 & 0.18 &  \\ 
  2000-2009 & Y02 & 0.18 & 0.09 & 0.09 & 0.18 & 0.18 & 0.18 & 0.18 & 0.09 & 0.18 \\ 
   \hline
2010-2019 & Science & 0.88 &  &  &  &  &  &  &  &  \\ 
  2010-2019 & WIPO & 0.09 & 0.13 &  &  &  &  &  &  &  \\ 
  2010-2019 & USPTO & 0.13 & 0.09 & 0.09 &  &  &  &  &  &  \\ 
  2010-2019 & All & 0.13 & 0.09 & 0.09 & 0.13 &  &  &  &  &  \\ 
  2010-2019 & G06 & 0.13 & 0.13 & 0.13 & 0.88 & 0.09 &  &  &  &  \\ 
  2010-2019 & H04W & 0.09 & 0.09 & 0.09 & 0.09 & 0.13 & 0.09 &  &  &  \\ 
  2010-2019 & C12 & 0.09 & 0.09 & 0.13 & 0.09 & 0.13 & 0.13 & 0.09 &  &  \\ 
  2010-2019 & B82 & 0.09 & 0.09 & 0.88 & 0.09 & 0.13 & 0.09 & 0.13 & 0.88 &  \\ 
  2010-2019 & Y02 & 0.09 & 0.09 & 0.09 & 0.13 & 0.13 & 0.13 & 0.13 & 0.13 & 0.13 \\ 
   \hline
\hline
\end{tabular}

\endgroup
\caption*{Notes: Entries show the p-value of a two-sided paired Wilcoxon signed rank test for the hypothesis that the compared pair ranks equal.}

\end{table}
\begin{table}[H]
\centering
\caption{Summary Statistics for Share of CPC Classes at 3-Digit Level}\label{tab:sum_complementarity_shareCPC3}
\begingroup\scriptsize

\begin{tabular}{l|cccccccccc}
  \hline
\hline
 & Keyword & Science & WIPO & USPTO & All & G06 & H04W & C12 & B82 & Y02 \\ 
  \hline
Mean  
1990-2019 & 0.61 & 0.6 & 0.59 & 0.82 & 0.93 & 0.7 & 0.3 & 0.51 & 0.54 & 0.87 \\ 
  Median  
1990-2019 & 0.6 & 0.59 & 0.57 & 0.83 & 0.93 & 0.71 & 0.28 & 0.49 & 0.57 & 0.88 \\ 
  St.dev.  
1990-2019 & 0.15 & 0.16 & 0.12 & 0.06 & 0 & 0.13 & 0.17 & 0.12 & 0.14 & 0.03 \\ 
   \hline
Mean  
1990-1999 & 0.47 & 0.43 & 0.48 & 0.76 & 0.93 & 0.56 & 0.13 & 0.4 & 0.37 & 0.85 \\ 
  Median  
1990-1999 & 0.45 & 0.43 & 0.51 & 0.75 & 0.93 & 0.56 & 0.12 & 0.4 & 0.36 & 0.85 \\ 
  St.dev.  
1990-1999 & 0.08 & 0.1 & 0.07 & 0.05 & 0 & 0.06 & 0.03 & 0.06 & 0.06 & 0.02 \\ 
   \hline
Mean  
2000-2009 & 0.59 & 0.59 & 0.58 & 0.82 & 0.93 & 0.7 & 0.27 & 0.48 & 0.57 & 0.87 \\ 
  Median  
2000-2009 & 0.6 & 0.58 & 0.57 & 0.83 & 0.93 & 0.71 & 0.28 & 0.49 & 0.57 & 0.88 \\ 
  St.dev.  
2000-2009 & 0.04 & 0.03 & 0.04 & 0.02 & 0 & 0.04 & 0.05 & 0.04 & 0.05 & 0.02 \\ 
   \hline
Mean  
2010-2019 & 0.78 & 0.77 & 0.71 & 0.87 & 0.93 & 0.85 & 0.49 & 0.64 & 0.68 & 0.9 \\ 
  Median  
2010-2019 & 0.79 & 0.77 & 0.69 & 0.86 & 0.93 & 0.85 & 0.46 & 0.69 & 0.67 & 0.91 \\ 
  St.dev.  
2010-2019 & 0.08 & 0.06 & 0.08 & 0.03 & 0 & 0.04 & 0.11 & 0.09 & 0.04 & 0.02 \\ 
   \hline
\hline
\end{tabular}

\endgroup
\end{table}
\begin{table}[H]
\centering
\caption{Share of CPC Classes at 4-Digit Level} \label{tab:wilcoxon_complementarity_shareCPC4}

\begingroup\scriptsize

\begin{tabular}{l|lccccccccc}
  \hline
\hline
Period & 4-digit & Keyword & Science & WIPO & USPTO & All & G06 & H04W & C12 & B82 \\ 
  \hline
1990-2019 & Science & 0.83 &  &  &  &  &  &  &  &  \\ 
  1990-2019 & WIPO & 0.83 & 0.33 &  &  &  &  &  &  &  \\ 
  1990-2019 & USPTO & 0.00 & 0.00 & 0.00 &  &  &  &  &  &  \\ 
  1990-2019 & All & 0.00 & 0.00 & 0.00 & 0.00 &  &  &  &  &  \\ 
  1990-2019 & G06 & 0.00 & 0.00 & 0.00 & 0.00 & 0.00 &  &  &  &  \\ 
  1990-2019 & H04W & 0.00 & 0.00 & 0.00 & 0.00 & 0.00 & 0.00 &  &  &  \\ 
  1990-2019 & C12 & 0.00 & 0.00 & 0.00 & 0.00 & 0.00 & 0.00 & 0.00 &  &  \\ 
  1990-2019 & B82 & 0.00 & 0.00 & 0.00 & 0.00 & 0.00 & 0.00 & 0.00 & 0.31 &  \\ 
  1990-2019 & Y02 & 0.00 & 0.00 & 0.00 & 0.00 & 0.00 & 0.00 & 0.00 & 0.00 & 0.00 \\ 
   \hline
1990-1999 & Science & 0.77 &  &  &  &  &  &  &  &  \\ 
  1990-1999 & WIPO & 0.48 & 0.42 &  &  &  &  &  &  &  \\ 
  1990-1999 & USPTO & 0.09 & 0.09 & 0.09 &  &  &  &  &  &  \\ 
  1990-1999 & All & 0.09 & 0.09 & 0.09 & 0.09 &  &  &  &  &  \\ 
  1990-1999 & G06 & 0.09 & 0.09 & 0.09 & 0.09 & 0.09 &  &  &  &  \\ 
  1990-1999 & H04W & 0.09 & 0.09 & 0.09 & 0.09 & 0.09 & 0.09 &  &  &  \\ 
  1990-1999 & C12 & 0.09 & 0.48 & 0.09 & 0.09 & 0.09 & 0.09 & 0.09 &  &  \\ 
  1990-1999 & B82 & 0.09 & 0.09 & 0.09 & 0.09 & 0.09 & 0.09 & 0.09 & 0.09 &  \\ 
  1990-1999 & Y02 & 0.09 & 0.09 & 0.09 & 0.09 & 0.09 & 0.09 & 0.09 & 0.09 & 0.09 \\ 
   \hline
2000-2009 & Science & 0.09 &  &  &  &  &  &  &  &  \\ 
  2000-2009 & WIPO & 0.92 & 0.09 &  &  &  &  &  &  &  \\ 
  2000-2009 & USPTO & 0.09 & 0.09 & 0.09 &  &  &  &  &  &  \\ 
  2000-2009 & All & 0.09 & 0.09 & 0.09 & 0.09 &  &  &  &  &  \\ 
  2000-2009 & G06 & 0.09 & 0.09 & 0.09 & 0.09 & 0.09 &  &  &  &  \\ 
  2000-2009 & H04W & 0.09 & 0.09 & 0.09 & 0.09 & 0.09 & 0.09 &  &  &  \\ 
  2000-2009 & C12 & 0.09 & 0.09 & 0.09 & 0.09 & 0.09 & 0.09 & 0.09 &  &  \\ 
  2000-2009 & B82 & 0.09 & 0.09 & 0.09 & 0.09 & 0.09 & 0.09 & 0.09 & 0.09 &  \\ 
  2000-2009 & Y02 & 0.09 & 0.09 & 0.09 & 0.09 & 0.09 & 0.09 & 0.09 & 0.09 & 0.09 \\ 
   \hline
2010-2019 & Science & 0.48 &  &  &  &  &  &  &  &  \\ 
  2010-2019 & WIPO & 0.09 & 0.22 &  &  &  &  &  &  &  \\ 
  2010-2019 & USPTO & 0.09 & 0.09 & 0.09 &  &  &  &  &  &  \\ 
  2010-2019 & All & 0.09 & 0.09 & 0.09 & 0.09 &  &  &  &  &  \\ 
  2010-2019 & G06 & 0.09 & 0.09 & 0.09 & 0.22 & 0.09 &  &  &  &  \\ 
  2010-2019 & H04W & 0.09 & 0.09 & 0.09 & 0.09 & 0.09 & 0.09 &  &  &  \\ 
  2010-2019 & C12 & 0.09 & 0.09 & 0.09 & 0.09 & 0.09 & 0.09 & 0.22 &  &  \\ 
  2010-2019 & B82 & 0.09 & 0.09 & 0.09 & 0.09 & 0.09 & 0.09 & 0.25 & 0.46 &  \\ 
  2010-2019 & Y02 & 0.09 & 0.09 & 0.09 & 0.09 & 0.09 & 0.09 & 0.09 & 0.09 & 0.09 \\ 
   \hline
\hline
\end{tabular}

\endgroup
\caption*{Notes: Entries show the p-value of a two-sided paired Wilcoxon signed rank test for the hypothesis that the compared pair ranks equal.}
\end{table}
\begin{table}[H]
\centering
\caption{Summary Statistics for the Share of CPC Classes at 4-Digit Level}\label{tab:sum_complementarity_shareCPC4}

\begingroup\scriptsize

\begin{tabular}{l|cccccccccc}
  \hline
\hline
 & Keyword & Science & WIPO & USPTO & All & G06 & H04W & C12 & B82 & Y02 \\ 
  \hline
Mean  
1990-2019 & 0.37 & 0.37 & 0.36 & 0.6 & 0.96 & 0.49 & 0.17 & 0.27 & 0.29 & 0.74 \\ 
  Median  
1990-2019 & 0.33 & 0.35 & 0.33 & 0.6 & 0.96 & 0.48 & 0.14 & 0.25 & 0.31 & 0.72 \\ 
  St.dev.  
1990-2019 & 0.14 & 0.13 & 0.12 & 0.1 & 0.01 & 0.15 & 0.13 & 0.08 & 0.09 & 0.06 \\ 
   \hline
Mean  
1990-1999 & 0.24 & 0.24 & 0.25 & 0.48 & 0.96 & 0.33 & 0.06 & 0.2 & 0.18 & 0.68 \\ 
  Median  
1990-1999 & 0.24 & 0.23 & 0.27 & 0.48 & 0.96 & 0.33 & 0.06 & 0.2 & 0.17 & 0.68 \\ 
  St.dev.  
1990-1999 & 0.06 & 0.07 & 0.05 & 0.05 & 0 & 0.05 & 0.02 & 0.03 & 0.03 & 0.03 \\ 
   \hline
Mean  
2000-2009 & 0.33 & 0.36 & 0.33 & 0.6 & 0.96 & 0.47 & 0.14 & 0.24 & 0.3 & 0.72 \\ 
  Median  
2000-2009 & 0.33 & 0.35 & 0.33 & 0.6 & 0.96 & 0.48 & 0.14 & 0.24 & 0.31 & 0.72 \\ 
  St.dev.  
2000-2009 & 0.02 & 0.02 & 0.02 & 0.02 & 0 & 0.03 & 0.03 & 0.01 & 0.03 & 0.01 \\ 
   \hline
Mean  
2010-2019 & 0.54 & 0.53 & 0.5 & 0.71 & 0.97 & 0.66 & 0.32 & 0.36 & 0.39 & 0.81 \\ 
  Median  
2010-2019 & 0.56 & 0.53 & 0.5 & 0.7 & 0.97 & 0.67 & 0.31 & 0.38 & 0.39 & 0.82 \\ 
  St.dev.  
2010-2019 & 0.09 & 0.07 & 0.09 & 0.05 & 0.01 & 0.08 & 0.11 & 0.07 & 0.02 & 0.04 \\ 
   \hline
\hline
\end{tabular}

\endgroup
\end{table}


\begin{table}[H]
\centering
\caption{Average Diversity at 1-Digit Level}\label{tab:wilcoxon_complementarity_avg_1}
\begingroup\scriptsize

\begin{tabular}{l|lccccccccc}
  \hline
\hline
Period & 1-digit & Keyword & Science & WIPO & USPTO & All & G06 & H04W & C12 & B82 \\ 
  \hline
1990-2019 & Science & 1.00 &  &  &  &  &  &  &  &  \\ 
  1990-2019 & WIPO & 0.02 & 0.05 &  &  &  &  &  &  &  \\ 
  1990-2019 & USPTO & 0.00 & 0.00 & 0.00 &  &  &  &  &  &  \\ 
  1990-2019 & All & 0.00 & 0.02 & 1.00 & 0.00 &  &  &  &  &  \\ 
  1990-2019 & G06 & 0.00 & 0.06 & 1.00 & 0.00 & 1.00 &  &  &  &  \\ 
  1990-2019 & H04W & 0.00 & 0.01 & 0.03 & 0.00 & 0.00 & 0.00 &  &  &  \\ 
  1990-2019 & C12 & 0.00 & 0.00 & 0.00 & 0.00 & 0.00 & 0.00 & 0.00 &  &  \\ 
  1990-2019 & B82 & 0.00 & 0.00 & 0.00 & 0.00 & 0.00 & 0.00 & 0.00 & 0.00 &  \\ 
  1990-2019 & Y02 & 0.00 & 0.00 & 0.00 & 0.00 & 0.00 & 0.00 & 0.00 & 0.00 & 0.82 \\ 
   \hline
1990-1999 & Science & 1.00 &  &  &  &  &  &  &  &  \\ 
  1990-1999 & WIPO & 0.09 & 1.00 &  &  &  &  &  &  &  \\ 
  1990-1999 & USPTO & 0.09 & 0.84 & 0.15 &  &  &  &  &  &  \\ 
  1990-1999 & All & 0.25 & 1.00 & 0.09 & 0.09 &  &  &  &  &  \\ 
  1990-1999 & G06 & 0.09 & 1.00 & 0.15 & 0.09 & 0.45 &  &  &  &  \\ 
  1990-1999 & H04W & 1.00 & 1.00 & 0.84 & 0.09 & 1.00 & 1.00 &  &  &  \\ 
  1990-1999 & C12 & 0.09 & 0.09 & 0.09 & 0.09 & 0.09 & 0.09 & 0.09 &  &  \\ 
  1990-1999 & B82 & 0.09 & 0.09 & 0.09 & 0.09 & 0.09 & 0.09 & 0.09 & 0.09 &  \\ 
  1990-1999 & Y02 & 0.09 & 0.09 & 0.09 & 0.09 & 0.09 & 0.09 & 0.09 & 0.09 & 0.71 \\ 
   \hline
2000-2009 & Science & 0.09 &  &  &  &  &  &  &  &  \\ 
  2000-2009 & WIPO & 0.29 & 0.09 &  &  &  &  &  &  &  \\ 
  2000-2009 & USPTO & 0.09 & 0.09 & 0.09 &  &  &  &  &  &  \\ 
  2000-2009 & All & 0.09 & 0.09 & 0.83 & 0.09 &  &  &  &  &  \\ 
  2000-2009 & G06 & 0.09 & 0.09 & 0.83 & 0.09 & 0.83 &  &  &  &  \\ 
  2000-2009 & H04W & 0.09 & 0.09 & 0.09 & 0.29 & 0.09 & 0.09 &  &  &  \\ 
  2000-2009 & C12 & 0.09 & 0.09 & 0.09 & 0.09 & 0.09 & 0.09 & 0.09 &  &  \\ 
  2000-2009 & B82 & 0.09 & 0.09 & 0.09 & 0.09 & 0.09 & 0.09 & 0.09 & 0.09 &  \\ 
  2000-2009 & Y02 & 0.09 & 0.09 & 0.09 & 0.09 & 0.09 & 0.09 & 0.09 & 0.09 & 0.29 \\ 
   \hline
2010-2019 & Science & 1.00 &  &  &  &  &  &  &  &  \\ 
  2010-2019 & WIPO & 1.00 & 1.00 &  &  &  &  &  &  &  \\ 
  2010-2019 & USPTO & 0.09 & 0.09 & 0.09 &  &  &  &  &  &  \\ 
  2010-2019 & All & 0.58 & 0.11 & 0.84 & 0.09 &  &  &  &  &  \\ 
  2010-2019 & G06 & 1.00 & 1.00 & 1.00 & 0.09 & 0.84 &  &  &  &  \\ 
  2010-2019 & H04W & 0.09 & 0.09 & 0.09 & 0.27 & 0.09 & 0.09 &  &  &  \\ 
  2010-2019 & C12 & 0.09 & 0.09 & 0.09 & 0.09 & 0.09 & 0.09 & 0.09 &  &  \\ 
  2010-2019 & B82 & 0.09 & 0.09 & 0.09 & 0.09 & 0.09 & 0.09 & 0.09 & 0.09 &  \\ 
  2010-2019 & Y02 & 0.09 & 0.09 & 0.09 & 0.09 & 0.09 & 0.09 & 0.09 & 0.09 & 0.09 \\ 
   \hline
\hline
\end{tabular}

\endgroup
\caption*{Notes: Entries show the p-value of a two-sided paired Wilcoxon signed rank test for the hypothesis that the compared pair ranks equal.}
\end{table}
\begin{table}[H]
\centering
\caption{Summary Statistics for Average Diversity at 1-Digit Level}\label{tab:sum_complementarity_avg_1}
\begingroup\scriptsize
\begin{tabular}{l|cccccccccc}
  \hline
\hline
 & Keyword & Science & WIPO & USPTO & All & G06 & H04W & C12 & B82 & Y02 \\ 
  \hline
Mean  
1990-2019 & 1.39 & 1.4 & 1.36 & 1.27 & 1.36 & 1.36 & 1.32 & 1.8 & 2.48 & 2.47 \\ 
  Median  
1990-2019 & 1.36 & 1.42 & 1.33 & 1.24 & 1.33 & 1.32 & 1.29 & 1.78 & 2.48 & 2.46 \\ 
  St.dev.  
1990-2019 & 0.09 & 0.1 & 0.12 & 0.08 & 0.08 & 0.11 & 0.08 & 0.15 & 0.11 & 0.05 \\ 
   \hline
Mean  
1990-1999 & 1.35 & 1.3 & 1.27 & 1.24 & 1.31 & 1.29 & 1.33 & 1.64 & 2.39 & 2.43 \\ 
  Median  
1990-1999 & 1.35 & 1.28 & 1.27 & 1.24 & 1.32 & 1.29 & 1.29 & 1.65 & 2.37 & 2.43 \\ 
  St.dev.  
1990-1999 & 0.03 & 0.08 & 0.02 & 0.01 & 0.02 & 0.01 & 0.08 & 0.08 & 0.08 & 0.03 \\ 
   \hline
Mean  
2000-2009 & 1.35 & 1.43 & 1.33 & 1.24 & 1.33 & 1.32 & 1.26 & 1.79 & 2.48 & 2.46 \\ 
  Median  
2000-2009 & 1.36 & 1.43 & 1.33 & 1.24 & 1.33 & 1.32 & 1.26 & 1.78 & 2.48 & 2.45 \\ 
  St.dev.  
2000-2009 & 0.02 & 0.02 & 0.03 & 0.01 & 0.01 & 0.01 & 0.02 & 0.04 & 0.03 & 0.01 \\ 
   \hline
Mean  
2010-2019 & 1.47 & 1.47 & 1.48 & 1.35 & 1.45 & 1.47 & 1.37 & 1.97 & 2.58 & 2.52 \\ 
  Median  
2010-2019 & 1.48 & 1.47 & 1.47 & 1.35 & 1.44 & 1.48 & 1.36 & 1.99 & 2.55 & 2.51 \\ 
  St.dev.  
2010-2019 & 0.11 & 0.08 & 0.14 & 0.11 & 0.09 & 0.12 & 0.09 & 0.08 & 0.1 & 0.06 \\ 
   \hline
\hline
\end{tabular}

\endgroup
\end{table}
\begin{table}[H]
\centering
\caption{Average Diversity at 3-Digit Level}\label{tab:wilcoxon_complementarity_avg_3}

\begingroup\scriptsize

\begin{tabular}{l|lccccccccc}
  \hline
\hline
Period & 3-digit & Keyword & Science & WIPO & USPTO & All & G06 & H04W & C12 & B82 \\ 
  \hline
1990-2019 & Science & 0.65 &  &  &  &  &  &  &  &  \\ 
  1990-2019 & WIPO & 0.38 & 0.87 &  &  &  &  &  &  &  \\ 
  1990-2019 & USPTO & 0.00 & 0.00 & 0.00 &  &  &  &  &  &  \\ 
  1990-2019 & All & 0.00 & 0.00 & 0.00 & 0.00 &  &  &  &  &  \\ 
  1990-2019 & G06 & 0.79 & 0.79 & 0.46 & 0.00 & 0.00 &  &  &  &  \\ 
  1990-2019 & H04W & 0.00 & 0.00 & 0.00 & 0.79 & 0.00 & 0.00 &  &  &  \\ 
  1990-2019 & C12 & 0.00 & 0.00 & 0.00 & 0.00 & 0.00 & 0.00 & 0.00 &  &  \\ 
  1990-2019 & B82 & 0.00 & 0.00 & 0.00 & 0.00 & 0.00 & 0.00 & 0.00 & 0.00 &  \\ 
  1990-2019 & Y02 & 0.00 & 0.00 & 0.00 & 0.00 & 0.00 & 0.00 & 0.00 & 0.00 & 0.00 \\ 
   \hline
1990-1999 & Science & 1.00 &  &  &  &  &  &  &  &  \\ 
  1990-1999 & WIPO & 0.09 & 1.00 &  &  &  &  &  &  &  \\ 
  1990-1999 & USPTO & 0.09 & 0.09 & 0.09 &  &  &  &  &  &  \\ 
  1990-1999 & All & 0.09 & 1.00 & 0.76 & 0.09 &  &  &  &  &  \\ 
  1990-1999 & G06 & 1.00 & 1.00 & 0.09 & 0.09 & 0.09 &  &  &  &  \\ 
  1990-1999 & H04W & 0.09 & 1.00 & 0.49 & 1.00 & 1.00 & 0.09 &  &  &  \\ 
  1990-1999 & C12 & 0.09 & 0.09 & 0.09 & 0.09 & 0.09 & 0.09 & 0.09 &  &  \\ 
  1990-1999 & B82 & 0.09 & 0.09 & 0.09 & 0.09 & 0.09 & 0.09 & 0.09 & 0.09 &  \\ 
  1990-1999 & Y02 & 0.09 & 0.09 & 0.09 & 0.09 & 0.09 & 0.09 & 0.09 & 0.09 & 0.09 \\ 
   \hline
2000-2009 & Science & 0.09 &  &  &  &  &  &  &  &  \\ 
  2000-2009 & WIPO & 0.09 & 0.09 &  &  &  &  &  &  &  \\ 
  2000-2009 & USPTO & 0.09 & 0.09 & 0.09 &  &  &  &  &  &  \\ 
  2000-2009 & All & 0.09 & 0.09 & 0.09 & 0.09 &  &  &  &  &  \\ 
  2000-2009 & G06 & 0.92 & 0.09 & 0.09 & 0.09 & 0.09 &  &  &  &  \\ 
  2000-2009 & H04W & 0.09 & 0.09 & 0.09 & 0.86 & 0.09 & 0.09 &  &  &  \\ 
  2000-2009 & C12 & 0.09 & 0.09 & 0.09 & 0.09 & 0.09 & 0.09 & 0.09 &  &  \\ 
  2000-2009 & B82 & 0.09 & 0.09 & 0.09 & 0.09 & 0.09 & 0.09 & 0.09 & 0.09 &  \\ 
  2000-2009 & Y02 & 0.09 & 0.09 & 0.09 & 0.09 & 0.09 & 0.09 & 0.09 & 0.09 & 0.09 \\ 
   \hline
2010-2019 & Science & 1.00 &  &  &  &  &  &  &  &  \\ 
  2010-2019 & WIPO & 0.09 & 0.10 &  &  &  &  &  &  &  \\ 
  2010-2019 & USPTO & 0.09 & 0.09 & 0.09 &  &  &  &  &  &  \\ 
  2010-2019 & All & 0.09 & 0.09 & 0.09 & 0.09 &  &  &  &  &  \\ 
  2010-2019 & G06 & 0.32 & 1.00 & 0.09 & 0.09 & 0.09 &  &  &  &  \\ 
  2010-2019 & H04W & 0.09 & 0.09 & 0.09 & 0.10 & 0.09 & 0.09 &  &  &  \\ 
  2010-2019 & C12 & 0.09 & 0.09 & 0.09 & 0.09 & 0.09 & 0.09 & 0.09 &  &  \\ 
  2010-2019 & B82 & 0.09 & 0.09 & 0.09 & 0.09 & 0.09 & 0.09 & 0.09 & 0.09 &  \\ 
  2010-2019 & Y02 & 0.09 & 0.09 & 0.09 & 0.09 & 0.09 & 0.09 & 0.09 & 0.09 & 0.09 \\ 
   \hline
\hline
\end{tabular}

\endgroup
\caption*{Notes: Entries show the p-value of a two-sided paired Wilcoxon signed rank test for the hypothesis that the compared pair ranks equal.} 
\end{table}
\begin{table}[H]
\centering
\caption{Summary Statistics for Average Diversity at 3-Digit Level}\label{tab:sum_complementarity_avg_3}
\begingroup\scriptsize

\begin{tabular}{l|cccccccccc}
  \hline
\hline
 & Keyword & Science & WIPO & USPTO & All & G06 & H04W & C12 & B82 & Y02 \\ 
  \hline
Mean  
1990-2019 & 1.61 & 1.64 & 1.64 & 1.43 & 1.54 & 1.62 & 1.43 & 2.32 & 3.03 & 2.85 \\ 
  Median  
1990-2019 & 1.55 & 1.65 & 1.58 & 1.39 & 1.48 & 1.56 & 1.39 & 2.35 & 2.99 & 2.81 \\ 
  St.dev.  
1990-2019 & 0.15 & 0.14 & 0.21 & 0.12 & 0.12 & 0.15 & 0.11 & 0.16 & 0.19 & 0.12 \\ 
   \hline
Mean  
1990-1999 & 1.55 & 1.51 & 1.5 & 1.39 & 1.48 & 1.57 & 1.43 & 2.16 & 2.88 & 2.77 \\ 
  Median  
1990-1999 & 1.56 & 1.48 & 1.48 & 1.4 & 1.48 & 1.57 & 1.4 & 2.16 & 2.86 & 2.79 \\ 
  St.dev.  
1990-1999 & 0.04 & 0.1 & 0.04 & 0.02 & 0.03 & 0.04 & 0.09 & 0.15 & 0.09 & 0.06 \\ 
   \hline
Mean  
2000-2009 & 1.54 & 1.69 & 1.58 & 1.36 & 1.48 & 1.54 & 1.35 & 2.36 & 3.01 & 2.81 \\ 
  Median  
2000-2009 & 1.54 & 1.68 & 1.59 & 1.36 & 1.48 & 1.54 & 1.35 & 2.36 & 3.01 & 2.81 \\ 
  St.dev.  
2000-2009 & 0.02 & 0.05 & 0.05 & 0.02 & 0.01 & 0.02 & 0.03 & 0.04 & 0.07 & 0.03 \\ 
   \hline
Mean  
2010-2019 & 1.74 & 1.73 & 1.85 & 1.53 & 1.67 & 1.75 & 1.49 & 2.46 & 3.2 & 2.96 \\ 
  Median  
2010-2019 & 1.73 & 1.71 & 1.82 & 1.52 & 1.65 & 1.74 & 1.46 & 2.48 & 3.15 & 2.91 \\ 
  St.dev.  
2010-2019 & 0.2 & 0.14 & 0.26 & 0.17 & 0.15 & 0.21 & 0.14 & 0.08 & 0.22 & 0.14 \\ 
   \hline
\hline
\end{tabular}

\endgroup
\end{table}
\begin{table}[H]
\centering
\caption{Average Diversity at 4-Digit Level}\label{tab:wilcoxon_complementarity_avg_4}
\begingroup\scriptsize

\begin{tabular}{l|lccccccccc}
  \hline
\hline
Period & 4-digit & Keyword & Science & WIPO & USPTO & All & G06 & H04W & C12 & B82 \\ 
  \hline
1990-2019 & Science & 0.01 &  &  &  &  &  &  &  &  \\ 
  1990-2019 & WIPO & 0.00 & 0.10 &  &  &  &  &  &  &  \\ 
  1990-2019 & USPTO & 0.00 & 0.00 & 0.00 &  &  &  &  &  &  \\ 
  1990-2019 & All & 0.00 & 0.00 & 0.00 & 0.00 &  &  &  &  &  \\ 
  1990-2019 & G06 & 0.00 & 0.00 & 0.00 & 0.00 & 0.52 &  &  &  &  \\ 
  1990-2019 & H04W & 0.00 & 0.00 & 0.00 & 0.00 & 0.00 & 0.00 &  &  &  \\ 
  1990-2019 & C12 & 0.00 & 0.00 & 0.00 & 0.00 & 0.00 & 0.00 & 0.00 &  &  \\ 
  1990-2019 & B82 & 0.00 & 0.00 & 0.00 & 0.00 & 0.00 & 0.00 & 0.00 & 0.00 &  \\ 
  1990-2019 & Y02 & 0.00 & 0.00 & 0.00 & 0.00 & 0.00 & 0.00 & 0.00 & 0.00 & 0.00 \\ 
   \hline
1990-1999 & Science & 1.00 &  &  &  &  &  &  &  &  \\ 
  1990-1999 & WIPO & 1.00 & 1.00 &  &  &  &  &  &  &  \\ 
  1990-1999 & USPTO & 0.09 & 0.09 & 0.09 &  &  &  &  &  &  \\ 
  1990-1999 & All & 0.09 & 1.00 & 0.09 & 0.09 &  &  &  &  &  \\ 
  1990-1999 & G06 & 0.09 & 1.00 & 0.09 & 0.09 & 1.00 &  &  &  &  \\ 
  1990-1999 & H04W & 0.09 & 0.09 & 0.09 & 0.09 & 0.09 & 0.09 &  &  &  \\ 
  1990-1999 & C12 & 0.09 & 0.09 & 0.09 & 0.09 & 0.09 & 0.09 & 0.09 &  &  \\ 
  1990-1999 & B82 & 0.09 & 0.09 & 0.09 & 0.09 & 0.09 & 0.09 & 0.09 & 0.09 &  \\ 
  1990-1999 & Y02 & 0.09 & 0.09 & 0.09 & 0.09 & 0.09 & 0.09 & 0.09 & 0.09 & 0.10 \\ 
   \hline
2000-2009 & Science & 0.09 &  &  &  &  &  &  &  &  \\ 
  2000-2009 & WIPO & 0.09 & 0.26 &  &  &  &  &  &  &  \\ 
  2000-2009 & USPTO & 0.09 & 0.09 & 0.09 &  &  &  &  &  &  \\ 
  2000-2009 & All & 0.09 & 0.09 & 0.09 & 0.09 &  &  &  &  &  \\ 
  2000-2009 & G06 & 0.09 & 0.09 & 0.09 & 0.09 & 0.28 &  &  &  &  \\ 
  2000-2009 & H04W & 0.09 & 0.09 & 0.09 & 0.09 & 0.09 & 0.09 &  &  &  \\ 
  2000-2009 & C12 & 0.09 & 0.09 & 0.09 & 0.09 & 0.09 & 0.09 & 0.09 &  &  \\ 
  2000-2009 & B82 & 0.09 & 0.09 & 0.09 & 0.09 & 0.09 & 0.09 & 0.09 & 0.09 &  \\ 
  2000-2009 & Y02 & 0.09 & 0.09 & 0.09 & 0.09 & 0.09 & 0.09 & 0.09 & 0.09 & 0.09 \\ 
   \hline
2010-2019 & Science & 0.25 &  &  &  &  &  &  &  &  \\ 
  2010-2019 & WIPO & 0.09 & 0.09 &  &  &  &  &  &  &  \\ 
  2010-2019 & USPTO & 0.09 & 0.09 & 0.09 &  &  &  &  &  &  \\ 
  2010-2019 & All & 0.09 & 0.09 & 0.09 & 0.09 &  &  &  &  &  \\ 
  2010-2019 & G06 & 0.09 & 0.09 & 0.09 & 0.09 & 0.64 &  &  &  &  \\ 
  2010-2019 & H04W & 0.09 & 0.09 & 0.70 & 0.09 & 0.09 & 0.09 &  &  &  \\ 
  2010-2019 & C12 & 0.09 & 0.09 & 0.09 & 0.09 & 0.09 & 0.09 & 0.09 &  &  \\ 
  2010-2019 & B82 & 0.09 & 0.09 & 0.09 & 0.09 & 0.09 & 0.09 & 0.09 & 0.09 &  \\ 
  2010-2019 & Y02 & 0.09 & 0.09 & 0.09 & 0.09 & 0.09 & 0.09 & 0.09 & 0.09 & 0.15 \\ 
   \hline
\hline
\end{tabular}

\endgroup
\flushleft\caption*{Notes: Entries show the p-value of a two-sided paired Wilcoxon signed rank test for the hypothesis that the compared pair ranks equal.}
\end{table}
\begin{table}[H]
\centering
\caption{Summary Statistics for Average Diversity at 4-Digit Level} \label{tab:sum_complementarity_avg_4}
\begingroup\scriptsize

\begin{tabular}{l|cccccccccc}
  \hline
\hline
 & Keyword & Science & WIPO & USPTO & All & G06 & H04W & C12 & B82 & Y02 \\ 
  \hline
Mean  
1990-2019 & 1.88 & 1.97 & 2.05 & 1.64 & 1.8 & 1.81 & 2.26 & 2.93 & 3.5 & 3.39 \\ 
  Median  
1990-2019 & 1.77 & 1.97 & 1.96 & 1.56 & 1.71 & 1.71 & 2.2 & 2.89 & 3.41 & 3.36 \\ 
  St.dev.  
1990-2019 & 0.26 & 0.25 & 0.38 & 0.21 & 0.21 & 0.25 & 0.21 & 0.3 & 0.27 & 0.23 \\ 
   \hline
Mean  
1990-1999 & 1.77 & 1.73 & 1.76 & 1.55 & 1.68 & 1.69 & 2.16 & 2.64 & 3.31 & 3.2 \\ 
  Median  
1990-1999 & 1.77 & 1.67 & 1.75 & 1.56 & 1.68 & 1.71 & 2.14 & 2.65 & 3.3 & 3.21 \\ 
  St.dev.  
1990-1999 & 0.05 & 0.13 & 0.04 & 0.03 & 0.04 & 0.04 & 0.07 & 0.2 & 0.14 & 0.1 \\ 
   \hline
Mean  
2000-2009 & 1.75 & 2 & 1.95 & 1.53 & 1.7 & 1.69 & 2.16 & 2.9 & 3.46 & 3.35 \\ 
  Median  
2000-2009 & 1.75 & 1.99 & 1.98 & 1.54 & 1.7 & 1.69 & 2.17 & 2.89 & 3.44 & 3.36 \\ 
  St.dev.  
2000-2009 & 0.03 & 0.06 & 0.09 & 0.02 & 0.01 & 0.03 & 0.09 & 0.07 & 0.1 & 0.05 \\ 
   \hline
Mean  
2010-2019 & 2.12 & 2.19 & 2.43 & 1.83 & 2.01 & 2.05 & 2.45 & 3.24 & 3.72 & 3.63 \\ 
  Median  
2010-2019 & 2.09 & 2.16 & 2.39 & 1.82 & 2 & 2.02 & 2.4 & 3.31 & 3.66 & 3.58 \\ 
  St.dev.  
2010-2019 & 0.35 & 0.26 & 0.44 & 0.28 & 0.24 & 0.32 & 0.26 & 0.2 & 0.33 & 0.22 \\ 
   \hline
\hline
\end{tabular}

\endgroup
\end{table}

\FloatBarrier

\newpage
\subsection{Additional results}

\subsubsection{Volume and time trends}\label{app:tables_moved_down_volume_and_timetrends}
\begin{figure}[H]
\centering
    \caption{AI patents by year (1990-2019)}
    \label{app:fig:timeseries}
    \begin{subfigure}[b]{0.45\textwidth}
        \centering
        \includegraphics[width=0.95\textwidth]{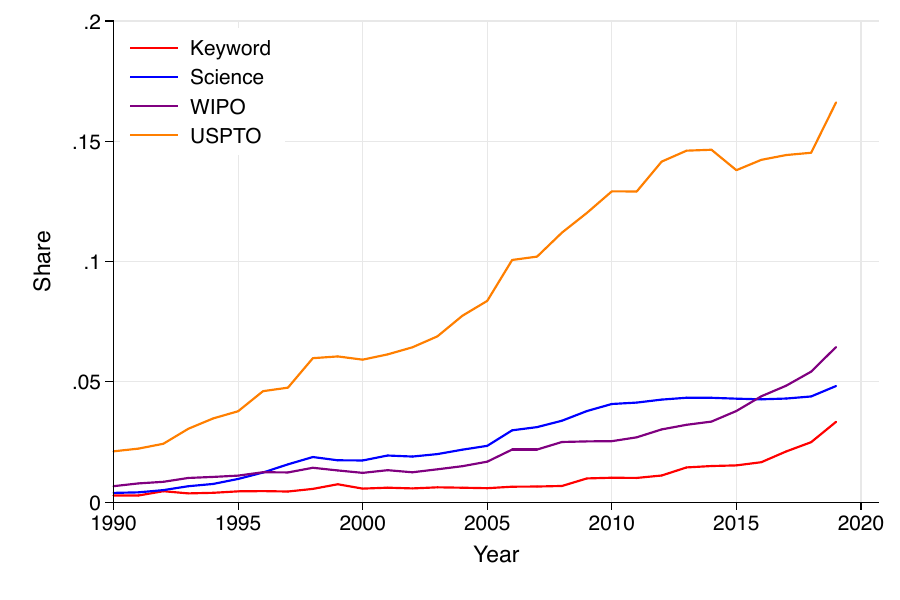}
        \caption{Share of AI patents}
         \label{fig:share_all_patents}
    \end{subfigure}

     \footnotesize\justifying
    Notes: This figure shows the evolution of AI patents over time as identified by the four different approaches, as a share of all US patents granted in the same year. 
\end{figure}


\subsubsection{Generality}\label{app:tables_moved_down_generality}


\begin{table}[H]
\centering
\caption{Average number of citing CPCs (1990-2019): cited patents} 
\label{app:tab:avg_citations_cited}
\begin{tabular}{lcccc}
  \hline
  \hline
  & Keyword & Science & WIPO & USPTO \\[0.25em]
  \hline
  1 digit & 2.71 & 2.42 & 2.41 & 2.26 \\[0.25em]
  3 digit & 5.99 & 5.00 & 5.23 & 4.74 \\[0.25em]
  4 digit & 9.92 & 8.54 & 9.05 & 8.21 \\[0.25em]
  \hline\hline
     \end{tabular}
     \footnotesize 

\footnotesize \justifying
Notes: This table shows the numbers of different CPC classes making a citation to an average patent of the respective group, conditional on the patent being cited at least once. Citations within the same class are excluded.


\end{table}

\begin{table}[H]
\centering
\caption{Average Generality Index (1990-2019) } 
\label{app:tab:generality}
\begin{tabular}{lcccc}
  \hline
  \hline
  & Keyword & Science & WIPO & USPTO \\[0.25em]
  \hline
  1 digit & 0.76 & 0.73 & 0.72 & 0.68 \\[0.25em]
  3 digit & 0.91 & 0.87 & 0.87 & 0.84 \\[0.25em]
  4 digit & 0.96 & 0.95 & 0.94 & 0.93 \\[0.25em]
  \hline\hline
     \end{tabular}
    \footnotesize \justifying
    
    Notes: The generality index is defined as share of citations to patents in different CPC classes at different aggregation levels (see \ref{sec:methods_generality}). Citations within the same class are excluded. 
\end{table}

\begin{table}[H]
\centering
\caption{Average Number of Citing CPCs (1990-2019)} 
\label{app:tab:avg_citations_all}
\begin{tabular}{lcccc}
  \hline
  \hline
  & Keyword & Science & WIPO & USPTO \\[0.25em]
  \hline
  1 digit & 2.15 & 1.83 & 1.82 & 1.68 \\[0.25em]
  3 digit & 5.18 & 4.14 & 4.39 & 3.91 \\[0.25em]
  4 digit & 8.90 & 7.41 & 8.04 & 7.13 \\[0.25em]
  \hline\hline
  \end{tabular}
    \footnotesize \justifying
    
    Notes: This table shows the numbers of different CPC classes making a citation to an average patent of the respective group. Citations within the same class are excluded.
    
\end{table}

\begin{figure}[H]
    \centering
    \caption{Average Number of Classes Citing AI}
    \label{app:fig:extra:avg_classes_citing_AI}   

    \begin{subfigure}{0.45\textwidth}
        \centering
        \includegraphics[width=0.95\textwidth]{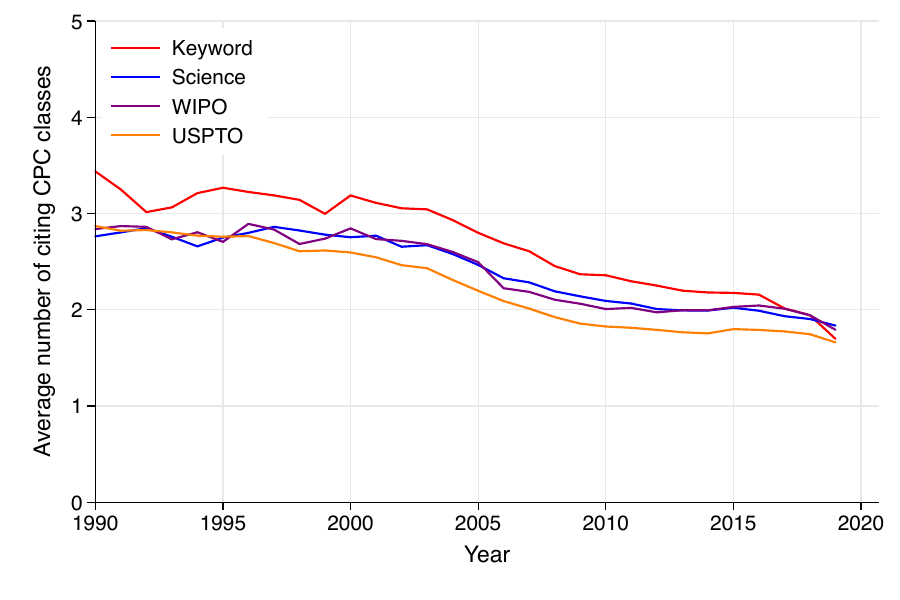}
        \caption{Subset of cited patents}
         \label{fig:avg_class_cit_AI_cited}
    \end{subfigure}
    \begin{subfigure}{0.45\textwidth}
        \centering
        \includegraphics[width=0.95\textwidth]{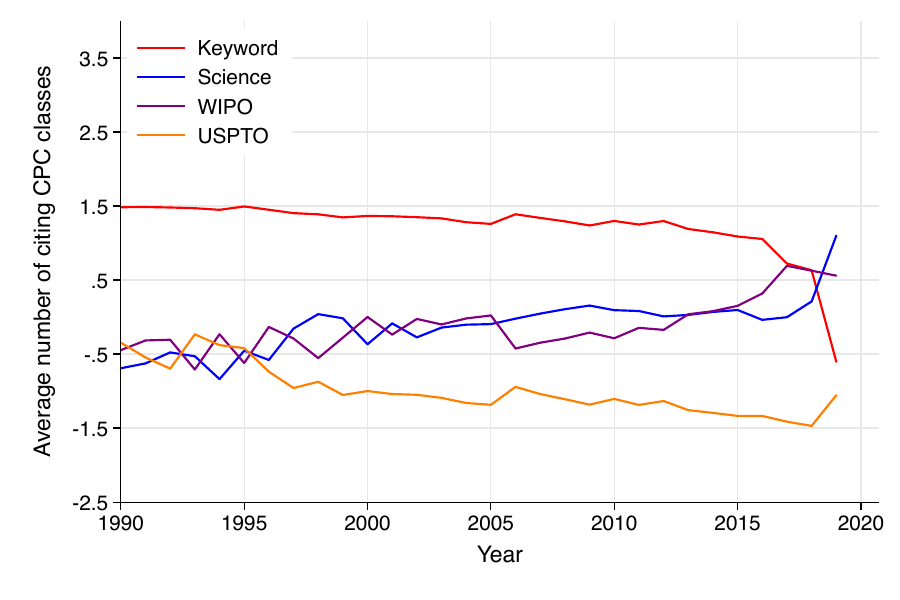}
        \caption{Subset of cited patents (z-score scaled)}
         \label{fig:avg_class_cit_AI_scaled_cited}
    \end{subfigure} 
    
    \footnotesize\justifying \noindent
    Notes: The z-scored value equals the level of the generality index minus its average across the four approaches divided by the standard deviation for each year. 
\end{figure}

\subsubsection{Generality of AI descendants}
\label{app:add_results_generality}
In this section, we report additional results for the wide-ranging usefulness of technological descendants of AI, i.e. those patents that cite an AI patent but are not AI themselves. This serves as an additional indicator of the widespread of AI in a range of different products and processes. The results confirm the persistence of the ranking, indicating the highest generality of keyword patents across different indicators.  

\begin{table}[H]
\centering
\caption{Average Generality Index: AI Descendants}
\begin{tabular}{lcccc}
  \hline
  \hline 
 & Citing Keyword & Citing Science & Citing WIPO & Citing USPTO \\ 
  \hline
  1 digit & 0.74 & 0.73 & 0.72 & 0.72 \\ 
  3 digit & 0.89 & 0.87 & 0.87 & 0.88 \\ 
  4 digit & 0.96 & 0.95 & 0.95 & 0.95 \\ 

   \hline \hline
\end{tabular}
\caption*{Notes: Generality is measured as $G =   1-\sum (s^2)$ with s as share of citations to patents in different CPC classes at different aggregation levels.  Citations within the same class are excluded.} 
\end{table}

\begin{table}[H]
\centering
\caption{Average Number of Citing CPCs: AI Descendants}
\begin{tabular}{lcccc}
  \hline \hline
 & Citing Keyword & Citing Science & Citing WIPO & Citing USPTO \\ 
 \hline
 1 digit & 1.32 & 1.15 & 1.23 & 1.16 \\
 3 digit & 2.75 & 2.27 & 2.57 & 2.33 \\
 4 digit & 4.82 & 3.99 & 4.57 & 4.06 \\

   \hline \hline
\end{tabular}
\caption*{Notes: This table shows the number of different CPC classes making a citation to an average patent of the respective group. Citations within the same class are excluded.} 
\end{table}

\begin{table}[H]
\centering
\caption{Average Number of Citing CPCs (Cited): AI Descendants}
\begin{tabular}{lcccc}
  \hline \hline
 & Citing Keyword & Citing Science & Citing WIPO & Citing USPTO \\ 
 \hline
 1 digit & 2.35 & 2.17 & 2.24 & 2.17 \\
 3 digit & 4.60 & 4.04 & 4.39 & 4.10 \\
 4 digit & 7.51 & 6.60 & 7.29 & 6.63 \\
   \hline \hline
\end{tabular}
\caption*{Notes: The table shows the numbers of different CPC classes making a citation to an average patent of the respective group that receives at least one citation. Citations within the same class are excluded.} 
\end{table}

\begin{table}[ht]
\centering
\caption{Average Citation Lags by Group of AI Citing Patents}
\label{tab:avg_citation_lags_AI_citing}
\begin{tabular}{lcccc}
  \hline \hline
 Period & Citing Keyword & Citing Science & Citing WIPO & Citing USPTO \\ 
  \hline
   1990-1999 & 12.79 & 12.59 & 12.66 & 12.47 \\ 
   2000-2009 & 8.95 & 9.03 & 9.01 & 8.84 \\ 
   2010-2019 & 4.30 & 4.29 & 4.22 & 4.28 \\ 
   \hline
   \hline
\end{tabular}
\caption*{Notes: This table shows the average number of years it takes until a patent in the sample is cited. The average number of years is lower during the more recent decade as the maximal time lag is truncated since our data ends in 2019. Note that the group of AI citing includes all patents that cite AI but are not identified as AI by the respective approach.}
\end{table}

\subsubsection{Complementarity}\label{app:tables_moved_down_complementarity}

\begin{table}[H]
\centering
\caption{Average Number of 1-, 3- and 4-digit CPCs per Patent
}\label{app:cpc_134sum}
        \begin{tabular}{l*{4}{c}}
\hline\hline
            &\multicolumn{1}{c}{Keyword}&\multicolumn{1}{c}{Science}&\multicolumn{1}{c}{WIPO}&\multicolumn{1}{c}{USPTO}\\
\hline
1 digit&        1.39&        1.40&        1.36&        1.27\\
3 digit&        1.61&        1.64&        1.64&        1.43\\
4 digit&        1.88&        1.97&        2.05&        1.64\\
\hline\hline
\end{tabular}

\footnotesize \justifying

Note: The table shows the average of annual average number of technology classes by 1-, 3- or 4-digit CPC per patent. 
\end{table}

\begin{figure}[H]
     \centering
      \caption{Patent-Level Diversity - Average Technology Classes}
        \label{app:diverse_perpatent}
     \begin{subfigure}[b]{0.45\textwidth}
         \centering
         \includegraphics[width=\textwidth]{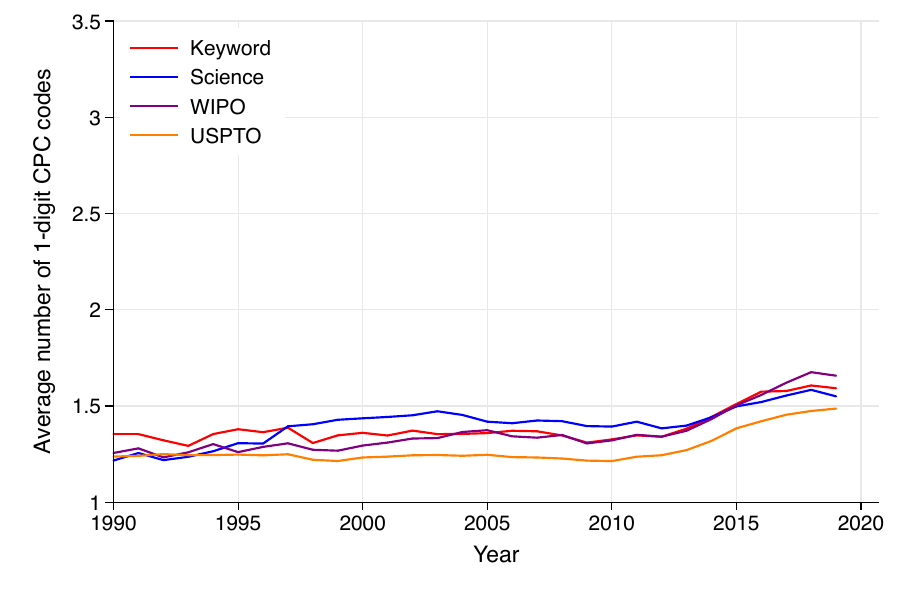}
         \caption{Average number of 1-digit CPC}
         \label{fig:diverse_perpatent_1d}
     \end{subfigure}
     \hfill
\end{figure}  
\FloatBarrier
\subsubsection{Concentration by technology field}
\label{app:concentration_by_field}

\begin{landscape}

\begin{table}[]
\tiny
\begin{tabular}{@{}p{2cm}p{3cm}p{3cm}p{3cm}p{3cm}p{3cm}p{3cm}@{}}
\toprule
 & \multicolumn{1}{c}{{\ul \textbf{CPC A}}} & \multicolumn{1}{c}{{\ul \textbf{CPC B}}} & \multicolumn{1}{c}{{\ul \textbf{CPC C}}} & \multicolumn{1}{c}{{\ul \textbf{CPC G}}} & \multicolumn{1}{c}{{\ul \textbf{CPC H}}} & \multicolumn{1}{c}{{\ul \textbf{CPC Y}}} \\
 \midrule
Keyword & Intuitive Surgical Inc & Fanuc Corp & Applied Materials Inc & \textbf{IBM Corp} & \textbf{IBM Corp} & Fanuc Corp \\
 & iRobot Corp & Honda Motor & Hitachi & \textbf{Microsoft} & Applied Materials Inc & Samsung \\
 & Siemens & Seiko Epson & SYMYX TECH INC & \textbf{Google} & Samsung & iRobot Corp \\
 & Samsung & Yaskawa Elect Corp & EBARA CORP & Samsung & \textbf{Microsoft} & \textbf{IBM Corp} \\
 & TECH HOLDINGS CORP & iRobot Corp & General Electric & Fanuc Corp & \textbf{Google} & Boeing \\
{\color[HTML]{FF0000} \textbf{HHI  (Keyword)}} & {\color[HTML]{FF0000} \textbf{0,0139}} & {\color[HTML]{FF0000} \textbf{0,0079}} & {\color[HTML]{FF0000} \textbf{0,0102}} & {\color[HTML]{FF0000} \textbf{0,0100}} & {\color[HTML]{FF0000} \textbf{0,0079}} & {\color[HTML]{FF0000} \textbf{0,0071}} \\
\midrule
Science & GENENTECH INC & SILVERBROOK RESEARCH PTY LTD & GENENTECH INC & \textbf{IBM Corp} & \textbf{Microsoft} & \textbf{IBM Corp} \\
 & HUMAN GENOME SCIENCES INC & iRobot Corp & HUMAN GENOME SCIENCES INC & \textbf{Microsoft} & \textbf{IBM Corp} & SILVERBROOK RESEARCH PTY LTD \\
 & Siemens & AUTOMOTIVE TECH INT INC & ZYMOGENETICS INC & \textbf{Google} & \textbf{Google} & \textbf{Microsoft} \\
 & ZYMOGENETICS INC & FORD GLOBAL TECH LLC & MILLENNIUM PHARMACEUTICALS INC & Apple & AT\&T & DEXCOM INC \\
 & \textbf{Microsoft} & MAGNA ELECT INC & EI DU PONT DE NEMOURS \& CO & Sony & Sony & MONSANTO TECH LLC \\
{\color[HTML]{FF0000} \textbf{HHI (Science)}} & {\color[HTML]{FF0000} \textbf{0,0035}} & {\color[HTML]{FF0000} \textbf{0,0080}} & {\color[HTML]{FF0000} \textbf{0,0044}} & {\color[HTML]{FF0000} \textbf{0,0161}} & {\color[HTML]{FF0000} \textbf{0,0108}} & {\color[HTML]{FF0000} \textbf{0,0051}} \\
\midrule
WIPO & Siemens & Toyota & SILVERBROOK RESEARCH PTY LTD & \textbf{IBM Corp} & \textbf{IBM Corp} & \textbf{IBM Corp} \\
 & Sony & FORD GLOBAL TECH LLC & Applied Materials Inc & \textbf{Microsoft} & \textbf{Microsoft} & Toyota \\
 & Samsung & Honda Motor & UBIOME INC & \textbf{Google} & \textbf{Google} & SILVERBROOK RESEARCH PTY LTD \\
 & \textbf{Microsoft} & Genral Motors & ROCHE MOLECULAR SYSTEMS INC & Canon & Sony & Genral Motors \\
 & \textbf{IBM Corp} & SILVERBROOK RESEARCH PTY LTD & NANOGEN INC & Samsung & Samsung & \textbf{Google} \\
{\color[HTML]{FF0000} \textbf{HHI (WIPO)}} & {\color[HTML]{FF0000} \textbf{0,0049}} & {\color[HTML]{FF0000} \textbf{0,0088}} & {\color[HTML]{FF0000} \textbf{0,0190}} & {\color[HTML]{FF0000} \textbf{0,0117}} & {\color[HTML]{FF0000} \textbf{0,0084}} & {\color[HTML]{FF0000} \textbf{0,0069}} \\
\midrule
USPTO & Siemens & Genral Motors & AFFYMETRIX INC & \textbf{IBM Corp} & \textbf{IBM Corp} & \textbf{IBM Corp} \\
 & Philips & FORD GLOBAL TECH LLC & Applied Materials Inc & \textbf{Microsoft} & \textbf{Microsoft} & Intel \\
 & General Electric & Toyota & EI DU PONT DE NEMOURS \& CO & \textbf{Google} & AT\&T & General Electric \\
 & MEDTRONIC INC & Robert Bosch & SYNGENTA PARTICIPATIONS AG & Intel & \textbf{Google} & Genral Motors \\
 & Sony & Honda Motor & MONSANTO TECH LLC & Hewlett Packard & Samsung & FORD GLOBAL TECH LLC \\
{\color[HTML]{FF0000} \textbf{HHI (USPTO)}} & {\color[HTML]{FF0000} \textbf{0,0059}} & {\color[HTML]{FF0000} \textbf{0,0056}} & {\color[HTML]{FF0000} \textbf{0,0020}} & {\color[HTML]{FF0000} \textbf{0,0172}} & {\color[HTML]{FF0000} \textbf{0,0102}} & {\color[HTML]{FF0000} \textbf{0,0097}} \\ \bottomrule
\end{tabular}

\caption{List of top-5 firms and HHI (by CPC section)}
\label{tab:top5_by_CPC1}
\end{table}

\end{landscape}

\begin{landscape}
\begin{table}[]
\tiny
\begin{tabular}{@{}p{2cm}p{3cm}p{3cm}p{3cm}p{3cm}p{3cm}@{}}
\toprule
 & \textbf{H04 (ELECTRIC COMMUNICATION)} & \textbf{G06 (COMPUTING)} & \textbf{A61 (MEDICAL \& VETERINARY SCIENCE)} & \textbf{G01 (MEASURING; TESTING)} & \textbf{Y02 (CLIMATE TECHNOLOGIES)} \\
 
 \midrule
Keyword & \textbf{IBM Corp} & \textbf{IBM Corp} & Intuitive Surgical Inc & Samsung & Fanuc Corp \\
 & \textbf{Microsoft} & \textbf{Microsoft} & Siemens & Siemens & Samsung \\
 & Samsung & \textbf{Google} & ETHICON LLC & Boeing & iRobot Corp \\
 & \textbf{Google} & Amazon & HANSEN MEDICAL INC & General Electric & Boeing \\
 & Intel & Intel & Philips & \textbf{IBM Corp} & \textbf{IBM Corp} \\
{\color[HTML]{FF0000} \textbf{HHI  (Keyword)}} & {\color[HTML]{FF0000} \textbf{0,01167}} & {\color[HTML]{FF0000} \textbf{0,01699}} & {\color[HTML]{FF0000} \textbf{0,02382}} & {\color[HTML]{FF0000} \textbf{0,00337}} & {\color[HTML]{FF0000} \textbf{0,00727}} \\

 \midrule
Science & \textbf{Microsoft} & \textbf{IBM Corp} & GENENTECH INC & \textbf{Microsoft} & SILVERBROOK RESEARCH PTY LTD \\
 & \textbf{IBM Corp} & \textbf{Microsoft} & HUMAN GENOME SCIENCES INC & Siemens & \textbf{IBM Corp} \\
 & \textbf{Google} & \textbf{Google} & Siemens & \textbf{IBM Corp} & \textbf{Microsoft} \\
 & Sony & Apple & ZYMOGENETICS INC & \textbf{Google} & DEXCOM INC \\
 & AT\&T & Sony & Philips & GENENTECH INC & MONSANTO TECH LLC \\
{\color[HTML]{FF0000} \textbf{HHI (Science)}} & {\color[HTML]{FF0000} \textbf{0,01207}} & {\color[HTML]{FF0000} \textbf{0,02173}} & {\color[HTML]{FF0000} \textbf{0,00392}} & {\color[HTML]{FF0000} \textbf{0,00264}} & {\color[HTML]{FF0000} \textbf{0,00522}} \\

 \midrule
WIPO & \textbf{IBM Corp} & \textbf{IBM Corp} & Siemens & \textbf{Google} & \textbf{IBM Corp} \\
 & \textbf{Microsoft} & \textbf{Microsoft} & Philips & \textbf{IBM Corp} & Toyota \\
 & \textbf{Google} & \textbf{Google} & \textbf{IBM Corp} & Toyota & SILVERBROOK RESEARCH PTY LTD \\
 & Sony & Canon & CARDIAC PACEMAKERS INC & Boeing & Genral Motors \\
 & AT\&T & Sony & Samsung & Honeywell & Google \\
{\color[HTML]{FF0000} \textbf{HHI (WIPO)}} & {\color[HTML]{FF0000} \textbf{0,00928}} & {\color[HTML]{FF0000} \textbf{0,01345}} & {\color[HTML]{FF0000} \textbf{0,00575}} & {\color[HTML]{FF0000} \textbf{0,00396}} & {\color[HTML]{FF0000} \textbf{0,00707}} \\

 \midrule
USPTO & \textbf{IBM Corp} & \textbf{IBM Corp} & Siemens & \textbf{IBM Corp} & \textbf{IBM Corp} \\
 & \textbf{Microsoft} & \textbf{Microsoft} & Philips & Siemens & Intel \\
 & AT\&T & \textbf{Google} & General Electric & General Electric & Genral Motors \\
 & \textbf{Google} & Intel & MEDTRONIC INC & SCHLUMBERGER TECH CORP & General Electric \\
 & Samsung & Hewlett Packard & CARDIAC PACEMAKERS INC & Hitachi & FORD GLOBAL TECH LLC \\
{\color[HTML]{FF0000} \textbf{HHI (USPTO)}} & {\color[HTML]{FF0000} \textbf{0,01109}} & {\color[HTML]{FF0000} \textbf{0,02263}} & {\color[HTML]{FF0000} \textbf{0,00876}} & {\color[HTML]{FF0000} \textbf{0,00409}} & {\color[HTML]{FF0000} \textbf{0,00970}}\\
\bottomrule
\end{tabular}
\caption{List of top-5 firms and HHI (by selected top 3-digit CPC)}
\label{tab:top5_by_CPC3}
\end{table}
\end{landscape}

\end{document}